\documentclass[fleqn,usenatbib]{mnras}

\usepackage{newtxtext,newtxmath}

\usepackage[T1]{fontenc}

\DeclareRobustCommand{\VAN}[3]{#2}
\let\VANthebibliography\thebibliography
\def\thebibliography{\DeclareRobustCommand{\VAN}[3]{##3}\VANthebibliography}

\usepackage{graphicx}	
\usepackage{amsmath}	

\usepackage{pdflscape}
\DeclareUnicodeCharacter{2212}{-}
\DeclareUnicodeCharacter{2032}{'}
\usepackage{comment}
\usepackage{subcaption}
\usepackage{wrapfig}
\usepackage{float}
\usepackage{blindtext}
\usepackage{array}

\usepackage{amsmath}
\let\oldAA\AA
\renewcommand{\AA}{\text{\normalfont\oldAA}}
\usepackage{tablefootnote}
\usepackage[referable]{threeparttablex}
\usepackage{hyperref}
\usepackage{multirow}

\newcommand{\gs}{\mathrel{\lower0.6ex\hbox{$\buildrel {\textstyle >}
 \over {\scriptstyle \sim}$}}}
\newcommand{\ls}{\mathrel{\lower0.6ex\hbox{$\buildrel {\textstyle <}
 \over {\scriptstyle \sim}$}}}

\newcommand{\lta}{\mathrel{\spose{\lower 3pt\hbox{$\mathchar"218$}}
     \raise 2.0pt\hbox{$\mathchar"13C$}}}
\newcommand{\gta}{\mathrel{\spose{\lower 3pt\hbox{$\mathchar"218$}}
     \raise 2.0pt\hbox{$\mathchar"13E$}}}

\newcommand{\oiiia}{\mbox{[O\,{\textsc{iii}}]}\,}

\newcommand{\nii}{\mbox{[N\,{\sc ii]}}\,}
\newcommand{\niil}{\mbox{[N\,{\sc ii]\sc{$\lambda$6548}}}\,}
\newcommand{\niih}{\mbox{[N\,{\sc ii]\sc{$\lambda$6585}}}\,}

\title[Search for H$\alpha$ emitters at $z > 6$]{The \emph{JWST} Emission Line Survey (JELS): An untargeted search for H$\alpha$ emission line galaxies at $z > 6$ and their physical properties}

\author[C. A. Pirie et al.]{C. A. Pirie,$^{1}$\thanks{E-mail: corey.pirie@ed.ac.uk (CAP)}
P. N. Best,$^{1}$
K. J. Duncan,$^{1}$
D. J. McLeod,$^{1}$
R. K. Cochrane,$^{1,2,3}$
M. Clausen,$^{1}$
J. S. Dunlop,$^{1}$\newauthor
S. R. Flury,$^{1}$
J. E. Geach,$^{4}$
C. L. Hale,$^{5,1}$
E. Ibar,$^{6,7}$
R. Kondapally,$^{1,8}$
 Zefeng Li,$^{8}$
J. Matthee,$^{9}$
R. J. McLure,$^{1}$\newauthor
L. Ossa-Fuentes,$^{6}$
A. L. Patrick,$^{1}$
Ian Smail,$^{8}$
D. Sobral,$^{10,11}$
H. M. O. Stephenson,$^{12}$ 
J. P. Stott,$^{12}$ and \newauthor
A. M. Swinbank$^{8}$
\vspace{0.2cm}\\
$^{1}$Institute for Astronomy, University of Edinburgh, Royal Observatory, Blackford Hill, Edinburgh, EH9 3HJ, UK\\
$^{2}$Department of Astronomy, Columbia University, New York, NY 10027, USA\\
$^{3}$Jodrell Bank Centre for Astrophysics, Alan Turing Building, University of Manchester, Oxford Road, Manchester M13 9PL, UK\\
$^{4}$Centre for Astrophysics Research, School of Physics, Engineering and Computer Science, University of Hertfordshire, College Lane, Hatfield AL10 9AB, UK\\
$^{5}$Astrophysics, Department of Physics, University of Oxford, Denys Wilkinson Building, Keble Road, Oxford, OX1 3RH, UK\\
$^{6}$Instituto de F\'isica y Astronom\'ia, Universidad de Valpara\'iso, Avda. Gran Breta\~na 1111, Valpara\'iso, Chile\\
$^{7}$Millennium Nucleus for Galaxies (MINGAL)\\
$^{8}$Centre for Extragalactic Astronomy, Department of Physics, Durham University, South Road, Durham DH1 3LE, UK\\
$^{9}$Institute of Science and Technology Austria (ISTA), Am Campus 1, 3400 Klosterneuburg, Austria\\
$^{10}$Departamento de F\'isica, Faculdade de Ci\`encias, Universidade de Lisboa, Edif\'icio C8, Campo Grande, PT1749-016 Lisbon, Portugal\\
$^{11}$BNP Paribas Corporate \& Institutional Banking, Torre Ocidente Rua Galileu Galilei, 1500-392 Lisbon, Portugal\\
$^{12}$Department of Physics, Lancaster University, Lancaster LA1 4YB, UK\\
}

\date{Accepted XXX. Received YYY; in original form ZZZ}

\pubyear{2025}

\begin{document}
\label{firstpage}
\pagerange{\pageref{firstpage}--\pageref{lastpage}}
\maketitle

\begin{abstract}
We present the first results of the \emph{JWST} Emission Line Survey (JELS). Utilising the first NIRCam narrow-band imaging at 4.7$\mu$m, over 63 arcmin$^{2}$ in the PRIMER/COSMOS field, we have identified 609 emission line galaxy candidates. From these, we robustly selected 35 H$\alpha$ star-forming galaxies at $z \sim 6.1$, with H$\alpha$ star-formation rates ($\rm{SFR_{H\alpha}}$) of $\sim0.9-15\ \rm{M_{\odot} \ yr^{-1}}$. Combining our unique H$\alpha$ sample with the exquisite panchromatic data in the field, we explored their physical properties and star-formation histories, and compared these to a broad-band selected sample at $z\sim 6$ which has offered vital new insights into the nature of high-redshift galaxies. UV-continuum slopes ($\beta$) were considerably redder for our H$\alpha$ sample ($\langle\beta\rangle\sim-1.92$) compared to the broad-band sample ($\langle\beta\rangle\sim-2.35$). This was not due to dust attenuation as our H$\alpha$ sample was relatively dust-poor (median $A_V=0.23$); instead, we argue that the reddened slopes could be due to nebular continuum. We compared $\rm{SFR_{H\alpha}}$ and the UV-continuum-derived $\rm{SFR_{UV}}$ to SED-fitted measurements averaged over canonical timescales of 10 and 100 Myr ($\rm{SFR_{10}}$ and $\rm{SFR_{100}}$). We found an increase in recent SFR for our sample of H$\alpha$ emitters, particularly at lower stellar masses ($<10^9 \ \rm{M_{\odot}}$). We also found that $\rm{SFR_{H\alpha}}$ strongly traces SFR averaged over 10 Myr timescales, whereas the UV-continuum over-predicts SFR on 100 Myr timescales at low stellar masses. These results point to our H$\alpha$ sample undergoing `bursty' star formation. Our F356W $z \sim 6$ sample showed a larger scatter in $\rm{SFR_{10}/SFR_{100}}$ across all stellar masses, which has highlighted how narrow-band photometric selections of H$\alpha$ emitters are key to quantifying the burstiness of star-formation activity.
\end{abstract}




\section{INTRODUCTION}
\label{sec:intro}

Arguably one of the most fundamental measurements in galaxy formation and evolution studies is the evolution of the cosmic star-formation rate density ($\rho_{\rm{SFR}}$) across cosmic history \citep[see comprehensive review from][]{2014ARA&A..52..415M}. In addition, it is important to study the distribution function of $\rho_{\rm{SFR}}$ amongst the galaxy population as a function of galaxy properties such as stellar mass, morphology and  metallicity \citep[e.g.][]{2014MNRAS.437.3516S,2020MNRAS.491..944C,2023ApJS..269...33N}, as well as galaxy environment, at different redshift epochs. Determining $\rho_{\rm{SFR}}$ at any given epoch require selections of large and unbiased samples of star-forming galaxies, with accurately determined star-formation rates (SFR). 

In the decade prior to the launch of the James Webb Space Telescope (\emph{JWST}), extragalactic surveys, combining observations from both the Hubble Space Telescope (\emph{HST}) and the Spitzer Space Telescope, mapped the evolution of star-forming galaxies and their contribution to $\rho_{\rm{SFR}}$ beyond the peak star-forming epoch at `cosmic noon' ($z \sim 2$), right the way out to redshifts $z$ $\sim$ 9 \citep[e.g][]{2013ApJ...763L...7E,2013MNRAS.432.2696M,2015ApJ...810...71F,2015MNRAS.450.3032M,2016MNRAS.459.3812M,2018ApJ...855..105O,2021AJ....162...47B,2022ApJ...940...55B}. Here, the selection of high-redshift galaxies utilised their rest-frame ultra-violet (UV) emission, either by direct Lyman continuum break colour selection or by using photometric redshift (photo-$z$) measurements which were strongly driven by the position of the Lyman break feature. As a star-formation tracer, the rest-UV wavelength regime traces emission from the brightest, short-lived ($\leq 100$\,Myr) stellar populations within galaxies and so traces recent star formation, but is also significantly attenuated by dust. 

At cosmic noon, star formation in galaxies, and hence measurements of the $\rho_{\rm{SFR}}$, are dust-obscured by around 85 per cent on average \citep[e.g.][]{2013MNRAS.434.3218I,2017MNRAS.466..861D,2017ApJ...838..119T,2024A&A...681A.118T} and this becomes particularly important at high stellar masses \citep[e.g.][]{2017ApJ...850..208W,2023ApJ...950....7S}. The impact of dust extinction appears to decline at higher redshifts, with the Universe transitioning from primarily obscured star formation at $z$ $\lesssim$ 4 to primarily dust-unobscured star formation at $z$ $\gtrsim$ 5 \citep{2017MNRAS.466..861D,2020ApJ...902..112B}. However, the impact of dust on our picture of galaxy formation and evolution at high redshift may still be significant \citep[e.g.][]{2014MNRAS.438.1267S,2020A&A...643A...8G,2022MNRAS.510.5088B,2023ApJ...943L...9Z,2023A&A...671A.105A,2023MNRAS.518.6142A}; where samples arising from rest-frame UV-driven photo-$z$ selections could be biased, with measurements depending on the prior assumptions on the UV-continuum slopes and emission line properties \citep{2023Natur.622..707A,2023ApJ...958..141L}. Therefore, this selection technique carries a high risk that systematic effects may affect both the selection of the galaxy sample and the determination of their physical properties \citep[c.f.][]{2015MNRAS.452.2018O}.

Since the launch of the \emph{JWST} \citep{2023AAS...24110001R}, the high-redshift frontier has been extended with the discovery of $z$ $>$ 10 star-forming galaxies through their rest-fame UV emission, with samples allowing the measurement of $\rho_{\rm{SFR}}$ out to $z \lesssim 15$ \citep[e.g.][]{2022ApJ...940L..14N,2023ApJS..265....5H,2023MNRAS.518.6011D,2023MNRAS.520.4554D,2024MNRAS.533.3222D,2024MNRAS.527.5004M,2024ApJ...965..169A}. These selected high-redshift star-forming galaxies have then been targeted for follow-up multi-object and IFU spectroscopy using the NIRSpec instrument \citep{2022A&A...661A..80J} to study their physical properties in great detail, including their ionised gas through measurements of the H$\alpha$ emission line up to $z$ $\sim$ 6.5 \citep[e.g.] []{2023A&A...677A.115C,2023ApJ...950L...1S,2025ApJ...980..242S,2023ApJ...955...54S,2024ApJ...962...24S,2024ApJ...976..193R}. The H$\alpha$ emission line is a well-calibrated SFR indicator, providing a clean and highly-sensitive selection criterion which is less attenuated by dust compared to the UV-continuum. This emission line traces the recombination of gas ionised by UV photons from star-forming regions and traces star-formation activity timescales of $\leq$ 10Myr. 

Studying the star-formation properties of galaxies by using their emission lines can provide a complementary view of cosmic star-formation, but only if the biases that might be associated with the pre-selection of rest-frame UV/optical selected galaxies for targeted spectroscopic surveys are avoided. To achieve this, untargeted selection of high-redshift emission line galaxies, such as those selected based on their H$\alpha$ emission, is optimal. \emph{JWST} has achieved this using a variety of different techniques.
The NIRCam instrument \citep{2005SPIE.5904....1R,2023PASP..135b8001R} has been used to perform both slitless grism spectroscopy \cite[e.g. ][]{2023ApJ...950...67M,2023MNRAS.525.2864O} and medium-band imaging surveys \citep[e.g.][]{2023ApJS..268...64W,2023ApJ...952..143R}. Both techniques have enabled the selection of H$\alpha$-emitting galaxies beyond cosmic noon and found them to be compact in nature, with some exhibiting signs of hosting an active galactic nucleus \citep[AGN; e.g.][]{2024arXiv240919205G}. In addition, the first measurements of the H$\alpha$ luminosity functions (LF) have now been made out to the Epoch of Reionization at $z > 6$ \citep{2023ApJ...953...53S,2025A&A...694A.178C,2025arXiv250303829F}. However, these surveys are restricted to sources with higher line fluxes or equivalent widths ($EW$). 

In this paper, we explore the complementary approach of a narrow-band survey, selecting emission line galaxies in a clean and sensitive way, utilising the ability to separate line emission from continuum emission. The narrow-band technique has been used to select emission line galaxies at lower-redshifts \citep[e.g the HiZELS Survey;] []{2008MNRAS.388.1473G,2013MNRAS.428.1128S,2013ASSP...37..235B}. Despite narrow-band selections probing smaller volumes than either grism spectroscopy or medium-band imaging, narrow-band surveys have several advantages. Firstly, narrow-band filter sensitivities allow measurements of fainter line fluxes (reaching a factor of 2-4 fainter than other approaches). Secondly, the filter widths allow a clean selection of emission line galaxies across cosmic time based on line strength only (which traces the SFR) without bias towards sources with high-$EW$. Thirdly, the narrow-band images provide a direct, spatially-resolved map of ionised gas (star formation) across the galaxies.

The capability to conduct wide-area and high-sensitivity observations has enabled ground-based narrow-band imaging to select robust samples of 100s of $\rm{H\alpha}$ emitters out to $z \sim 2.2$ and map out their evolution over cosmic time \citep[e.g.][]{2008MNRAS.388.1473G,2010A&A...509L...5H,2011MNRAS.411..675S,2013MNRAS.428.1128S,2014MNRAS.437.3516S,2017MNRAS.469.2913C,2018MNRAS.475.3730C,2020ApJ...892...30H}. Narrow-band selections of $\rm{H\alpha}$ emission line galaxies have proven successful, with follow-up spectroscopic observations confirming the identity and redshifts of these emission line galaxies, as well as providing further information on their physical properties \citep[e.g.][]{2012MNRAS.426..935S,2013MNRAS.436.1130S,2014MNRAS.443.2695S,2017MNRAS.466..892M,2021MNRAS.503.2622C}. Importantly, \emph{JWST} offers the opportunity to extend this approach back to the Epoch of Reionization and gain a complete picture of the evolution of emission line galaxies across cosmic time. The NIRCam instrument is able to overcome the redshift limitation from the ground, thanks to the long-channel narrow-band filters, which probe wavelengths out to 4.7\,$\rm{\mu m}$: F466N and F470N. Observations in these filters are part of the `\emph{JWST} Emission Line Survey' (JELS; GO 2321; PI: Philip Best). 

The JELS narrow-band observations were taken in the F466N, F470N and F212N filters \citep[as outlined in][]{2024arXiv241009000D} to select H$\alpha$ emission line galaxies at $z$ = 6.1 and $z$ = 2.2 in an untargeted manner across cosmic history. Multi-wavelength ancillary imaging from the public \emph{JWST} Treasury Programme, Public Release IMaging for Extragalactic Research \citep[PRIMER; GO 1837; PI:][]{2021JWST.prop.1837D}, aided in the physical characterisation of these emission line-selected sources. This wealth of multi-wavelength ancillary data crucially provides access to the rest-frame UV-continuum; together with the H$\alpha$ line measurements, this enables studies of the timescales of star-formation activity in these narrow-band selected galaxies. 

This paper focuses on the $z = 6.1$ JELS population, and in particular utilises both H$\alpha$-to-UV comparisons and spectral energy distribution (SED) fitting to study the physical properties, star formation histories, and dust properties of the galaxies. Previous studies of the H$\alpha$-to-UV luminosity ratio in star-forming galaxies have found that higher ratios are indicative of higher dust extinction and, potentially, show a dependence on metallicity \citep[e.g.][]{2019ApJ...871..128T}. However, this ratio also depends on recent `burstiness' of star-formation activity, due to the different response timescales of the two indicators to changes in instantaneous SFR \citep[e.g.][]{2012ApJ...744...44W,2019ApJ...884..133F,2019ApJ...881...71E,2022MNRAS.511.4464A}. The `burstiness' of star-forming galaxies at high-redshift likely impacts the inferred galaxy properties \citep{2023MNRAS.524.2312E,2024MNRAS.527.6139S} and observed populations \citep[e.g. UV luminosity functions;][]{2023MNRAS.526.2665S,2023ApJ...955L..35S}.
Therefore, comparing the selections and properties of H$\alpha$ emission line selected star-forming galaxies to those selected from rest-frame UV-continuum based indicators is important when considering the whole star-forming galaxy population over cosmic time, and JELS provides the opportunity to investigate this.

The paper is laid out as follows. In Section~\ref{sec:obs}, we outline the datasets utilised in this analysis, including the JELS narrow-band observations and the additional multi-wavelength ancillary data sets from both \emph{JWST} and \emph{HST} instruments. In Section~\ref{sec:narrowband_cat}, we describe the methodology for creating the multi-wavelength detection catalogues, including performing forced photometry using the ancillary multi-wavelength imaging, the catalogue cleaning steps, and the photometric redshift analysis for the detected sources. We present the narrow-band excess source selection criteria and the resulting sample of emission line galaxy candidates, focusing on the $z$ $>$ 6 $\rm{H\alpha}$ emission line galaxy sample, in Section~\ref{sec:line_cat}. In Section~\ref{sec:physical_properties}, we explore the physical properties of the H$\alpha$ emission line sample, such as their stellar masses, star-formation rates (SFRs), and dust content, and compare these against a broad-band photo-$z$-selected comparison sample at the same epoch. We then discuss the nature of these H$\alpha$ emission line galaxies in the context of previous results from observations and simulations, focusing on impact of dust attenuation and star-formation activity, in Section~\ref{sec:discussion}. Finally, we draw conclusions in Section~\ref{sec:conclusion}. We adopt the following cosmological parameters in all cosmological calculations: $H_{\rm{0}}$ = 70 km s$^{−1}$ Mpc$^{−1}$, $\Omega_{\rm{M}}$ = 0.3 and $\Omega_{\rm{\Lambda}}$ = 0.7. All magnitudes are in the AB system \citep{1974ApJS...27...21O,1983ApJ...266..713O}.

\begin{figure}
    \centering
    \includegraphics[width=0.47\textwidth]{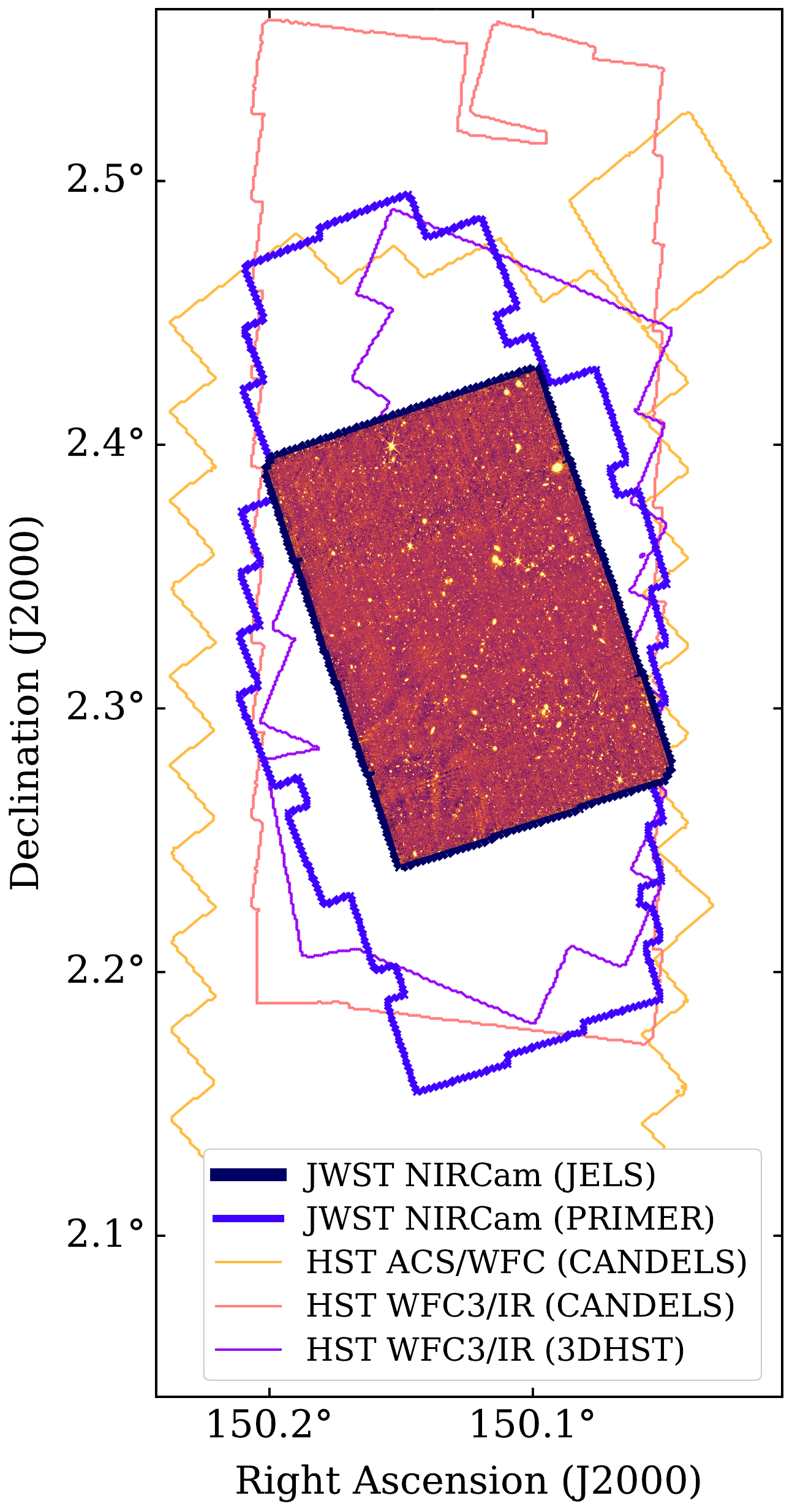}

    \caption{The JELS narrow-band image survey footprint and corresponding F466N image in the COSMOS field taken from \emph{JWST}/NIRCam at 4.7$\rm{\mu}$m. In addition, the multi-wavelength ancillary imaging survey footprints include: i) PRIMER utilising \emph{JWST}/NIRCam imaging in eight filters from 0.9 to 4.5$\rm{\mu}$m, ii) 3D-\emph{HST} utilising \emph{HST} WFC3/IR imaging at 1.4$\rm{\mu}$m, iii) CANDELS utilising \emph{HST} WFC3/IR imaging in three filters from 1.25 to 1.6$\rm{\mu}$m and iv) CANDELS utilising \emph{HST} ACS/WFC imaging in two filters from 0.6 to 0.8$\rm{\mu}$m (see survey and filter descriptions in Table \ref{table:multiwav_data}).}
    
    \label{fig:jels_footprint}

\end{figure}

\begin{figure*}
    \centering
    \includegraphics[width=1.0\textwidth]{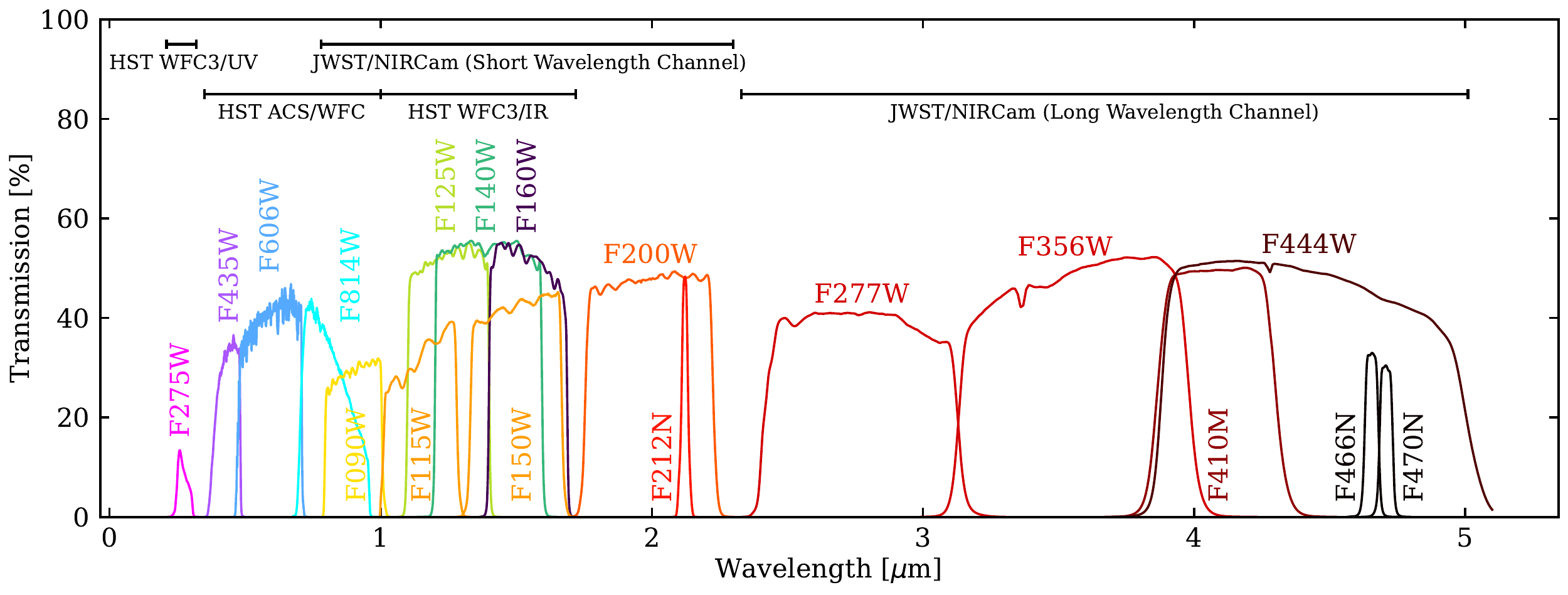}

    \caption{The broad and narrow-band filter profiles and associated transmission for the \emph{JWST} and \emph{HST} filters from the COSMOS-CANDELS, PRIMER and JELS surveys. Note, the F212N filter (used for the short-wavelength channel JELS observations) is included for completeness but was not utilised in this analysis.}
    
    \label{fig:filters}

\end{figure*}

\begin{table*}
\centering
  \caption{Key properties of the multi-wavelength data in the COSMOS-CANDELS field. For each filter, we include the pivot wavelength $\lambda_{\rm{pivot}}$, effective width $\Delta \lambda$, median PSF FWHM (assuming a Moffat profile), 5$\sigma$ depths (estimated from the variance of empty, source free 0.3 arcsec diameter apertures), the filter dependent Galactic extinction values $A_{\rm{filter}}$/$E(B−V)$, and the approximate area covered by each survey.}
  \label{table:multiwav_data}
  \begin{tabular}{c c c c c c c c c}
    \hline
    \hline
    Survey/Detector & Filter & $\lambda_{\rm{pivot}}$ & $\Delta \lambda$ & PSF FWHM & 5$\sigma$ Depth & $A_{\mathrm{filter}}$/$E$($B$ − $V$) & Area\\
    & & [$\rm{\mu}$m] & [$\rm{\mu}$m] & [arcsec] & [mag] & [mag] & [arcmin$^{2}$]\\
    \hline
    \emph{HST} \ WFC3/UVIS & & & & & & & \\
    \hline
    & & & & & & & \\
    UVCANDELS & F275W & 0.2702 & 0.0416 & 0.09 & 27.08 & 6.19 & 117\\
    & & & & & & & \\
    \hline
    \emph{HST} \ ACS/WFC & & & & & & & \\
    \hline
    & & & & & & & \\
    UVCANDELS & F435W & 0.4330 & 0.0822 & 0.11 & 28.08 & 4.09 & 143\\
    & & & & & & & \\
    CANDELS + 3D-\emph{HST} & F606W & 0.5922 & 0.1772 & 0.09 & 28.00 & 2.71 & 278\\
     & F814W & 0.8046 & 0.1889 & 0.10 & 27.98 & 1.68\\
     & & & & & & & \\
    \hline
    \emph{HST} \ WFC3/IR & & & & & & & \\
    \hline
    & & & & & \\
    CANDELS & F125W & 1.2486 & 0.2674 & 0.12 & 27.65 & 0.80 & 203\\
     & F160W & 1.5370 & 0.2750 & 0.18 & 27.72 & 0.57\\
     & & & & & & & \\
    3D-\emph{HST} & F140W & 1.3923 & 0.3570 & 0.15 & 27.02 & 0.66 & 124\\
    & & & & & & & \\
    \hline
    \emph{JWST}/NIRCam & & & & & & &\\
    \hline
    & & & & & & & \\
    PRIMER & F090W & 0.9021 & 0.1773 & 0.05 & 27.95 & 1.40 & 144\\
    & F115W & 1.1543 & 0.2055 & 0.06 & 27.96 & 0.92\\
    & F150W & 1.5007 & 0.2890 & 0.06 & 28.15 & 0.59\\
    & F200W & 1.9886 & 0.4190 & 0.07 & 28.34 & 0.39\\
    & F277W & 2.7617 & 0.6615 & 0.11 & 28.71 & 0.25\\
    & F356W & 3.5684 & 0.7239 & 0.13 & 28.87 & 0.19\\
    & F410M & 4.0822 & 0.4263 & 0.15 & 28.17 & 0.16\\
    & F444W & 4.4043 & 1.0676 & 0.15 & 28.51 & 0.14\\
    & & & & & \\
    JELS & F466N & 4.6541 & 0.0535 & 0.17 & 26.28 & 0.14 & 63\\
    & F470N & 4.7078 & 0.0510 & 0.17 & 26.24 & 0.14\\
    & & & & & \\
    \hline
  \end{tabular}
\centering
\end{table*}

\section{Observations and current datasets}
\label{sec:obs}

The JELS narrow-band observations (see survey footprint in Fig. \ref{fig:jels_footprint}) are described in full in \citet{2024arXiv241009000D}. In summary, we utilised the \emph{JWST}/NIRCam long-wavelength filters F466N and F470N (see Table \ref{table:multiwav_data} for filter properties and Fig. \ref{fig:filters} for filter transmission curves). In parallel, we observed the same field in the short-wavelength channel using the F212N ($\lambda_{\rm{pivot}}$ = 2.1213 $\rm{\mu}$m and $\Delta \lambda$ = 0.0274 $\rm{\mu}m$) and F200W filters, but these data were not considered in this paper \citep[see discussion in][]{2024arXiv241009000D}. The JELS observations used a $3\times3$ mosaic strategy with 57 per cent overlap between columns, and adopted the `Medium8' observing strategy with 9 groups for the F466N filter and 10 groups with the F470N filter, which gives $\sim$ 1000s on-sky per observation. A 3-point intramodule dithering pattern, with two sub-pixel dithers at each location, was then used to account for bad pixels and cosmic rays. This observation setup gave continuous coverage over an area of 63 arcmin$^{2}$ of the Cosmic Evolution Survey (COSMOS) field with central coordinates (RA, Dec) = (150.125, 2.333) deg. The on-sky integration time was $\sim$ 6 ks over the full mosaic (see Fig. \ref{fig:jels_footprint}) with double-depth imaging ($\sim$ 12 ks) over the central $\sim$ 40 per cent of the mosaic. This totalled to 43.0 hours of programme time.

As shown in Fig. \ref{fig:jels_footprint}, the narrow-band observations overlapped in area with high quality \emph{HST} imaging data, mainly from the Cosmic Assembly Near-IR Deep Extragalactic Legacy Survey \citep[CANDELS;][]{2011ApJS..197...35G,2011ApJS..197...36K} observed using the WFC3 and ACS instruments, which provides multi-wavelength coverage from the UV through to 1.6 $\mu$m. Further imaging over a smaller area was also available from the 3D-\emph{HST} survey \citep{2012ApJS..200...13B} and UVCANDELS \citep{2018hst..prop15647T}. In addition, this same field was selected for the PRIMER survey which provides \emph{JWST}/NIRCam imaging in 8 filters. This combined dataset provided high-resolution space-based imaging from UV to IR wavelengths. Fig. \ref{fig:filters} shows the filters used from both \emph{HST} and \emph{JWST} observations and Table \ref{table:multiwav_data} shows the filter pivot wavelengths ($\lambda_{\rm{pivot}}$), effective widths ($\Delta \lambda$), point-spread function (PSF) full-width half-maximum (FWHM) values, 5$\sigma$ global depths (see description in Section~\ref{sec:phot_err}), filter dependent Galactic extinction values and survey areas for all the multi-wavelength imaging in addition to narrow-band imaging. The complete JELS survey area (63 arcmin$^{2}$; see Fig \ref{fig:jels_footprint}) overlapped with the combined \emph{HST} and PRIMER footprints, resulting in nearly $\sim100$ per cent of sources identified in the JELS imaging having rich ancillary multi-wavelength data. In addition, the high resolution of the \emph{HST} and \emph{JWST} imaging meant we could perform point spread function (PSF) homogenisation (see Section~\ref{sec:psf}) across a wide wavelength range and study multi-wavelength resolved properties of the galaxies even at high-redshift.

Data reduction of the JELS and PRIMER NIRCam imaging was performed using the PRIMER Enhanced NIRCam Image Processing Library (\textsc{PENCIL}; Dunlop et al., in preparation) software. The pipeline outputted the main science images along with the weight maps (exposure time maps) and RMS images (which take the standard deviation of the science image and incorporate the Poisson and read noise). The reduced imaging was astrometrically aligned to GAIA DR3 \citep{2023A&A...674A...1G} and stacked to the same pixel scale of 0.03 arcsec. The \emph{HST} ancillary imaging was re-scaled to the JELS-PRIMER area with matched pixel scale. Both narrow-band mosaics were affected by scattered light contamination during observation of the COSMOS field: the F470N had four and the F466N had one of the constituent pointings affected, respectively. This was accounted for in the image reduction stage \citep[see][]{2024arXiv241009000D} through subtraction of scattered light templates generated from the JELS imaging. This was successful in removing most of the scattered light contamination (that could otherwise be picked up as source detections) in both the F466N and F470N mosaics. There was, however, still some low-level residual scattered light contamination left over in both mosaics (particularly in the F470N image) and so visual inspection was required for final samples of line emission galaxy candidates (see Section~\ref{sec:halpha_sample} for how this was carried out for the H$\alpha$ emission line galaxy sample at $z > 6$).

\subsection{PSF Homogenisation}
\label{sec:psf}

To accurately characterise the narrow-band selected emission line galaxies (including the population of H$\alpha$ emission line galaxies at $z > 6$) and determine of their physical properties, we homogenised all the available \emph{HST} and \emph{JWST} imaging to a common PSF. Empirical PSFs were generated through the stacking of bright and unsaturated stars in the relevant field for a given filter. For the \emph{HST} filters, stars were identified from the Gaia DR3 catalogue \citep{2023A&A...674A...1G}, using the mean $g$-band magnitude measurements to better select stars with prominent UV-optical emission. For the \emph{JWST} filters, selecting stars from the Gaia catalogue was not suitable due to the large wavelength difference between these observations and the $g$-band. Therefore, stars were identified by selecting sources with small half-light radii and bright magnitude measurements in a preliminary narrow-band detected catalogue built from the native-resolution \emph{JWST} NIRCam imaging. For each filter, a magnitude range was then adopted to remove saturated stars that could broaden our PSF measurements; this was further enforced by visual inspection. Once the star samples were selected for each filter, the stars were centred (which involved re-pixelating to a smaller scale to minimise centroiding issues) and stacked using a bootstrapping method to generate PSF stacks in each filter. The PSF FWHM values were measured by fitting a Moffat profile to the PSF stacks for each filter; these can be found in Table \ref{table:multiwav_data}. 

Our measured empirical PSF FWHMs were comparable to the empirical measurements made during commissioning \citep{2023PASP..135d8001R} and to simulated models from WebbPSF \citep{2014SPIE.9143E..3XP} though were typically larger by $\sim$ 0.01 arcsec; this was due to a combination of centering errors in the empirical stacks, smearing due to dithering and mosaicing of the individual frames, and the \emph{JWST} PSF not being perfectly mapped by a Moffat model (due to the airy disk nature and prominent diffraction spikes). Note that the Moffat models were only fitted to obtain an estimate for the PSF FWHM in each filter and were not used in the image convolutions. 

\begin{figure}
    \centering
    \includegraphics[width=0.48\textwidth]{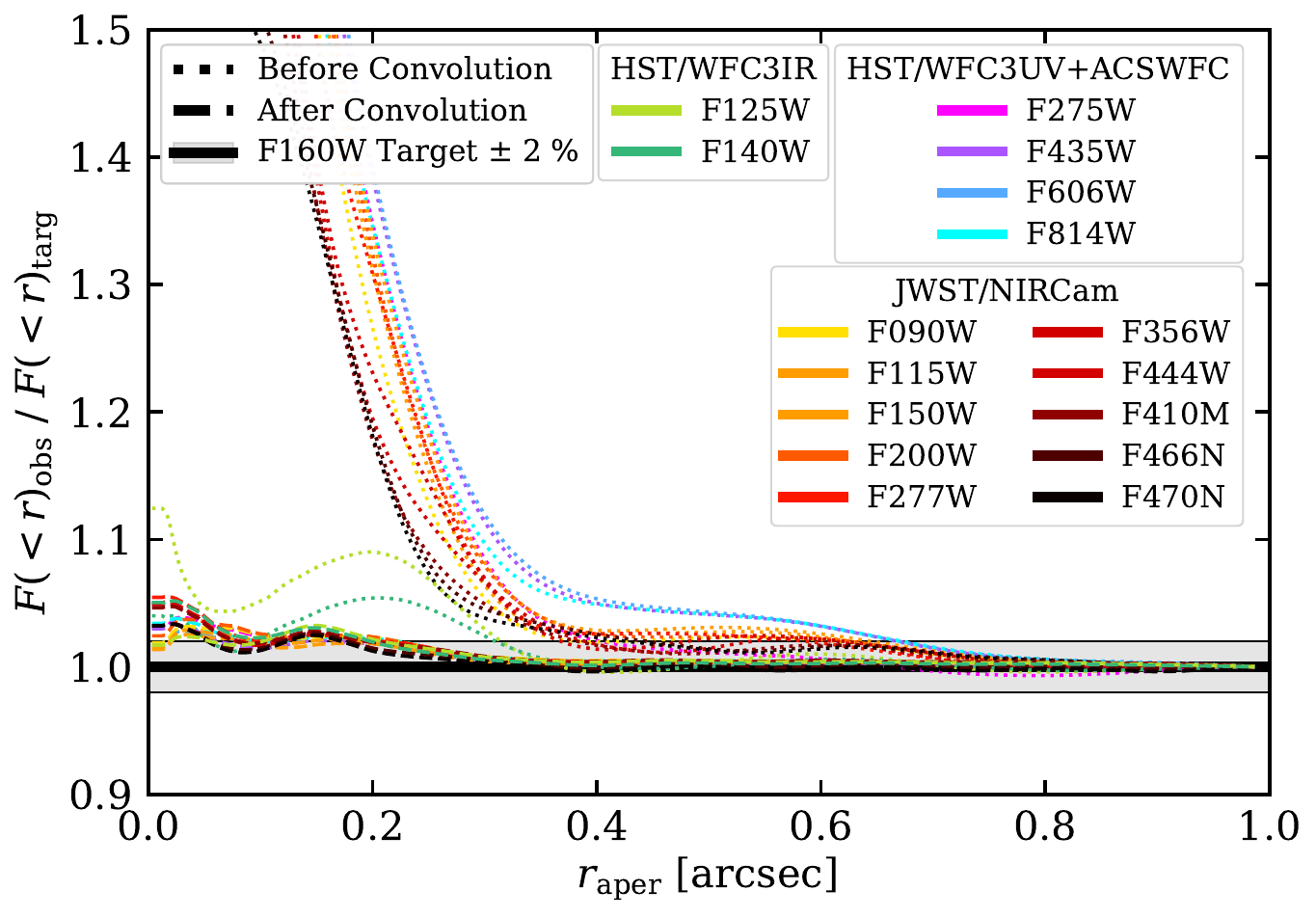}

    \caption{The fractional enclosed flux as a function of aperture radius (i.e. the curve of growth) for the filters used in this analysis ($F(< \ r)_{\rm{obs}}$) relative to that of the target empirical F160W PSF ($F(< \ r)_{\rm{targ}}$), which is the filter with the broadest PSF and hence lowest resolution. We assume that 100 per cent of the flux is enclosed in a 1.0 arcsec radius. Here, the dotted and dashed lines show the ratio for each filter before and after convolution, respectively. The solid black line shows the ratio to the target PSF (unity) and the shaded region shows $\pm$2 per cent of the target PSF.}
    
    \label{fig:cog_ratio}

\end{figure}

We then employed the \textsc{pypher} package \citep{2016A&A...596A..63B} to generate convolution kernels for each filter, using the empirical PSF stacks and the circularised average of the target PSF. Our target PSF was the \emph{HST}/WFC3 IR F160W filter, due to it exhibiting the broadest PSF ($\rm{FWHM} \sim 0.18$; see Table \ref{table:multiwav_data}). These kernels were then used to convolve the relevant images. Fig. \ref{fig:cog_ratio} shows that the PSFs of the convolved images agree to within 2 per cent of the target within a 0.2 arcsec radius, demonstrating that our convolved PSF distributions (and hence source photometric measurements) were consistent across all filters.

\section{Multi-Wavelength Catalogues}
\label{sec:narrowband_cat}

Narrow-band imaging is crucial for selecting star-forming galaxies based on their emission lines at different redshift epochs. However, PSF-homogenised photometry is crucial for identifying which emission line is being detected and obtaining the multi-wavelength properties of the galaxies. For H$\alpha$ emission line galaxies at $z>6$, this included obtaining constraints on the rest-frame UV and optical spectrum to measure physical properties such as dust attenuation and their star-formation activity on different timescales (see Section~\ref{sec:physical_properties}).

In this section, we discuss the creation of multi-wavelength catalogues which were built for sources detected on the native resolution JELS narrow-band (F466N and F470N) and PRIMER F356W imaging, with forced photometry then performed on the full set of PSF-homogenised multi-wavelength imaging (omitting the narrow-bands for the F356W detected catalogue). From these narrow-band catalogues, we selected narrow-band excess sources (see Section~\ref{sec:selection}) corresponding to emission line galaxy candidates, and analysed their multi-wavelength properties. The creation of our F356W detected catalogue was then instrumental in allowing the physical properties of the H$\alpha$ emission line galaxies to be compared to those selected photometrically on their rest-frame UV/optical broad-band photometry at the same epoch (see Section~\ref{sec:f356w_z=6_sources}). The F356W filter was chosen for the task since it was the most sensitive filter out of the PRIMER datasets \citep[see Table 1 of][]{2024MNRAS.533.3222D}.

\begin{table}
\centering
  \caption{Key \textsc{SExtractor} background, detection and deblending parameters used for the F466N, F470N and F356W detection images.}
  \label{table:sextractor_params}
  \begin{tabular}{c c}
    \hline
    \hline
    Parameter & Value\\
    \hline
    BACK$\_$SIZE & 128\\
    BACK$\_$FILTERSIZE & 5\\
    DETECT$\_$THRES & 1.7\\
    DETECT$\_$MINAREA & 9.0\\
    FILTER & gauss$\_$3.0$\_$7x7\\
    DEBLEND$\_$NTHRES & 16\\
    DEBLEND$\_$MINCONT & 0.001\\
    \hline
  \end{tabular}
\centering
\end{table}

\subsection{Source Detection}
\label{sec:sextractor}

Source detection was performed using \textsc{SExtractor} \citep{1996A&AS..117..393B}. The software was run in `dual mode', using a native-resolution source detection image (JELS F466N, JELS F470N or PRIMER F356W) and then we performed forced photometry across all the convolved multi-wavelength imaging, producing F466N, F470N and F356W detected catalogues respectively. Optimal \textsc{SExtractor} parameters (see Table \ref{table:sextractor_params}) were found by running a range of source extractions on smaller cutouts of the native-resolution F466N, F470N and F356W mosaics and testing the background, detection and de-blending parameters. To validate the astrometric accuracy of the JELS imaging, we cross-matched the narrow-band detected catalogues with the F356W detected catalogue and found a global offset of just $\sim$0.5 pixels (0.015 arcsec) confirming the world coordinate system (WCS) between the JELS and PRIMER observations were matched to high level of accuracy.

\subsection{Photometric Measurements}
\label{sec:phot}

We measured source fluxes from all the convolved images from the \emph{HST} and \emph{JWST} filter set for sources detected in the F466N, F470N and F356W images, using four aperture diameter values of 0.3, 0.6, 0.9 and 2.0 arcsec. The choice of 0.3 arcsec diameter aperture measurements was to maximise the signal-to-noise ratio (SNR) when applying a criteria for source detection (see Section~\ref{sec:low_sig_det}), for measuring high SNR source colours when identifying narrow-band excess sources (see Section~\ref{sec:selection}), and to accurately constrain source photo-$z$'s (see Section~\ref{sec:photo_z}). The 0.6 arcsec aperture diameter measurement was chosen to increase the light collection area when calculating source line fluxes (see Section~\ref{sec:selection}), and used to perform SED fitting (see Section~\ref{sec:sed_fitting}) and thus obtain accurate physical parameter estimates (for example, stellar mass). The 0.9 and 2.0 arcsec aperture diameter measurements were chosen to accommodate more extended sources which require larger apertures to acquire total fluxes. Note that only the 0.3 and 0.6 arcsec aperture diameter measurements were utilised in the analysis of this paper; these apertures capture $\sim$50 per cent and $\sim$82 per cent, respectively, of the total light assuming a point source.  In this paper, we did not apply aperture corrections to our photometry, or to the empirical or SED-derived physical properties for our H$\alpha$ emission line galaxies at $z>6$ (see Section~\ref{sec:selection}). However, because we used the PSF-homogenised photometry with consistent aperture sizes for all of our analyses, all properties (both empirical and SED-derived) were determined consistently and so could be directly compared. We note that the convolved PSFs vary by only $\sim$2 per cent for 0.3 arcsec diameter apertures, which is lower than the minimum 5 per cent errors we enforced for the photometric redshift and SED fitting (see also Section \ref{sec:sed_fitting}), and that in any case the photometric redshift fitting accounted for any remaining systematic offsets (see Section. \ref{sec:photo_z}). 

\subsubsection{Galactic extinction corrections}
\label{sec:extinction}
We used the position of each detected source to compute Galactic extinction corrections using the \citet{1998ApJ...500..525S} map and the \textsc{dustmaps} package \citep{2018JOSS....3..695Gcf}. The output $E$($B \ - \ V$) reddening value for each source was then multiplied by a filter dependent factor (see Table \ref{table:multiwav_data}) derived from the given filter transmission curve and the Milky Way extinction curve \citep{1989ApJ...345..245C}. The PSF-homogenised photometry was then corrected for extinction using the method described in Appendix A of \citet{1999PASP..111...63F}.

\begin{figure}
    \centering
    \includegraphics[width=0.47\textwidth]{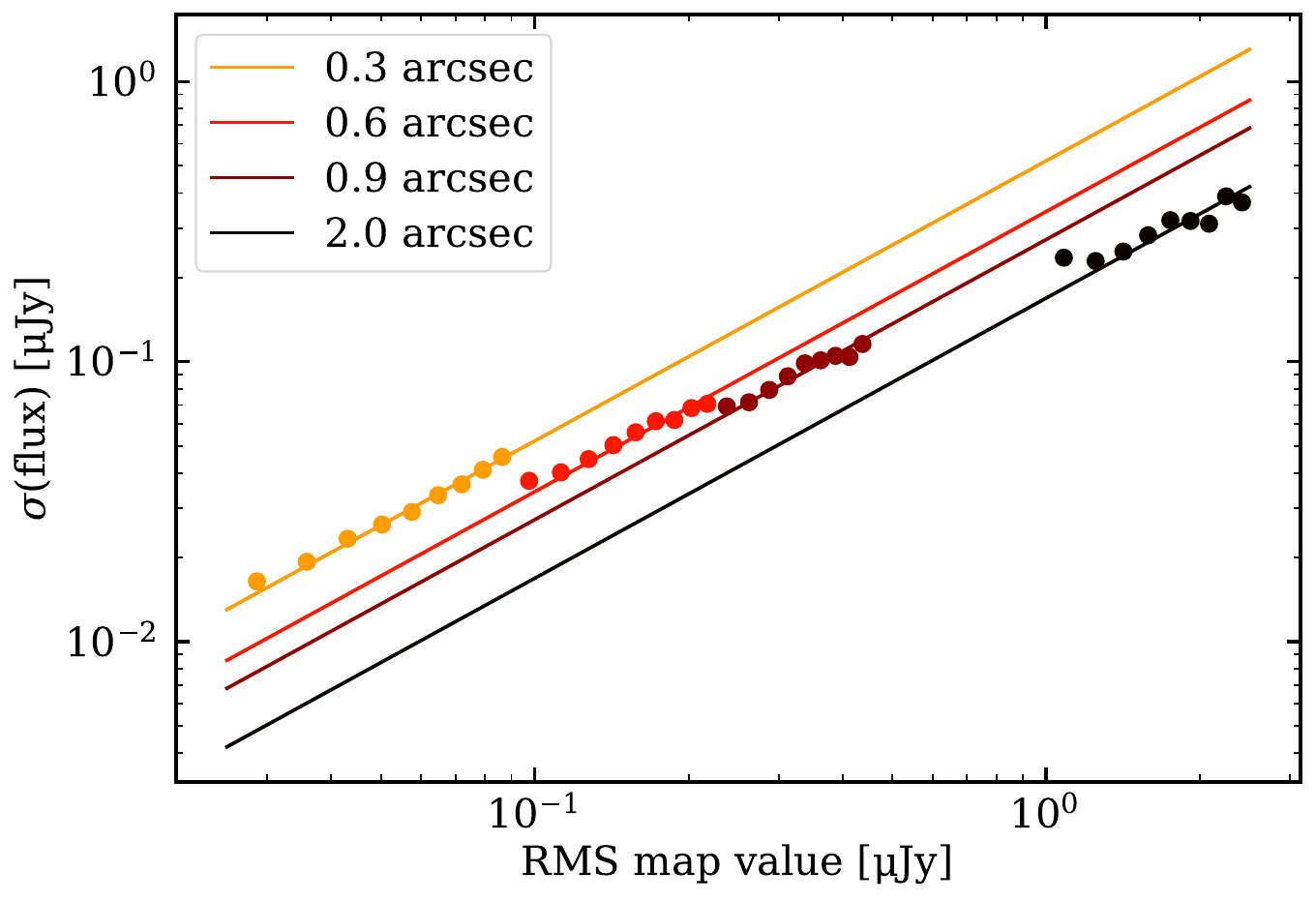}

    \caption{Plot showing the standard deviation of the flux values measured in the main science image for empty apertures of a given size, as a function of the value measured through the same aperture on the F466N RMS map outputted from \textsc{PENCIL}. The aperture diameters for the 0.3, 0.6, 0.9 and 2.0 arcsec measurements of the main science images and RMS maps are shown in orange, red, brown and black respectively. The weighted linear fitted relation allows the flux uncertainty due to background subtraction to be estimated for each source, based on the RMS map value at its location. This uncertainty was then added in quadrature to the flux uncertainty outputted from \textsc{SExtractor} to give a more accurate flux uncertainty on a source-by-source basis.}
    
    \label{fig:flux_sd_vs_rms_map_bin}

\end{figure}

\subsubsection{Computation of Photometric Errors}
\label{sec:phot_err}

The flux uncertainties reported by \textsc{SExtractor} typically underestimate the total uncertainties; this is a well-known issue. \textsc{SExtractor} only accounts for the photon and detector noise and not any background subtraction errors or correlated noise from re-sampling the pixel scale during image co-addition. Therefore, an additional flux uncertainty due to the variation in the background noise needed to be calculated and then combined in quadrature with the flux errors outputted from \textsc{SExtractor} \citep[e.g.][]{2012A&A...545A..23B,2016ApJS..224...24L,2021A&A...648A...3K} to provide more accurate flux uncertainties. For uniform-noise images, this can be achieved globally by placing apertures (of the same size used for the photometric measurements) across source-free regions of the image; measuring the RMS scatter (standard deviation) between them then gives the global 1$\sigma$ variation, or image depth. Our images were not uniform in depth, so we adopted a more sophisticated approach.

To account for the varying depth within the JELS and PRIMER images, we made use of the RMS maps produced in the image reduction stage (described in Section~\ref{sec:obs}) and the segmentation map outputted from \textsc{SExtractor}. Firstly, fluxes were measured in random isolated apertures (matching the size for the relevant aperture photometric measurement) for a given mosaic and filter. Apertures that were in areas of genuine sources (identified using the segmentation map) were then excluded. Secondly, aperture values were then extracted in the same manner as the science images but now for the RMS images. For a given aperture size, the fluxes measured from the science images were then binned by RMS value obtained from the RMS map. The standard deviation of the aperture flux measurements per RMS bin was then taken. Plotting the binned RMS value against the measured standard deviation of the aperture fluxes in that given bin gives a near-linear relation for each filter (see Fig. \ref{fig:flux_sd_vs_rms_map_bin} for an example of this fit for the F466N image). The total photometric uncertainty for a given source
was then calculated by taking the flux uncertainty reported by \textsc{SExtractor} and adding in quadrature the flux uncertainty which corresponds to the RMS measured from the RMS map at the same position for the given aperture size. These flux errors were also then translated into errors in magnitude. Note, the combination of the dither pattern and drizzling \citep[e.g.][]{2002PASP..114..144F} in the image reduction stage meant that the standard deviation between pixels (and hence photometric uncertainty) was probably underestimated due to correlated noise. However, this was likely on the $\sim$10 per cent level and was accounted for by implementing a minimum of 5 per cent error on photometric measurements for photo-$z$ and SED fitting (again, see Section. \ref{sec:photo_z} and Section \ref{sec:sed_fitting}).

\subsection{Catalogue cleaning}
\label{sec:cleaning}

\subsubsection{Cleaning low significance detections and artefacts}
\label{sec:low_sig_det}

To reduce the artefact contamination in our catalogues, we removed sources in our F466N, F470N and F356W detected catalogues which have SNR $<$ 5 in the detection filter for 0.3 arcsec diameter aperture photometric measurements. Inspection and examination of the fraction of these sources detected in other filters suggested a greatly increasing fraction of spurious sources and artefacts going below this SNR threshold. Even if these sources were real, they were too faint for reliable scientific exploitation.

It is important to realise that spurious detections in the F466N or F470N images that were picked up with a SNR $>$ 5 were very likely to be identified as narrow-band excess selected sources due to the likely non-detection in the complementary filter. Therefore, there could be a high-contamination fraction in the excess source catalogue (see Section~\ref{sec:line_cat}) even if the contamination fraction was lower in the parent multi-wavelength catalogue. Cosmic rays were an example of contamination that appeared to be high-SNR compact sources (see examples in Section~\ref{sec:halpha_sample}) that were found predominantly at the edges of the mosaics due to there being fewer overlapping exposures from the dither pattern of the JELS observations (see Section~\ref{sec:obs}). To remove these artefacts, we enforced that sources in the catalogue must have a half-light radius $>$ 1.5 pixels (measured using \textsc{SExtractor}); visual inspection of sources below this threshold revealed that they were all cosmic ray-like artefacts.

To further combat catalogue contamination, we also required that sources in our narrow-band selected catalogues have SNR $>$ 5 in their F356W photometric measurements (this was the most sensitive PRIMER filter -- see Section~\ref{sec:narrowband_cat}). This SNR cut in the F356W filter did pose a risk of missing genuine and very high-$EW$ sources which do not have a continuum detection; such extreme emission line sources have been observed into the Epoch of Reionization \citep[e.g.][]{2023MNRAS.524.2312E,2024A&A...691A..59L}. However, this was not a significant risk for our H$\alpha$ emission line galaxy sample (see Section~\ref{sec:halpha_sample}) since at $z$ $\sim$ 6 where H$\alpha$ is detected in the JELS narrow-bands, the \oiiia emission line falls within the F356W broad-band: any source with significant H$\alpha$ $EW$ was likely to be a strong \oiiia emitter and therefore detected in the F356W filter. To test this expectation, we visually inspected the narrow-band selected sources (with narrow-band SNR $>$ 5) and photometric redshifts above $z=5.5$ (see Section~\ref{sec:photo_z}) which had no significant detection in other filters in the mosaic and found that all of these sources appeared to be artefacts. For the lower redshift sample of emission line galaxies (such as the Paschen-$\alpha$ and $\beta$ emitters), we expected lower-$EW$ emission lines, and the peak of the strong stellar continuum `bump' at rest-frame 1.6 $\mu$m \citep[e.g.][]{1988A&A...193..189J} would contribute strongly to the F356W filter and so the risk of losing genuine emitters remained low, although it might potentially affect  Paschen-$\alpha$ emitters with very young stellar populations, resulting in lower continuum emission around the emission line and hence high $EW$s (less likely for Paschen-$\beta$). It is, nevertheless, worth stressing that the emission line galaxy selection presented in this paper is deliberately conservative, to produce a very secure sample of line emitters, and that additional emission line galaxies may be present in the data at lower SNR, or with extreme properties.

\begin{table}
\centering
  \caption{Mask sizes chosen for stars identified in the JELS field binned by narrow-band magnitude.}
  \label{table:star_mask}
  \begin{tabular}{c c}
    \hline
    \hline
    Narrow-band magnitude range & Circle radius [arcsec]\\
    \hline
      <17   & Dedicated mask \\
    17 - 18 & 6.0\\
    18 - 19 & 5.0\\
    19 - 20 & 2.0\\
    20 - 21 & 1.0\\
    21 - 22 & 1.0\\
    \hline
  \end{tabular}
\centering
\end{table}

\subsubsection{Stellar contamination}
\label{sec:bright_mask}

We masked regions of the images affected by bright stars and their diffraction spikes. This was particularly important for the narrow-band images, where their filter profiles transmit a narrow wavelength range of stellar light creating a dotted diffraction spike pattern unlike the smoother distributions seen in broad-band imaging from \emph{HST} and \emph{JWST}. This meant the features in the diffraction spike arms were extracted as sources and, given that there were many in the JELS mosaics, a significant fraction of sources within the raw detection catalogues were actually bright-star-related contaminants.

A star mask was created to flag sources in the regions of stellar contamination. In addition to the brighter stars used to the create the PSF stacks (Section~\ref{sec:psf}), sources with a half-light radius (see Section~\ref{sec:low_sig_det}) between 1.5 and 4 pixels in size (0.045 to 0.12 arcsec) were flagged as point sources. This was to account for contamination from fainter stars or stars misidentified in the Gaia DR3 catalogue \citep{2023A&A...674A...1G}. These flagged sources were most likely stars but they could also be quasars whose central emission dominates over the galaxy light and so appear as point-like in imaging \citep[e.g.][]{2022A&A...668A..99H}. 

For the JELS narrow-band images, stars with magnitudes $>$ 17 had negligible diffraction spike extension and so these stars were masked using circles of appropriate sizes for the stellar flux distribution in a given magnitude bin (see Table \ref{table:star_mask} for the mask sizes implemented). Brighter stars were then identified (9 in total) using the Gaia DR3 catalogue $g$-band magnitude measurements to avoid source fragmentation present in the narrow-band detection catalogues. A dedicated mask for each of these bright stars was built using appropriately sized circles and rectangles to account for the emission from the stellar core and the extended diffraction spikes. Sources in areas of stellar contamination were then flagged and omitted from the narrow-band detected catalogues.

Due to the deeper F356W imaging compared to the narrow-band images, a higher fraction of the footprint area contained stellar artefacts, including diffraction spikes which contaminated the F356W catalogue. However, the same stellar mask was applied to the F356W detected catalogue for the following reasons: i) This catalogue was primarily utilised for sample comparisons to the narrow-band detected catalogue and for global astrometry checks. ii) The broader wavelength coverage led to a smoother, more continuous diffraction spike, which results in fewer spurious `sources' detected at the catalogue production stage. Therefore, there were fewer falsely detected sources. Given the above, we accepted a small fraction of stellar contamination would make it into the F356W detected catalogue but this was negligible compared to the genuine source fraction (given the greater image depth than the narrow-bands). In addition, the photometric redshift fitting (see Section~\ref{sec:photo_z}) should have eliminated most of the contamination when selecting samples for comparison.

\subsubsection{Final cleaned catalogues}
\label{sec:final_detection_cat}

After application of the narrow-band/broad-band SNR criteria and removal of cosmic ray artefacts and stellar contamination, the final multi-wavelength catalogues source counts were 5645, 6150, and 34168 for the F466N, F470N, and F356W detection images, respectively. Table \ref{table:source_cleaning} shows the evolution in catalogue number counts after applying the above cleaning criteria. As noted in Section~\ref{sec:obs}, there was a higher degree of contamination in the F470N filter due to residual scattered light, and this was evident from the initial catalogue number count in Table \ref{table:source_cleaning}.

\begin{table}
\centering
  \caption{Multi-wavelength detection catalogue source counts after each stage of the catalogue creation/cleaning: 1) Native-resolution image photometric detections with SNR > 5 criteria. 2) The SNR > 5 criteria for the F356W photometric measurements in the narrow-band detection catalogues (not applicable to the F356W detection catalogue). 3) The cosmic ray removal. 4) The removal of sources located in areas of the star mask contaminated by bright stars and/or diffraction spikes, leading to the final catalogue.}
  \label{table:source_cleaning}
  \begin{tabular}{c c c c}
    \hline
    \hline
    Catalogue cleaning stage &  & Source count & \\
     & F466N & F470N & F356W\\
    \hline
    SNR(detection filter) > 5 & 6828 & 7572 & 34677\\
    SNR(F356W) > 5 & 6514 & 6983 & 34677\\
    Cosmic ray removal & 6476 & 6950 & 34525\\
    Star mask removal: final catalogue & 5645 & 6150 & 34168\\
    \hline
  \end{tabular}
\centering
\end{table}

\subsection{Photometric redshift analysis}
\label{sec:photo_z}

We performed SED fitting using \textsc{EAZY-Py} \citep{2008ApJ...686.1503B} to derive photometric redshift (photo-$z$) estimates for all sources in our full multi-wavelength catalogues. We utilised all filter coverage available for a given object in the narrow-band and F356W-detected catalogues (see Table \ref{table:multiwav_data} for filter properties and Fig. \ref{fig:filters} for filter transmission curves). We explored a redshift range 0 $<$ $z$ $<$ 10 using three template sets: the default flexible stellar population synthesis (FSPS) models supplemented with the high-redshift optimised templates of \citet[][specifically the `Lya\_Reduced' subset]{2023ApJ...958..141L}, the `SFHZ' models supplemented with the obscured AGN template \citep{2024A&A...691A..52K}, and the `\textsc{EAZY} v1.3' template library. 
For each template set, zeropoint offsets to the photometry were derived by fitting the templates to the known spectroscopic redshift for a pre-\emph{JWST} literature sample \citep[][$\sim2200$ sources]{2023ApJ...942...36K}.
For all three template sets, the offsets were found to be $<5$ per cent for all filters.
In particular, we highlight that the zeropoint offsets derived for the F466N/F470N narrowband filters were $<2$ per cent for all template sets, indicating that there were no significant flux calibration offsets specific to the narrow-band filters and any remaining calibration uncertainties were comparable to those measured for the commonly-used broadband filters \citep[see e.g.][]{2022RNAAS...6..191B}. We note that disentangling errors in the photometric measurements due to PSF homogenisation (see Section \ref{sec:psf}) from other factors like NIRCam calibrations is tricky but the above suggests that the total of these cannot be more than a few per cent and the 5 per cent minimum uncertainties implemented should be capturing this effect.

\begin{figure}
    \centering
    \includegraphics[width=0.47\textwidth]{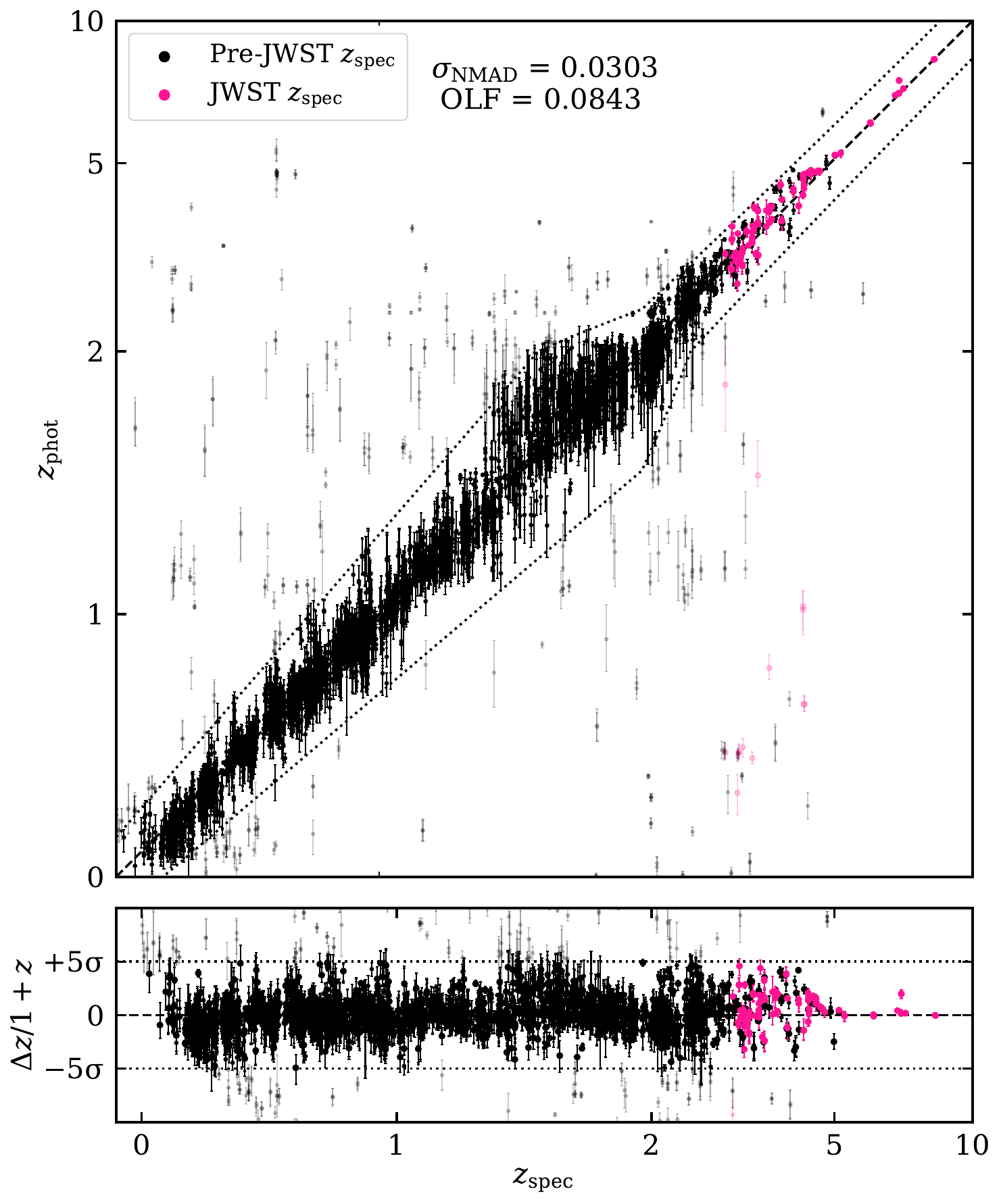}

    \caption{Top panel: Distribution of estimated photometric redshifts ($z_{\rm{phot}}$) outputted from \textsc{EAZY-Py} fits and measured from the 0.3 arcsec diameter aperture photometric data as a function of the spectroscopic redshift ($z_{\rm{spec}}$), for sources with spectroscopic data. The dashed black line shows where $z_{\rm{phot}}$ and $z_{\rm{spec}}$ are equivalent. The black and pink solid points shows cross-matches to spectroscopically confirmed sources obtained pre-\emph{JWST} and with \emph{JWST} respectively. The faded points with equivalent colours show sources with $|\Delta z| / ( 1 \ + \ z )$ > 5$\sigma_{\rm{NMAD}}$ (see definitions in Section~\ref{sec:photo_z}) with the dotted black lines showing where $\Delta z / ( 1 \ + \ z )$ = $\pm$5$\sigma_{\rm{NMAD}}$. Bottom panel: Distribution of $\Delta z / ( 1 \ + \ z )$ in units of $\sigma_{\rm{NMAD}}$ as a function of $z_{\rm{spec}}$. The point colours and transparency definitions are the same as in the top panel.}
    
    \label{fig:photo_z_vs_spec_z}

\end{figure}

\begin{figure*}
    \centering
    \includegraphics[width=\textwidth]{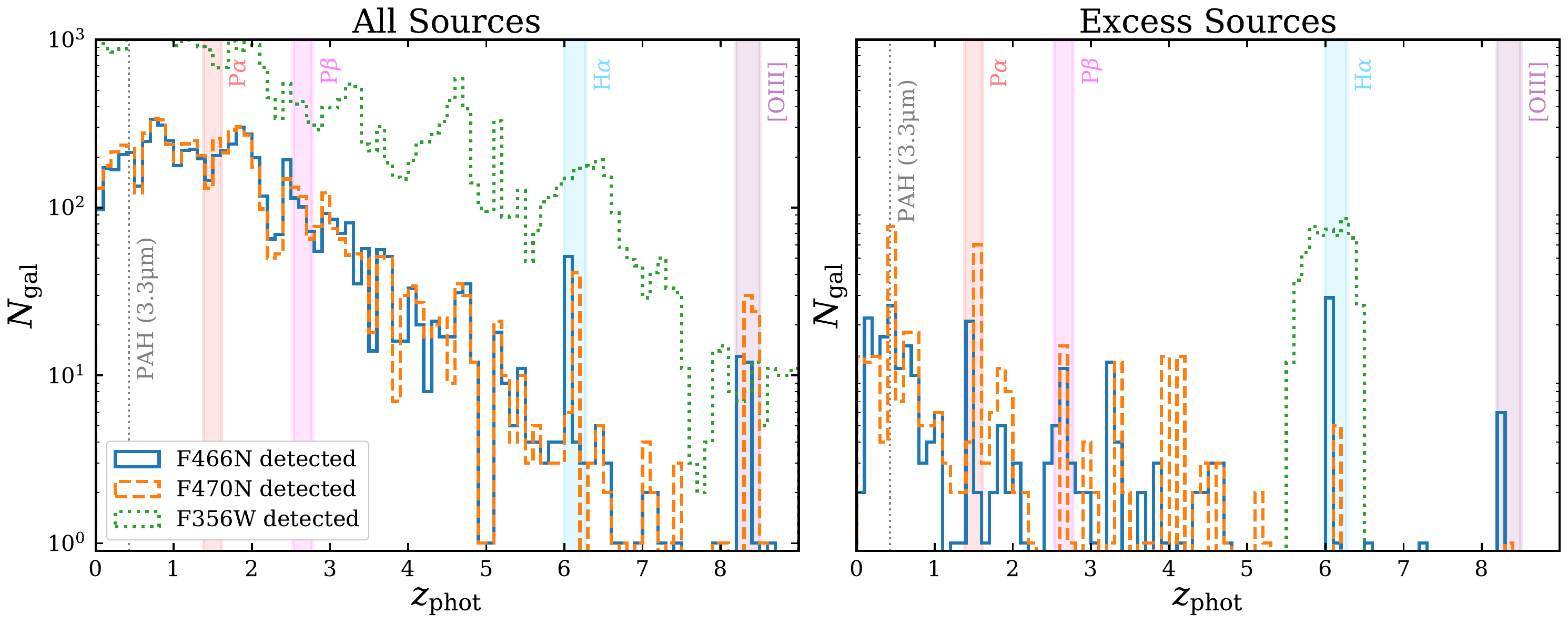}

    \caption{Left panel: Photometric redshift histograms for the full F466N (solid blue line), F470N (dashed orange line) and F356W (dotted green line) detected catalogues. Right panel: Photometric redshift histograms for the subset of these sources that are narrow-band excess selected as described in Section~\ref{sec:selection}, and for F356W detected sources for which integrated $P(z)$ $>$ 0.7 within the redshift range 5.5 $<$ $z$ $<$ 6.5. The same histogram colours and line styles apply. The grey dotted and colour shaded regions show the spectral features and emission lines observing regions in redshift space for the F466N and F470N filters. In addition, we performed visual inspection of the narrow-band `excess sources' with $z_{\rm{phot}}$ $>$ 5.5 (see description in Section~\ref{sec:halpha_sample}) and plotted only the visually confirmed sources in this histogram. The visual inspection of the lower-redshift narrow-band excess sources was not carried out in this analysis and so there may be a degree of contamination in the excess source histogram for $z_{\rm{phot}}$ $<$ 5.5 sources.}
    
    \label{fig:photo_z}

\end{figure*}

Photo-$z$ estimates for each template set were then calculated for all three detection catalogues (F466N, F470N, F356W) with the zeropoint offsets applied, including an additional 5 per cent flux error added in quadrature to account for remaining template and calibration uncertainty.
Consensus redshift estimates were then derived following a simplified version of the Hierarchical Bayesian combination procedure described in \citet[][see also \citeauthor{2013ApJ...775...93D}~\citeyear{2013ApJ...775...93D}]{2018MNRAS.473.2655D}, assuming moderate covariance between the individual estimates ($\beta=2$). 

Based on the consensus photo-$z$ posterior for each source, the corresponding `best' redshift was then determined by taking the median of the primary 90 per cent highest probability density (HPD) credible interval (CI) peak \citep[see e.g][]{2019A&A...622A...3D}. Fig.~\ref{fig:photo_z_vs_spec_z} demonstrates the overall quality of the photo-$z$ estimates for the pre-\emph{JWST} compilation used to derive zeropoint offsets (black points), as well as for additional $z > 3$ spectroscopic confirmations from \emph{JWST} observations (purple points).\footnote{Extracted from the DAWN \emph{JWST} Archive: \url{https://s3.amazonaws.com/msaexp-nirspec/extractions/nirspec_graded_v3.html}, with only spectroscopic redshifts graded as `robust' used for comparison.} Director’s Discretionary programme DD 6585 (PI: Coulter) managed to spectroscopically confirm four emission line galaxies in our JELS sample (included in Fig.~\ref{fig:photo_z_vs_spec_z}) -- see discussion in \citet{2024arXiv241009000D}.

We evaluated the resulting photo-$z$ performance by calculating bulk quality statistics for the spectroscopically confirmed sources, using normalised median absolute distribution, $\sigma_{\rm{NMAD}} = 1.48 \times \rm{median}(|\Delta z| / (1 + z_{\rm{spec}}))$, and the absolute outlier fraction, $\rm{OLF} = |\Delta z| / (1 + z_{\rm{spec}}) > 0.15$ \citep[following common literature definitions; e.g.][]{2013ApJ...775...93D}. Note, $\Delta z$ = $z_{\rm{phot}} \ - \ z_{\rm{spec}}$. When considering the full sample, we calculate $\sigma_{\rm{NMAD}}$ = 0.0303 and OLF = 0.0843 (as shown in Fig. \ref{fig:photo_z_vs_spec_z}). These values change to $\sigma_{\rm{NMAD}}$ = 0.0741 and OLF = 0.1806 when limiting the sample to $z_{\rm{phot}}$ $>$ 3, but still show good photo-$z$ performance across redshift space.

The left panel of Fig. \ref{fig:photo_z} presents the measured photo-$z$ distribution for each of the multi-wavelength catalogues. We note, however, that subsequent sample selections made use of the full photo-$z$ posterior in addition to the single point estimates.

\section{Narrow-band excess source catalogue and H$\alpha$ emitter sample}
\label{sec:line_cat}

\subsection{Narrowband excess selection}
\label{sec:selection}

The original premise of the JELS survey was to observe in the adjacent F466N and F470N narrow-band (NB) filters to detect emission line galaxies. Here, there should be negligible difference in continuum emission observed in the narrow-band filters and so any excess emission in one of the narrow-band filters was indicative of line emission. However, access to the PRIMER survey imaging (see Section~$\ref{sec:obs}$) meant we also had photometric data in the F444W filter, which was the complementary broad-band (BB) filter that overlaps with both the F466N and F470N filters. In the absence of emission lines, the NB and BB magnitudes should be very similar, whereas an emission line in the NB would lead to excess emission in the NB over the BB filter. Therefore, with access to F466N, F470N and F444W filters, we could perform selections for NB excess sources using both BB $-$ NB and NB $-$ NB colour selections (see Section~\ref{sec:bb_nb_selection} and Section~\ref{sec:nb_nb_selection}), for each of F466N and F470N filters.

\subsubsection{Accounting for source continuum colours}
\label{sec:cont_colour_corr}

\begin{figure}
    \centering
    \includegraphics[width=0.47\textwidth]{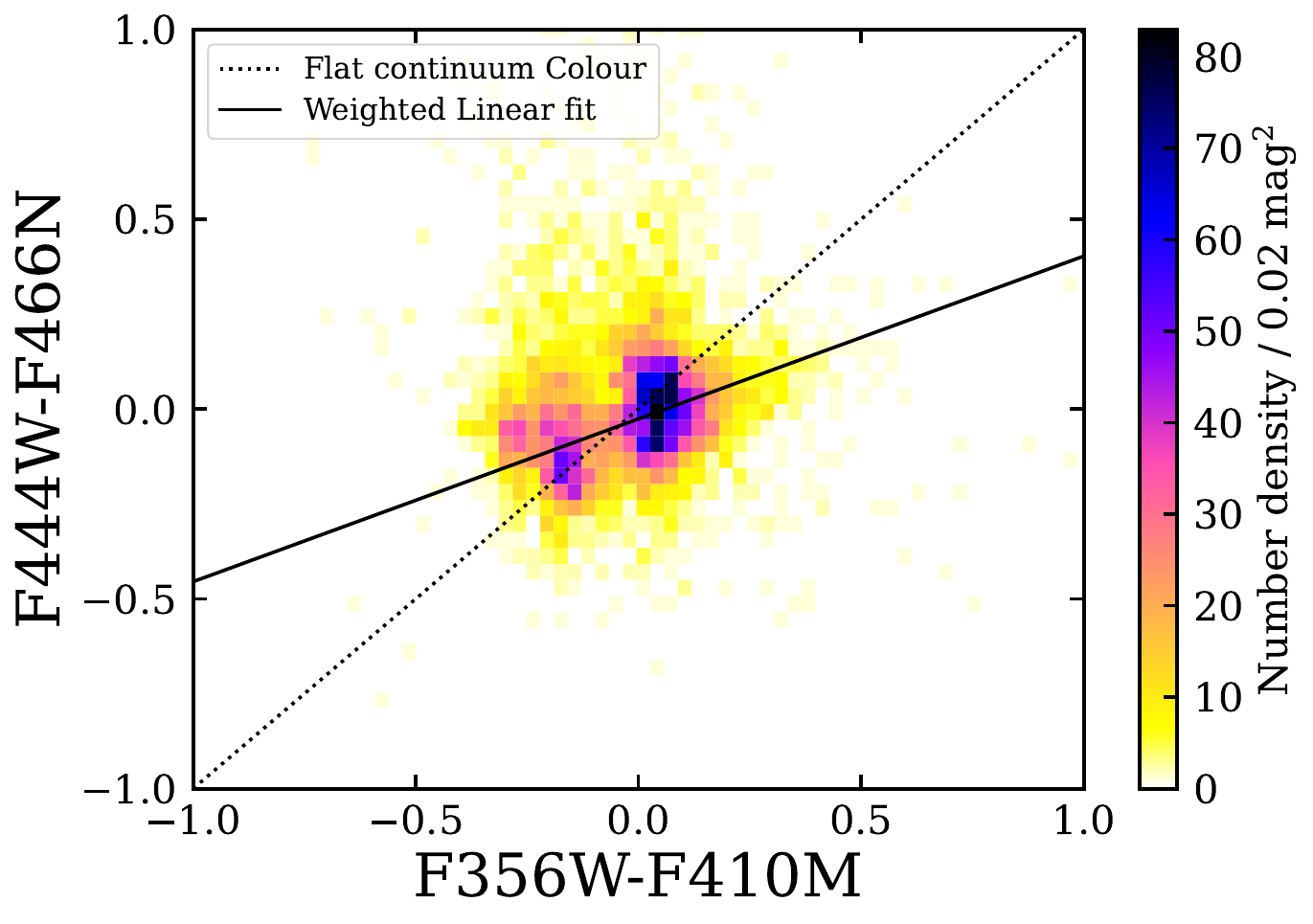}

    \caption{Number density plot showing the distribution of F466N detection catalogue sources in F356W $-$ F410M vs F444W $-$ F466N colour space using 0.02 mag$^{2}$ binning in the range: -1.0 $<$ colour $<$ +1.0. The dotted black line shows where F356W $-$ F410M = F444W $-$ F466N, corresponding to a flat continuum across the wavelength range spanning the filters shown. The solid black line shows a linear fit to the data weighted by the combined magnitude error for both the F356W $-$ F410M and F444W $-$ F466N colours (by adding the magnitude errors in the individual filters in quadrature). The weighted linear relation calculated for a source's given F356W $-$ F410M colour was then subtracted from the F444W $-$ F466N measured colour (see Eq. \ref{eq:colour_corr}) to correct for offsets due to significant continuum colours which could otherwise bias the emission line selection.}
    
    \label{fig:colur_colour_corr_plot}

\end{figure}

To account for the impact of colour across the broadband filter on the inferred narrow-band continuum flux (which for F466N/F470N compared to F444W was exacerbated by the relative wavelengths; see Fig. \ref{fig:filters}), we first calculated a continuum colour correction. To do this, we considered the distribution of F356W $-$ F410M versus BB $-$ NB colour for our narrow-band selected sources and fitted a weighted linear relation to the distribution in this parameter space. The F356W $-$ F410M  filter combination was chosen as it probes wavelengths close to the emission line without being contaminated by it. As discussed in Section~\ref{sec:narrowband_cat}, the F356W filter will contain the \oiiia emission line for our H$\alpha$ sample. We tested using F277W filter instead of F356W and found no impact on the final selection of emission line galaxy candidates (and no significant impact on inferred line fluxes). Therefore, we adopted the F356W $-$ F410M corrections to the BB $-$ NB and NB $-$ NB colours. We then corrected the BB $-$ NB colours using:

\begin{equation}
\label{eq:colour_corr}
({\rm{BB}} \ - \ {\rm{NB}})_{\rm{corr}}  = ({\rm{BB}} \ - \ {\rm{NB}})_{0} \ - \ [ A \ ({\rm{F356W}} \ - \ {\rm{F410M}}) \ + \ B ]
\end{equation}

\noindent where $(\rm{BB} \ - \ \rm{NB})_{0}$ is the originally measured BB $-$ NB colour, $A$ is the fitted gradient, and $B$ is the fitted y-intercept. Fig. \ref{fig:colur_colour_corr_plot} shows how this linear relation was fitted for performing the F444W $-$ F466N colour corrections and Table~\ref{table:colour_corr_fit_parameters} provides the fitted parameters. For sources with no F356W $-$ F410M measurement (due to non-detection in one or both filters), the median colour correction from the narrow-band selected sample was applied (see Table \ref{table:colour_corr_fit_parameters}). 

This colour correction method was also applied to the NB $-$ NB colours but, as can be seen in Table~\ref{table:colour_corr_fit_parameters}, the corrections were much smaller since the filters are very close in wavelength. Note that sources not detected in the F444W filter or the adjacent NB filter were assigned 1$\sigma$ upper limits when applying the continuum colour corrections.

\begin{table}
\centering
  \caption{Weighted linear fit parameters (gradient $A$ and y-intercept $B$, as defined in Eq.~\ref{eq:colour_corr}) for the BB $-$ NB and NB $-$ NB colour selections as a function of their F356W $-$ F410M colours. These parameters were then used to correct the relevant BB $-$ NB and NB $-$ NB colours using Eq. \ref{eq:colour_corr}. The final column quotes the median colour correction applied to the full sample.}
  \label{table:colour_corr_fit_parameters}
  \begin{tabular}{c c c c}
    \hline
    \hline
    Colour selection & $A$ & $B$ & Median Correction\\
    \hline
    F444W $-$ F466N & - 0.454 & + 0.022 & +0.039 \\
    F470N $-$ F466N & + 0.044 & - 0.005 & -0.007\\
    F444W $-$ F470N & - 0.573 & + 0.027 & +0.056\\
    F466N $-$ F470N & + 0.145 & + 0.008 & +0.001\\
    \hline
  \end{tabular}
\centering
\end{table}

\begin{table*}
\centering
  \caption{Excess source selection for both narrow-band detected catalogues showing: i) the NB detection filter; ii) the colour selection for that given detection filter; iii) the limiting colour excess criterion; iv) the corresponding observed-frame equivalent width limit, $EW_{\rm obs}$; v) the narrow-band excess significance criterion $\Sigma$; vi) the number of sources meeting the colour and $\Sigma$ criteria for the given colour selection; and vii) the final combined excess source sample for a given detection filter where sources meet the excess source criteria for at least one of the colour selections.}
  \label{table:excess_source_conditions}
  \begin{tabular}{c c c c c c c}
    \hline
    \hline
    Detection Filter & Colour Selection & Colour Criteria & $EW_{\rm{obs}}$ [$\rm{\AA}$] & $\Sigma$ Criteria & Excess Source Count & Combined Excess Sample\\
    \hline
    F466N  & F444W $-$ F466N & 0.30 & 135 & 3.0 & 177 & 241 \\
    & F470N $-$ F466N & 0.15 & 80 & 2.5 & 154 &  \\
    \hline
    F470N & F444W $-$ F470N & 0.35 & 165 & 3.0 & 298 & 368\\
    & F466N $-$ F470N & 0.20 & 103 & 2.5 & 246 &  \\
    \hline
  \end{tabular}
\centering
\end{table*}

\begin{figure*}
\centering
\begin{subfigure}{0.5\textwidth}
    \includegraphics[width=\linewidth]{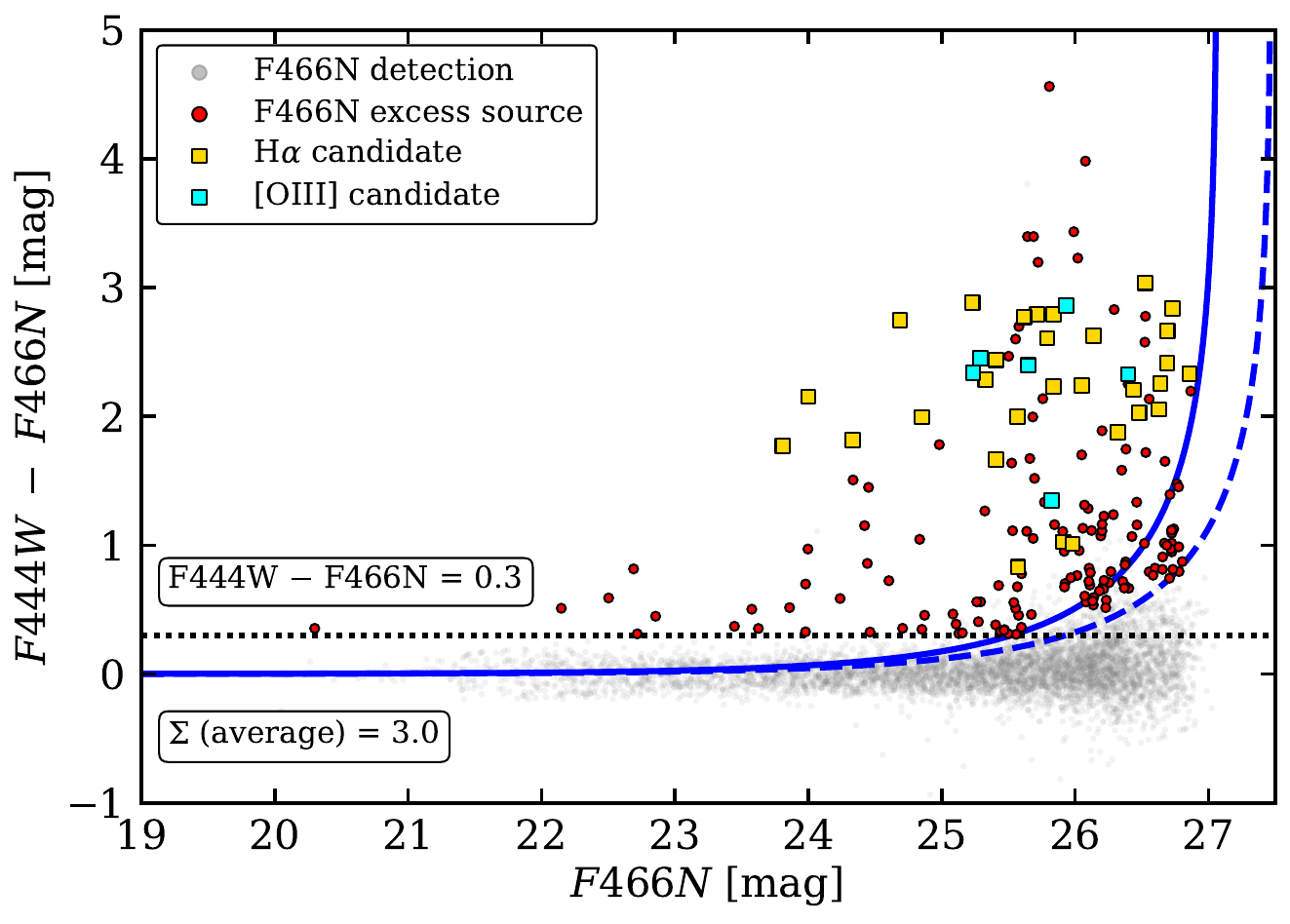}
\end{subfigure}\hfil
\begin{subfigure}{0.5\textwidth}
    \includegraphics[width=\linewidth]{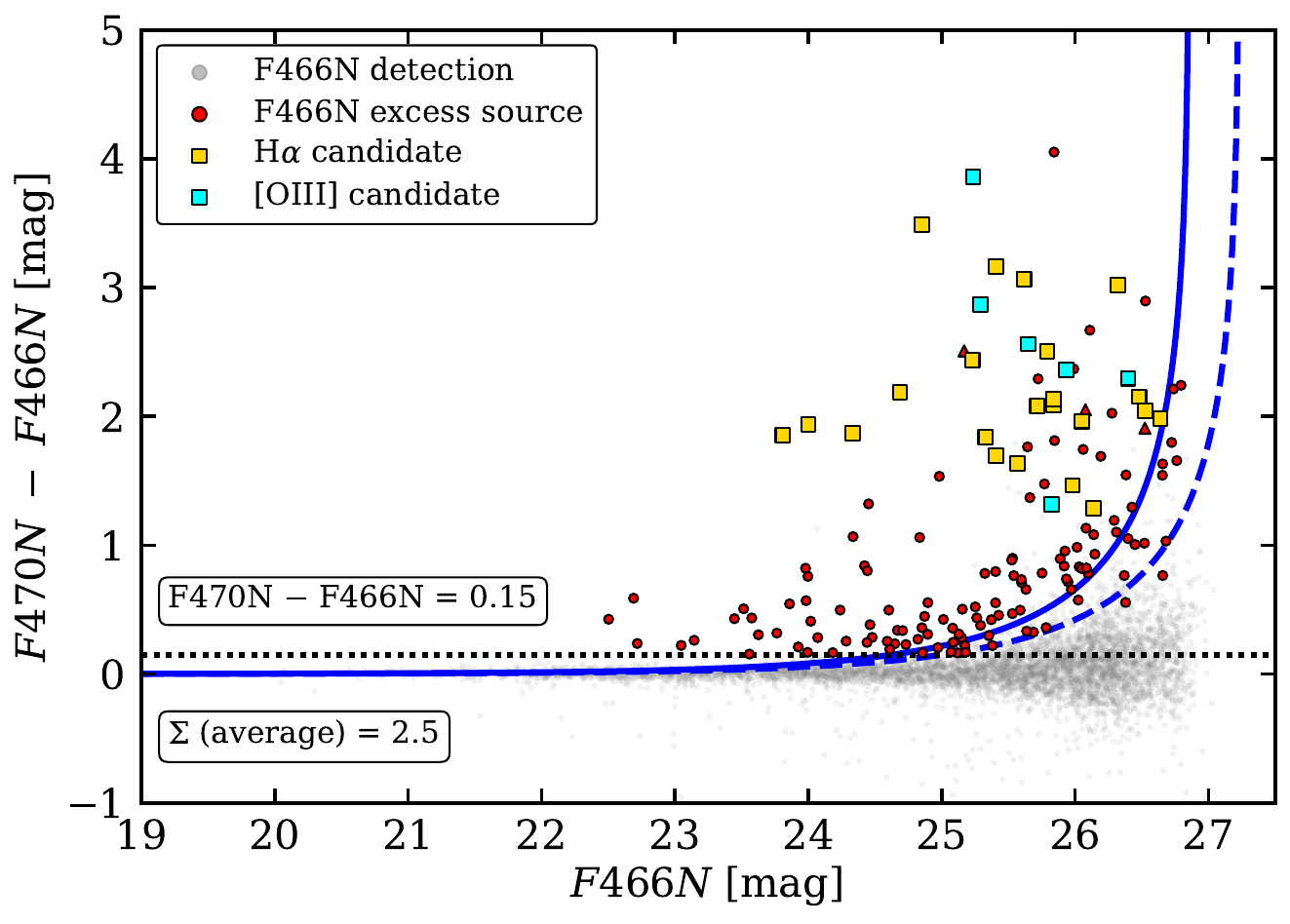}
\end{subfigure}\hfil

\medskip
\begin{subfigure}{0.5\textwidth}
  \includegraphics[width=\linewidth]{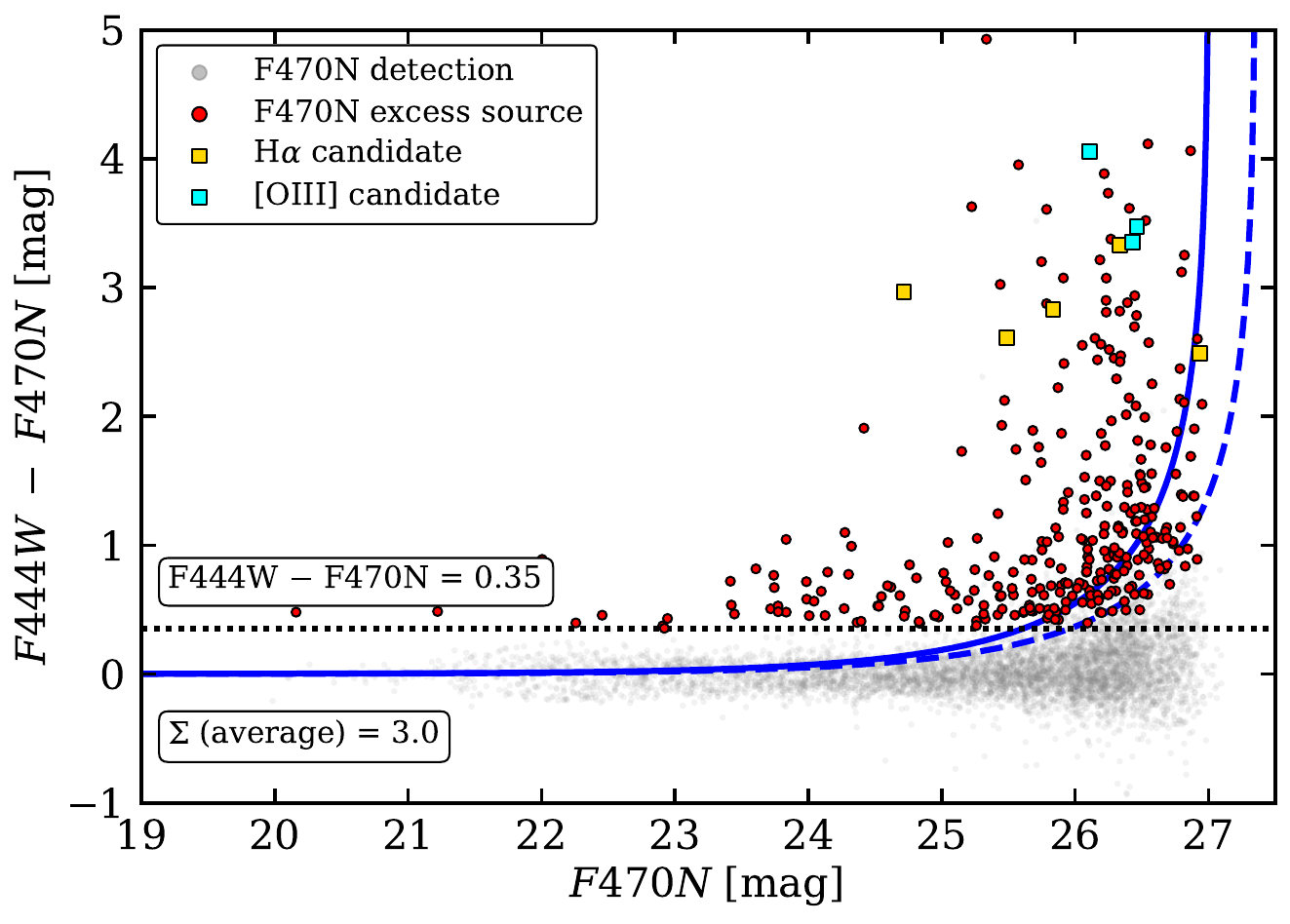}
\end{subfigure}\hfil
\begin{subfigure}{0.5\textwidth}
    \includegraphics[width=\linewidth]{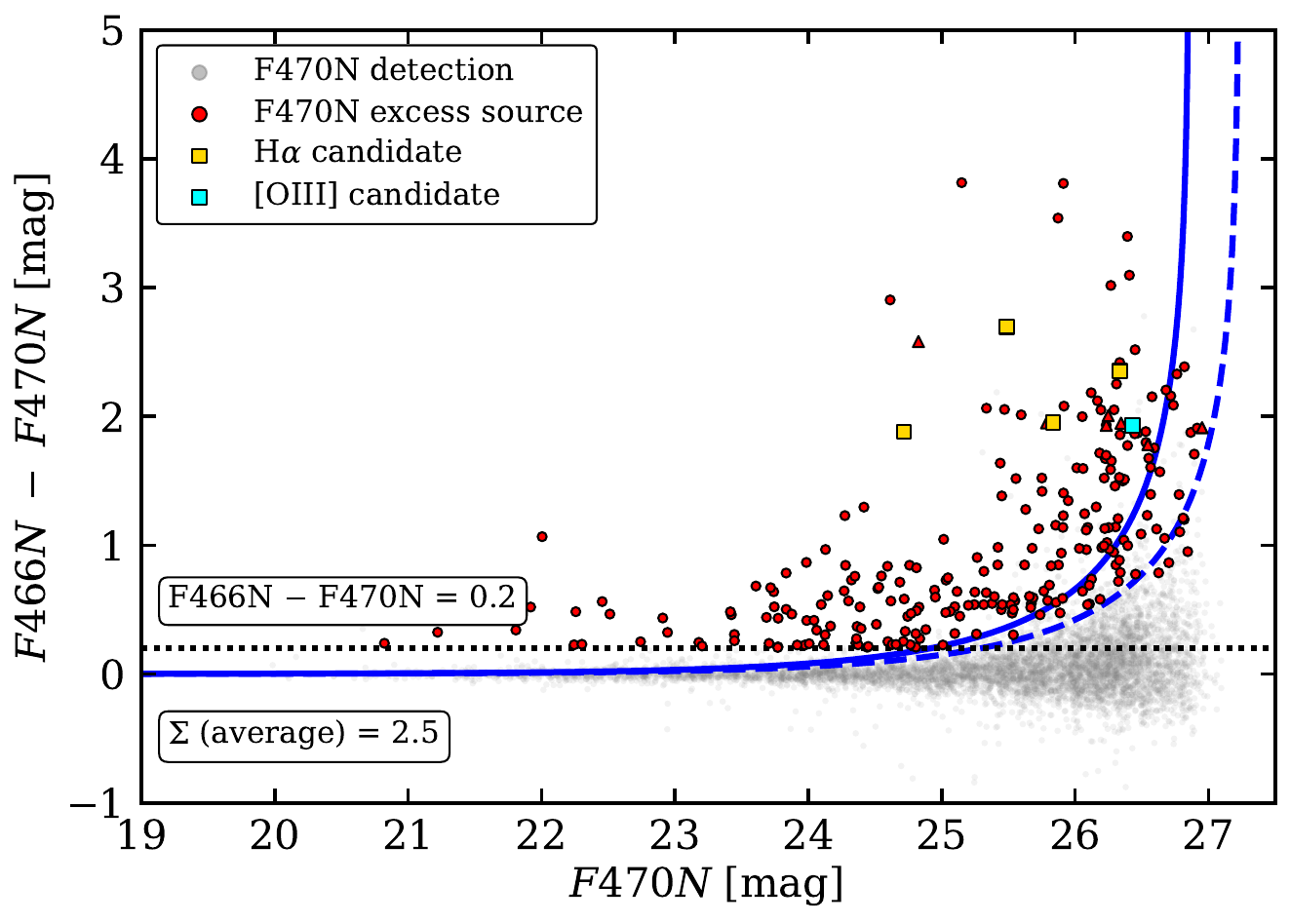}
\end{subfigure}

\caption{Colour-magnitude diagrams showing the different excess source selection criteria. Top row: the F466N detected sources selected based on their F444W $-$ F466N colour (left panel) and their F470N $-$ F466N colour (right panel). Bottom row: the F470N detected sources selected based on their F444W $-$ F470N colour (left panel) and their F466N $-$ F470N colour (right panel). For each colour-magnitude selection, sources considered to be excess sources (red points) are separated from the rest of narrow-band selected sources (grey points). This was done by applying a NB colour cut (black dotted lines) and a narrow band excess $\Sigma$ cut as described in Section~\ref{sec:selection}. $\Sigma$ was calculated for each source individually, but to guide the eye, the blue solid and dashed lines show the $\Sigma$ cut calculated for the average depths in the shallower and deeper regions, respectively, of the narrow band images. For all data points, upper limits are shown with triangle symbols. The yellow and cyan coloured squares correspond to H$\alpha$ and \oiiia emitter candidates respectively, which have also passed the visual inspection confirmation (Section~\ref{sec:halpha_sample}).}

\label{fig:colour_mag_plots}

\end{figure*}

\subsubsection{BB $-$ NB colour excess selection}
\label{sec:bb_nb_selection}

After correcting the BB $-$ NB colours for the different selections, there still remained some scatter around a colour excess of zero due to the uncertainties in the magnitude measurements, which increased towards fainter magnitudes. The degree of scatter depended on how the local noise varied from source to source and needed to be accounted for. Therefore, we defined the narrow-band excess parameter $\Sigma$, which quantifies the NB excess emission compared to the random scatter expected for a source with zero colour \citep{1995MNRAS.273..513B,2013MNRAS.428.1128S} and also accounts for the variable depths of the NB and BB imaging. For the BB $-$ NB selected sources, this is given by:

\begin{equation}
\label{eq:sig1}
\Sigma \ = \ \frac{1 - 10^{-0.4 (\rm{BB} \ - \ \rm{NB)}}}{10^{-0.4 (\rm{ZP} \ - \ \rm{NB}}) \sqrt{\sigma_{\rm{NB}}^{2} \ + \ \sigma_{\rm{BB}}^{2}}}
\end{equation}

\noindent where ZP is the zero point magnitude of the NB filter, which is set to 23.9 mag. $\sigma_{\rm{NB}}$ and $\sigma_{\rm{BB}}$ were photometric flux density errors (in $\rm{\mu}$Jy) for the NB and BB filters respectively for each source. Given that the JELS imaging contained distinct deeper and shallower regions (see Section~\ref{sec:obs}), sources that occupied identical colour-magnitude space could have differing values of $\Sigma$ depending primarily on the NB image depth (which was shallower than the BB image - see Table \ref{table:multiwav_data}) at their locations. This is illustrated in Fig. \ref{fig:colour_mag_plots}, where the dependence of $\Sigma$ on magnitude is plotted for the average depths of both shallow and deeper regions of the JELS footprint. In order to define a source as a reliable emission line candidate, we required that its narrow-band excess significance over the broad-band continuum was $\Sigma \ge 3$ (see Table~\ref{table:excess_source_conditions}).

As sources tend to bright magnitudes, the BB$-$NB colour corresponding to $\Sigma=3$ tends towards zero, but there was still intrinsic scatter in the BB $-$ NB measurements; this is due to systematic effects, such as the accuracy of the continuum colour correction given the scatter around the relation in Figure~\ref{fig:colur_colour_corr_plot}. Therefore, we also applied an $EW$ limit, which corresponds to a minimum BB $-$ NB colour for sources to be selected as an `excess source'. For each colour combination, we set the limiting colour criterion by assessing the scatter of the colours around zero at bright magnitudes; the selected criteria are given in Table~\ref{table:excess_source_conditions}. Sources that met our conditions on both $\Sigma$ and colour for the BB $-$ NB selected sources were then considered to be excess sources (see Table \ref{table:excess_source_conditions}).

The emission line flux, $F_{\rm{line}}$ and (observed) equivalent width $EW_{\rm{line}}$ of the selected line emitters were calculated following  \citet{2013MNRAS.428.1128S} as:

\begin{equation}
\label{eq:f_line_1}
F_{\rm{line}} \ = \ \Delta \lambda_{\rm{NB}} \ \frac{f_{\rm{NB}} \ - \ f_{\rm{BB}}}{1 \ - \ (\Delta \lambda_{\rm{NB}} / \Delta \lambda_{\rm{BB}})}
\end{equation}

\begin{equation}
\label{eq:ew_line_1}
EW_{\rm{line}} \ = \ \Delta \lambda_{\rm{NB}} \ \frac{f_{\rm{NB}} \ - \ f_{\rm{BB}}}{f_{\rm{BB}} \ - \ f_{\rm{NB}}(\Delta \lambda_{\rm{NB}} / \Delta \lambda_{\rm{BB}})}
\end{equation}

\noindent where $\Delta \lambda_{\rm{NB}}$ and $\Delta \lambda_{\rm{BB}}$ are the NB and BB filter widths, respectively, and where $f_{\rm{NB}}$ and $f_{\rm{BB}}$ were the measured flux densities in the NB and BB filters, respectively, with units of erg s$^{-1}$ cm$^{-2}$ $\rm{\AA}^{-1}$. Note, the above calculations account for emission line contamination in the BB filter when calculating NB excess and hence emission line flux.

\subsubsection{NB $-$ NB colour excess selection}
\label{sec:nb_nb_selection}

A similar procedure as described in Section~\ref{sec:bb_nb_selection} was applied to the NB $-$ NB selected sources. After performing the NB $-$ NB colour corrections for each source, the narrow-band excess parameter $\Sigma$ was calculated as described in Section~\ref{sec:bb_nb_selection} but replacing the BB magnitude measurement with the adjacent NB filter:

\begin{equation}
\label{eq:sig2}
\Sigma \ = \ \frac{1 \ - \ 10^{-0.4 (\rm{NB}_{2} \ - \ \rm{NB}_{1})}}{10^{-0.4 (\rm{ZP} \ - \ \rm{NB}_{1}}) \sqrt{\sigma_{\rm{NB_{1}}}^{2} \ + \ \sigma_{\rm{NB_{2}}}^{2}}}
\end{equation}

\noindent where the same definitions as in Eq. \ref{eq:sig1} apply, $\rm{NB}_{1}$ refers to the detection filter (showing excess flux due to the line emission), and $\rm{NB}_{2}$ is the complementary filter where you gain the colour excess information. As before, $\Sigma$ depends on the local noise for a given source image location and so this was accounted for in the criteria for excess source selection. We found that the values of $\Sigma$ were more robust for the NB$-$NB colours, perhaps due to lower continuum colour corrections being required, and hence assess that a lower significance threshold of $\Sigma \ge 2.5$ was viable for the selection of NB$-$NB emission line candidates. A colour cut was again applied to separate the colour-excess sources from the random scatter at bright magnitudes for sources with zero NB $-$ NB colour. However, due to the F466N and F470N filters being close in wavelength, there was again less scatter of sources around zero NB $-$ NB colour (see Fig. \ref{fig:colour_mag_plots}), allowing lower colour-excess criteria. Table~\ref{table:excess_source_conditions} gives the chosen $\Sigma$ and colour criteria for each NB $-$ NB selection.

The emission line flux, $F_{\rm{line}}$ and equivalent width $EW_{\rm{line}}$ were calculated as:

\begin{equation}
\label{eq:f_line_2}
F_{\rm{line}} \ = \ \Delta \lambda_{\rm{NB_{1}}} \ (f_{\rm{NB_{1}}} \ - \ f_{\rm{NB_{2}}})
\end{equation}

\begin{equation}
\label{eq:ew_line_2}
EW_{\rm{line}} \ = \ \frac{ F_{\rm{line}}}{f_{\rm{NB_{2}}} }
\end{equation}

\noindent where $\Delta \lambda_{\rm{NB_{1}}}$ and $\Delta \lambda_{\rm{NB_{2}}}$ are the NB filter widths and where $f_{\rm{NB}_{1}}$ and $f_{\rm{NB}_{2}}$ were the measured flux densities in the excess selected and complementary NB filters, respectively, with units of erg s$^{-1}$ cm$^{-2}$ $\rm{\AA}^{-1}$. Note that unlike for the BB $-$ NB selected sources, there was (generally) no emission line contribution to the adjacent NB filter for NB $-$ NB selected sources and hence the difference in the equations for $F_{\rm{line}}$ and $EW_{\rm{line}}$.

Where a source was detected by both BB$-$NB and NB$-$NB excess criteria, we recommend the measurements of $\Sigma$, $F_{\rm{line}}$ and $EW_{\rm{line}}$ made using BB $-$ NB selections. This was because these were typically higher fidelity, and because there were potential biases using NB $-$ NB selections due to the overlapping transmission between the two narrow-band filters \citep[e.g.][]{2014MNRAS.438.1377S} where for some redshifts line flux could be picked up in each filter. This latter effect also means that the NB $-$ NB selections may be biased against the selection of high-$EW$ emission lines transmitting in both NB filters, but these line emitters should still be selected from the BB $-$ NB criteria (see combined selection criteria in Section \ref{sec:combined_selection}). The narrow-band selected H$\alpha$ sample analysed in this paper (see Section~\ref{sec:halpha_sample}) were all BB $-$ NB selected, and so we utilised the more accurate calculations of the above quantities for all sources.

\subsubsection{Combined excess source selection}
\label{sec:combined_selection}

For the F466N and F470N selected sources, a source met the criteria of being an excess source (and so a potential emission line galaxy) if the $\Sigma$ and colour criteria were both met for either the BB $-$ NB or the NB$_{1}$ $-$ NB$_{2}$ selection. Table \ref{table:excess_source_conditions} shows the breakdown of sources that met individual colour selection requirements and the combined `excess source' sample for each detection filter. Note, the BB $-$ NB combinations required higher $EW$ for selection ($EW_{\rm{obs}} \gtrsim 150\AA$, compared to the NB $-$ NB selections with $EW_{\rm{obs}} \gtrsim 90\AA$) but also have lower combined flux density errors, causing less scatter at the faint end of the colour-magnitude distribution for a given $\Sigma$. This results in a higher value for the limiting NB $-$ NB colour excess source selection at faint magnitudes compared to BB $-$ NB colour excess source selection. This can be seen in Fig. \ref{fig:colour_mag_plots} where the colour cut for each BB $-$ NB selection criteria meets the $\Sigma$ condition $\sim$1 mag fainter compared to the NB $-$ NB selections (despite the higher $EW$ cuts). Despite this, examination of the continuum colours of the narrow-band excess sources identified by the two different selection techniques show these to be similar, and so there was no evidence that the two selections were identifying substantially different populations. 

Combining the results for the F466N and F470N detections yields a total of 609 excess source candidates, with 266 sources (43.7 per cent) selected in both colour selections for their given detection filter, 209 (34.3 per cent) BB $-$ NB only selections and 134 (22.0 per cent) NB $-$ NB only selections. The BB $-$ NB selections yielded a higher fraction of excess source candidates overall, but we still gained a significant sample of additional candidates from the NB $-$ NB selections, increasing our final sample size. Fig. \ref{fig:colour_mag_plots} shows the colour-magnitude diagrams for the sample of F466N and F470N detected sources, highlighting those that met the excess conditions and the identified high-redshift emission line galaxy candidates (see Section~\ref{sec:halpha_sample} for the H$\alpha$ emitter sample selection process), as well as the four sets of narrow-band excess and colour selection criteria.

\begin{figure*}
\centering
    \includegraphics[width=\linewidth]{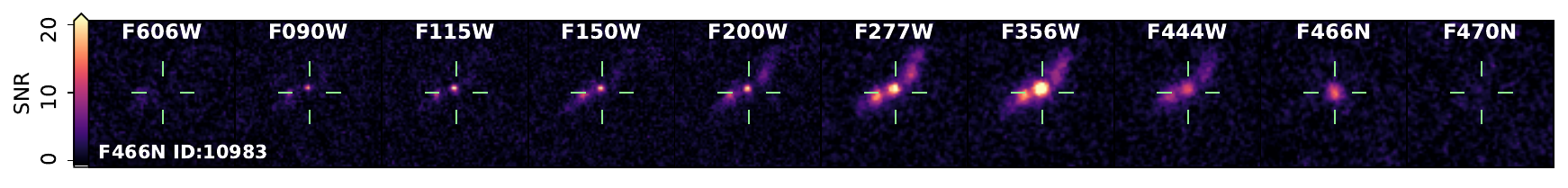}\vspace{-0.4cm}
  \includegraphics[width=\linewidth]{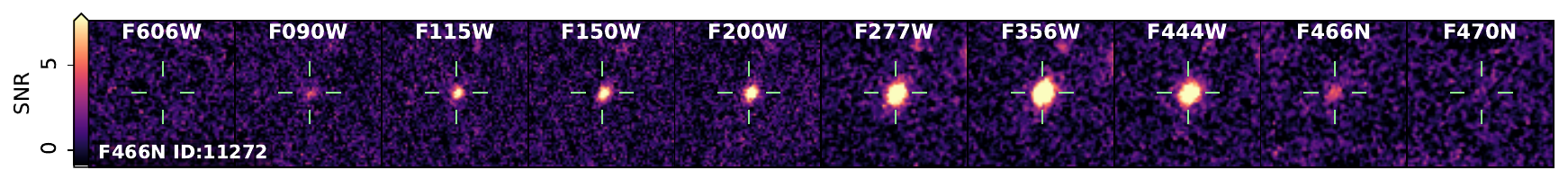}\vspace{-0.4cm}
  \includegraphics[width=\linewidth]{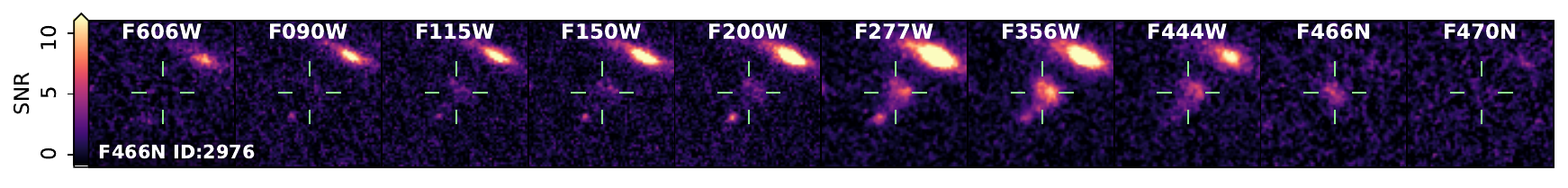}\vspace{-0.4cm}
  \includegraphics[width=\linewidth]{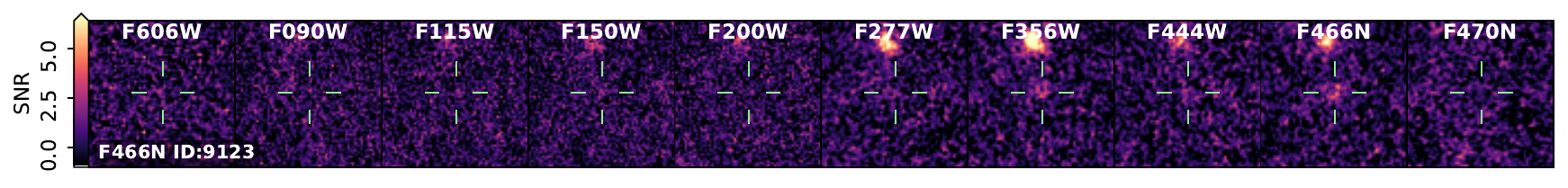}
\caption{Cutouts of the multi-wavelength imaging (2 $\times$ 2 arcsec) centred on narrow-band excess source candidates meeting the conditions described in Section~\ref{sec:line_cat}, meeting the H$\alpha$ photo-$z$ criteria described in Section~\ref{sec:halpha_sample}, and also passing visual inspection. The colour bar shows the range of SNR in the image cutouts for the following filters: F606W, F090W, F115W, F150W, F200W, F277W, F356W, F444W, F466N and F470N. The top two rows show robustly selected H$\alpha$ candidates with clear detections in the detection narrow-band and multi-wavelength imaging (including a dropout in the \emph{HST} F606W filter). Rows 3 and 4 show candidates close to the SNR detection threshold in the detection narrow-band, but with enough signal in 1 or more \emph{JWST} filters to be selected as a robust candidate.}
\label{fig:robust_sample}
\end{figure*}

\begin{figure*}
\centering
  \includegraphics[width=\linewidth]{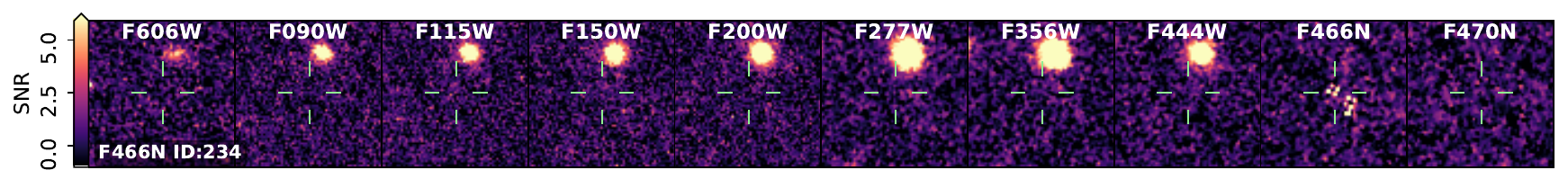}\vspace{-0.4cm}
    \includegraphics[width=\linewidth]{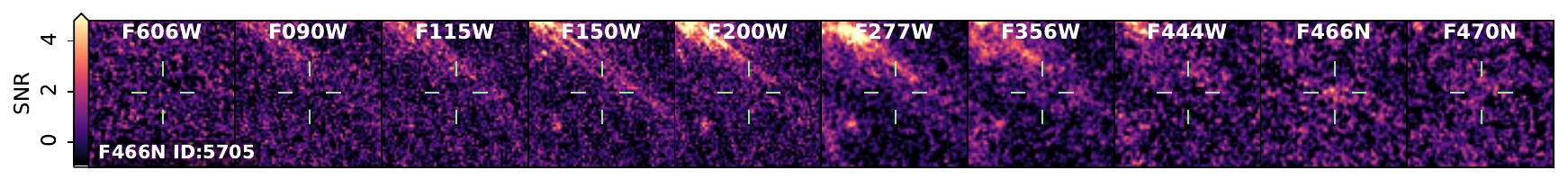}\vspace{-0.4cm}
    \includegraphics[width=\linewidth]{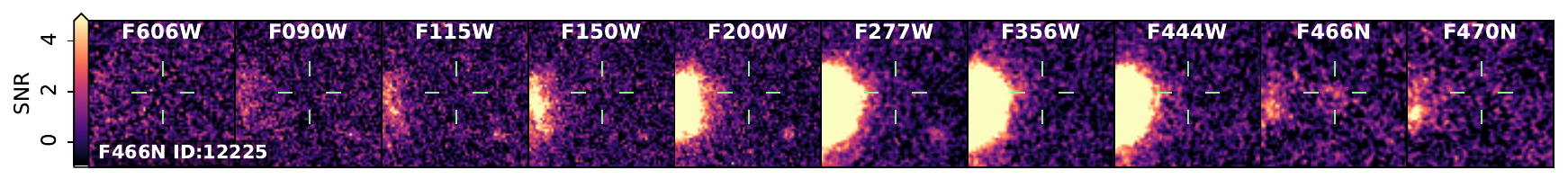}\vspace{-0.4cm}
  \includegraphics[width=\linewidth]{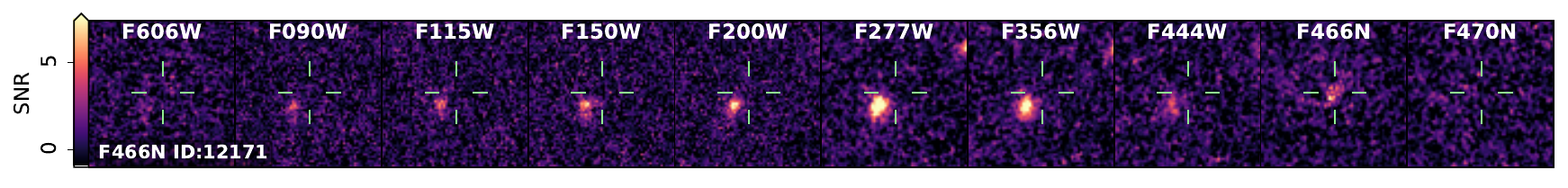}
\caption{Same as in Fig \ref{fig:robust_sample} but for sources failing the visual inspection. The first row shows clear contamination from strong cosmic ray hits in the narrow-band. The second row shows another cosmic ray hit in the narrow-band accompanied by contamination from diffraction spikes in the PRIMER broad-band images. The example in the third row appears to be diffuse light contamination left over from scattered light subtraction in the narrow-band image \citep[see details in][]{2024arXiv241009000D}. The final row shows an example of a questionable emitter, which was deemed insufficiently secure to make the robust sample analysed in the current paper: it was possibly genuine line emission, but it lies very close to a real source in the PRIMER images (with lower photometric redshift) with which it may be associated, and even if not, the nearby source contaminates the aperture photometric measurements, making the classification as an H$\alpha$ emitter insecure.}
\label{fig:rejected_sample}
\end{figure*}

\subsection{Selecting H$\alpha$ emitters at $z > 6$}
\label{sec:halpha_sample}

For our analysis, we wanted to select H$\alpha$ emission line galaxies at $z$ $>$ 6 from our excess source sample, to measure and compare their physical properties to other selections of star-forming galaxies. These measurements and comparisons required constraints on the multi-wavelength properties of our H$\alpha$ candidates from the rest-frame UV to rest-frame optical -- particularly for inferring the dust properties and star-formation activity of our sample. Therefore, we selected a robust sample of unambiguous H$\alpha$ emission line galaxy candidates from our excess source sample, requiring additional selection criteria to be met which were somewhat conservative. 

From our excess source sample, we first applied the following photometric redshift cut (taking the median of the primary peak in the redshift probability distribution described in Section~\ref{sec:photo_z}): 5.5 $\leq$ $z$ $\leq$ 6.5. Note that the narrowness of the photo-$z$ posteriors with the inclusion of the narrow-bands meant that this cut could, in fact, be much narrower (see Fig. \ref{fig:photo_z}) as all robust sources were constrained to $6.03 \ < \ z \ < \ 6.17$. We also required that the width of the primary redshift peak in the redshift probability distribution met the following condition: ($z_{\rm{1,max}} \ - \ z_{\rm{1,min}}) / (1 \ + \ z_{\rm{1,median}})$ < 0.4. Here, $z_{\rm{1,min}}$, $z_{\rm{1,max}}$ and $z_{\rm{1,median}}$ correspond to the lower bound, upper bound and the median of the primary 90 per cent highest probability density (HPD) credible interval (CI) peak respectively. This condition required our sample to have narrow and well-defined photo-$z$ posteriors, as expected for line emission driving strong narrow-band excess.

The sample of $z_{\rm{phot}}$ > 5.5 excess sources was still contaminated. In cases where contaminants showed prominent narrow-band detection and weak (or no) continuum detection, this would typically lead to a high-redshift photo-$z$ solution, meaning that much of the residual contamination in the narrow-band excess catalogue would fall in the H$\alpha$ and \oiiia samples. Therefore, we visually inspected all excess selected sources with $z_{\rm{phot}}$ > 5.5 that met the above $P(z)$ width criteria. To do this, the following cutouts were produced centred on each source: i) the full \emph{HST} and \emph{JWST} filter set at native resolution; ii) the narrow-band image for source detection, with convolution kernel applied (see convolution kernel described in Table \ref{table:sextractor_params}); and iii) a stack of all the PRIMER filters. From these images, we asked two questions: i) did the detection look genuine and/or significant? and ii) was the position aperture centred on the source detected in PRIMER? Each source was independently assessed by eight co-authors (CAP, PNB, KJD, DJM, RKC, ALP, HMOS and JPS) where each question was graded with a yes, no or maybe. The independent inspections showed mostly good agreement for the sample. For the sources with mixed feedback on their robustness, a subset of graders met together to explore the full multi-wavelength imaging available (including extended area and adaptable cut levels) and made a final consensus decision on the robustness of these sources. Fig. \ref{fig:robust_sample} shows examples of the visually inspected sources considered `robust' for our H$\alpha$ sample and Fig. \ref{fig:rejected_sample} shows sources rejected after visual inspection. The rejected sources were mainly obvious cases of cosmic ray hits or contamination by diffraction spikes. Others included convincing sources in the narrow-band that could be genuine emission line galaxies but were in close proximity to diffraction spikes or were offset in the broad-band imaging (like source 4 in Fig. \ref{fig:rejected_sample}). Note, the offsets in the broad-band imaging were much larger than the astrometric differences between the JELS and PRIMER imaging (see Section \ref{sec:sextractor}). Therefore, these sources were not deemed sufficiently robust to meet the conservative selection criteria for the current paper. We note that incompleteness as a result of this conservative selection will need to be accounted for in the analysis of the luminosity function.

\begin{table}
\centering
  \caption{Source counts for H$\alpha$ selected emission line galaxies after the following criteria was imposed: i) after cleaning the multi-wavelength detection catalogue in a given filter (final row in Table~\ref{table:source_cleaning}); ii) after applying the excess source criteria described in Section~\ref{sec:selection}; iii) after applying the photometric redshift selection criteria (discussed in Section~\ref{sec:halpha_sample}); and finally, iv) after visual inspection of the sources that passed the above criteria (see description in Section~\ref{sec:halpha_sample}).}
  \label{table:halpha_selection}
  \begin{tabular}{c c c}
    \hline
    \hline
    Selection Stage & Source Count & \\
     & F466N & F470N\\
    \hline
    Cleaned detection catalogue & 5645 & 6150 \\
    Pass excess source selection & 241 & 368 \\
    Pass photo-$z$ H$\alpha$ selection & 36 & 24 \\
    \hline
    Pass visual inspection and final count  & 30 & 5 \\
    \hline
  \end{tabular}
\centering
\end{table}

From this visual inspection, 35 out of 60 sources were then selected as the clean $z$ $>$ 6 H$\alpha$ emitter sample (30/36 selected from F466N and 5/24 selected from F470N; see Table~\ref{table:halpha_selection}). Multi-wavelength postage stamp images of all of these 35 sources are provided in Appendix~\ref{app:postage}. Comparing to Section~\ref{sec:combined_selection}, we found that all robust H$\alpha$ candidates met the BB $-$ NB conditions on $\Sigma$ and colour, whereas 26 out of the 35 (74.3 per cent) met the NB $-$ NB conditions. This shows that the BB $-$ NB selections captured more of our robust sample and that we would miss sources purely on a NB $-$ NB selection. In addition, we also found that the median primary photo-$z$ width ($z_{\rm{1,max}} \ - \ z_{\rm{1,min}}$; see definition in Section \ref{sec:photo_z}) for our H$\alpha$ sample was $\sim$ 0.04 and with a maximum value of 0.065 - both below the FWHM of both NB filters covering a redshift range $\Delta z \sim 0.08$. This highlights the high photo-$z$ accuracy achieved when  the emission line is placed within the narrow transmission function (and typically towards higher transmission). Note, as discussed in Section~\ref{sec:low_sig_det}, there was still residual scattered light contamination in the image reduction stage in both narrow-band mosaics, but this was more prevalent in the F470N mosaic; this may explain the lower success rate in visually confirming H$\alpha$ emitters from this filter. Nevertheless, the stark contrast in source density between the two filters after visual inspection could be evidence for significant clustering of H$\alpha$ emitters in the narrow redshift slice covered by the F466N filter compared to the F470N filter. This aligns with previous studies which showed significant clustering of spectroscopically confirmed $z \sim 6$ $i$-band drop-out galaxies in the COSMOS field \citep{2024MNRAS.527.6591B}. This result was also not a sensitivity effect as the `excess source' selections shown in the right panel of Fig. \ref{fig:photo_z} show a higher number density of Paschen line emitter candidates in the F470N filter compared to the F466N filter. Despite the above, caution must still be exercised given the small sample size of our H$\alpha$ emitters and the small area (and hence cosmic volume) probed with this one field.

\begin{table*}
\centering
  \caption{Key \textsc{Bagpipes} model parameters and priors used to fit our photometric data and thus determine the physical properties of our sample of H$\alpha$ emitters at $z$ $>$ 6. The model is described in Section~\ref{sec:sed_fitting}. Note, the oldest SFH bin edge is no older than the age of the Universe at the redshift estimated for the source. Logarithmic priors are all applied in base ten.}
  \label{table:bagpipes_params}
  \begin{threeparttable}  
  \begin{tabular}{c c c c c}
    \hline
    \hline
    Component/Model & Parameter & Symbol / Unit & Range & Prior\\
    \hline
    \hline
    General & Redshift & - & $\left( z_{\rm{EAZY, 16}}, z_{\rm{EAZY, 84}} \right)$ $\lor$ & Uniform\\
    & & & $\left( z_{\rm{EAZY, 50}}  -  0.1,  z_{\rm{EAZY, 50}}  +  0.1 \right)$\tnotex{tn:1} $^{,}$ \tnotex{tn:2} &  \\
    \hline
    SFH (Continuity) & Total stellar mass formed & $M_{\star}$ / M$_{\odot}$ & $\left( 10^{6}, 10^{13} \right)$ & Logarithmic\\
    & Stellar metallicity & $Z_{*}$ / Z$_{\odot}$ & (0.0005, 2.0) & Logarithmic\\
    & Continuity bins & $\rm{dSFR_{i}}$ / Myr & (0, 3, 10, 30, 100, 300,750) & Student's t\\ 
    \hline
    Nebular Emission & Ionisation Parameter & $\log (U)$ & (-4, -1) & Uniform\\
    \hline
    Dust (Salim) & V−band attenuation & $A_{\rm{V}}$ / mag & (0, 4) & Uniform\\
    & Deviation from Calzetti Slope & $\delta$ & (-1.2, 0.4) & Uniform\\
    & 2175$\AA$ bump strength & $B$ & 0 & -\\
    & Factor on $A_{\rm{V}}$ for stars in birth clouds & $\eta$ & (0.01, 4) & log-Gaussian\\
    & with respect to the ISM &  &  & \\
    \hline
  \end{tabular}

  \begin{tablenotes}
    \item[1] \label{tn:1} The 16th, 50th and 84th percentiles from the \textsc{EAZY-Py} photometric redshift probability distribution for a given source are denoted $z_{\rm{EAZY, 16}}$, $z_{\rm{EAZY, 84}}$ and $z_{\rm{EAZY, 84}}$.
    \item[2] \label{tn:2} The redshift prior was set between the 16th and 84th percentiles of the EAZY-Py photometric redshift probability distribution ($z_{\rm{EAZY, 16}}$ and $z_{\rm{EAZY, 84}}$ respectively) for a given source unless this range was less than $\pm$0.1. Otherwise, the prior was set to the 50th percentile ($z_{\rm{EAZY, 50}}$) $\pm$ 0.1.
  \end{tablenotes}  
  
  \end{threeparttable}  
\centering
\end{table*}

\subsection{Selecting $z$ $\sim$ 6 sources from the F356W detected catalogue}
\label{sec:f356w_z=6_sources}

Here, similar to Section~\ref{sec:halpha_sample}, we selected $z$ $\sim$ 6 candidates using the PRIMER F356W detection catalogue (see Section~\ref{sec:narrowband_cat}). This enabled comparisons of the selections and physical properties of our narrow-band selected sample of H$\alpha$ emission line galaxies to galaxies around the same epoch that were selected based on their photometric redshifts (which were driven by their rest-frame UV/optical broad-band photometric measurements). These F356W detected $z$ $\sim$ 6 sources were selected to have: i) their 50th percentile photometric redshift posterior in the range: 5.5 $<$ $z_{\rm{phot}}$ $<$ 6.5; and ii) the integrated redshift posterior distribution $P(z)$ $\geq$ 0.7 over the same redshift range. This is typical for robust selections of galaxies with broad-band photometric data and $z_{\rm{phot}}$ values primarily driven by the Lyman-break feature \citep[e.g.][]{2019A&A...622A...3D}. This selection yielded a sample of 568 galaxies, though note that the F356W $z$ $\sim$ 6 sample probes a wider redshift range compared to our narrow-band selected H$\alpha$ sample (F466N: $6.05 < z < 6.13$ and F470N: $6.14 < z < 6.21$ -- based on the effective widths for each filter). We note that some of the F356W-detected sources may show an excess in the narrow-band filters corresponding to H$\alpha$ emission line galaxies at $z \sim 6.1$ and that some of the narrow-band selected H$\alpha$ emitters may also be selected in the F356W-detected catalogue if they have strong enough optical continuum – meaning the two selections are not mutually exclusive. What is important here is that we mimic typical galaxy selections utilising single or stacked broad-band detections at a given redshift epoch.

\section{Physical Properties of the H$\alpha$ Emission Line Candidates}\label{sec:physical_properties}

In this section, we examined the physical properties of these $z$ $>$ 6 H$\alpha$ emitters and compare the nature of their star-formation activity and dust properties to those of the sample of objects at $z$ $\sim$ 6 selected photometrically from the F356W catalogue. SED fitting was performed to obtain estimations for a range of galaxy properties including stellar mass, SFR and dust attenuation. In addition, empirical measurements were made directly from the available photometry to calculate their UV-continuum slopes ($\beta$), absolute magnitudes ($M_{\rm{UV}}$) and UV-continuum and H$\alpha$ SFRs (the latter only relevant to the H$\alpha$ sample and not the F356W-detected sample with no narrow-band measurements to obtain H$\alpha$ emission line properties).

\subsection{Spectral energy distribution fitting and galaxy properties}
\label{sec:sed_fitting}

We performed SED fitting on our data using the \textsc{Bagpipes} spectral fitting code \citep{2018MNRAS.480.4379C} and utilised the \textsc{BPASS} \citep{2017PASA...34...58E,2018MNRAS.479...75S} stellar population synthesis (SPS) code (assuming a \citet{1993MNRAS.262..545K} IMF with a cutoff at 100 $\rm{M_{\odot}}$) and the \textsc{Cloudy} photo-ionisation code \citep{2017RMxAA..53..385F} to compute nebular emission lines. In addition, we implement the \citet{2018ApJ...859...11S} dust attenuation model and the \citet{2019ApJ...876....3L} continuity non-parametric star-formation history (SFH) model. The SED fitting models and priors used are summarised in Table \ref{table:bagpipes_params}. 

These models and prior assumptions were selected to stay consistent with the assumptions used in the empirical measurements (see Section ~\ref{sec:halpha_prop}) and to balance the number of free parameters (and hence number of degrees of freedom of the SED fitting) with the number and quality of photometric measurements available, so that we were not over-fitting and over-interpreting the data. From these SED fits, we extracted the posteriors for the stellar mass ($M_{\star}$), star-formation rates averaged over 10 Myr and 100 Myr timescales ($\rm{SFR_{10}}$ and $\rm{SFR_{100}}$ respectively), the ratio of these SFRs ($\rm{SFR_{10}}$/$\rm{SFR_{100}}$) and the dust attenuation from the stellar continuum in the $V$-band ($A_{V}$). 

Alternative SFH models were investigated, but this was found to have negligible impact on the stellar mass estimates of our sample. In addition, the non-parametric SFH model allowed freedom for sudden increases in the recent SFR, potentially important in our H$\alpha$ sample, and hence leading to less biased measurements of $\rm{SFR_{10}}$ (important when interpreting  $\rm{SFR_{10}}$/$\rm{SFR_{100}}$). Regarding the choice of dust models, observational studies of high-redshift star-forming galaxies present a conflicting picture on the nature of the attenuation curve at this epoch. Some studies have found that applying the SMC attenuation curve \citep{2003ApJ...594..279G} best re-produces the conditions in star-forming galaxies from cosmic noon towards the Epoch of Reionization \citep[e.g][]{2016A&A...587A.122A,2018ApJ...853...56R}, while others have found application of the \citet{2000ApJ...533..682C} attenuation slope was more appropriate \citep[e.g.][]{2018MNRAS.479.4355K,2018MNRAS.479...25M}. Given these results, we fit the photometry using the \citet{2018ApJ...859...11S} dust attenuation model, which fits the deviation from the \citet{2000ApJ...533..682C} attenuation law, the parameter $\delta$, thus enabling a range of slopes to be fitted. The fits for $\delta$ were relatively unconstrained for our sample but the stacked posteriors showed a median $\delta$ = $-$0.1 indicating that our sources have slopes that, on average, do not deviate significantly from the fiducial \citet{2000ApJ...533..682C} slope. The factor on $A_{V}$ for the stars in birth clouds with respect to the ISM \citep[$\eta$, equivalent to the factor between the emission line and continuum extinction but in the $V$-band; e.g.][]{2000ApJ...533..682C} was also unconstrained in the SED fitting. Therefore, our assumption for calculating the dust corrected H$\alpha$ luminosities and SFRs (see Section \ref{sec:halpha_prop}) was based on observational evidence of the nebular attenuation from previous studies \citep[e.g.][]{2020ApJ...902..123R}.

To investigate the robustness of our SED-fitted results, in Appendix~\ref{app:posteriors} we examined the stacked and normalised posterior distributions that we found for the SFR ratio ($\rm{SFR_{10}/SFR_{100}}$) and the dust attenuation factor $A_V$. We demonstrate that the resulting posterior distributions differ strongly from the input prior distribution, confirming that our results were robustly being driven by the data and not by the choice of priors.

\subsection{Empirical measurements of galaxy properties}
\label{sec:empirical}

\subsubsection{H$\alpha$ emission line luminosity, equivalent width and star-formation rate}
\label{sec:halpha_prop}

As discussed in Section~\ref{sec:halpha_sample}, all of our H$\alpha$ sources were BB $-$ NB selected and so we could calculate the line fluxes ($F_{\rm{H\alpha}}$) for these consistently using Eq. \ref{eq:f_line_1};  as discussed in Section~\ref{sec:nb_nb_selection}, this was also likely to be more robust than line fluxes calculated from the NB $-$ NB colours. The luminosity distance $D_{L}$ was then calculated for each source utilising their measured photometric redshifts and combined with their line fluxes to calculate their line luminosities: $L_{\rm{H\alpha}} = 4 \pi \ D^{2}_{L} \ F_{\rm{H\alpha}}$. The H$\alpha$ equivalent widths ($EW_{\rm{H\alpha}}$) for the sample were calculated by taking the observed frame measurements calculated using Eq. \ref{eq:ew_line_1} and dividing through by ($1 + z$) to obtain the rest-frame measurement for $EW_{\rm{H\alpha}}$. For objects where the SNR(F444W) $<$ 3, we take a 3$\sigma$ upper limit on the F444W flux density measurement to obtain a lower limit for $EW_{\rm{H\alpha}}$.

Although the H$\alpha$ emission line is less dust-affected than the UV-continuum, it was still essential to correct these H$\alpha$ luminosities for dust extinction. To do this, we made use of the $V$-band stellar continuum attenuation measurements, $A_{V}$, that we derive in our SED fitting (see Section~\ref{sec:sed_fitting}). These $A_{V}$ measurements needed to be converted into extinction in the H$\alpha$ emission line, $A_{\rm{H\alpha}}$ and this depended on the adopted dust attenuation law, but also on the relationship between the stellar continuum reddening, $E(B \ − \ V)_{\rm{cont}}$, and the nebular reddening, $E(B \ − \ V)_{\rm{neb}}$. The latter has been studied extensively in star-forming galaxies \citep[e.g.][]{2000ApJ...533..682C,2013ApJ...777L...8K,2014ApJ...788...86P,2015ApJ...806..259R} and shows inconsistent results, particularly when probing galaxies at high-redshift. In this work, we scale our $A_{V}$ measurements using a \citet{2000ApJ...533..682C} attenuation curve to obtain the attenuation in the stellar continuum at the H$\alpha$ wavelength of 6563$\AA$ and then multiply by 2.27 to get the extinction on the emission line, $A_{\rm{H\alpha}}$ \citep[assuming $E(B \ − \ V)_{\rm{cont}}$ = 0.44 $E(B \ − \ V)_{\rm{neb}}$ from][]{2000ApJ...533..682C}. We explored the impact of these assumptions in Section~\ref{sec:discussion}.

For H$\alpha$ emission line galaxies, the line flux measured from the narrow-band will include some contamination from \nii line emission where at least one of the \niil and \niih lines will fall withing the narrow-band filter \citep[e.g.][]{2012MNRAS.420.1926S}. Studies of narrow-band selected H$\alpha$ emitters at lower redshifts \citep[e.g.][]{2015MNRAS.451.2303S}, showed that the relationship of the \nii/H$\alpha$ ratio as a function of $EW$(H$\alpha + \nii$) derived in the local Universe from SDSS was applicable out to $z \sim 1$; if this holds to $z \sim 6$ then, given the $EW$ of our H$\alpha$ emitters, this would correspond to correction factors of $\sim$10--20 per cent (0.04-0.08 dex) for the H$\alpha$ luminosities. However, \citet{2023ApJ...950L...1S} demonstrated using early \emph{JWST} data that $z$ = 5.0 $−$ 6.5 star-forming galaxies show smaller typical ratios of $\log_{10}(\rm{[NII]6585/H\alpha})$ = −1.31 corresponding to 0.021 dex corrections to $L_{\rm{H\alpha}}$. A caveat here is that this sample was rest-frame UV selected and not emission line selected, and there is evidence that nitrogen over-abundances could be short-lived phases that take place during strong starbursts \citep[e.g.][]{2025ApJ...980..225T}, which H$\alpha$ emission will better trace. In addition, the discovery of nitrogen over-abundance observed at high-redshift \citep[e.g.][]{2023MNRAS.523.3516C} means future studies may have to consider \nii contamination as non-negligible if further spectroscopic follow-up shows this for a large sample of high-redshift galaxies. However, given the lack of robust measurements, and an expectation that the correction would be $\lesssim 0.05$ dex, we made no correction for potential \nii contamination. We then measured the SFR from the H$\alpha$ luminosity:

\begin{equation}
SFR_{\rm{H \alpha}} = \kappa_{\rm{H \alpha}} \ L_{\rm{H \alpha}}.
\label{eq:sfr_halpha}
\end{equation}

\noindent Here, we adopted $\kappa_{\rm{H\alpha}}$ = 10$^{-41.64}$ ($\rm{M_{\odot} \ yr^{-1}}$)/(erg s$^{-1}$) SFR calibration conversion factor following the analysis of \citet{2019ApJ...871..128T} which is a factor of $\sim$2.34 smaller than the \citet{2012ARA&A..50..531K} calibration. The calibration assumptions matched those implemented in our SED fitting (See Section~\ref{sec:sed_fitting}) which utilised the \textsc{BPASS} SPS models, assumed a \citet{1993MNRAS.262..545K} initial mass function (IMF) with an upper-mass limit of 100 $\rm{M_{\odot}}$ and a metallicity of $Z$ = 0.002. Our adopted $\kappa_{\rm{H\alpha}}$ is similar to other studies utilising low-metallicity \textsc{BPASS} SPS models \citep[e.g.][]{2022ApJ...926...31R,2023ApJ...950L...1S} and was guided by the evolving mass-metallicity relation \citep[e.g.][]{2021ApJ...914...19S}. This relation reflects the greater ionising photon production efficiencies in lower-metallicity massive stars in binary systems observed in lower stellar mass and high-redshift star-forming galaxies. 

\begin{figure}
    \centering
    \begin{subfigure}{0.45\textwidth}
        \includegraphics[width=0.97\linewidth]{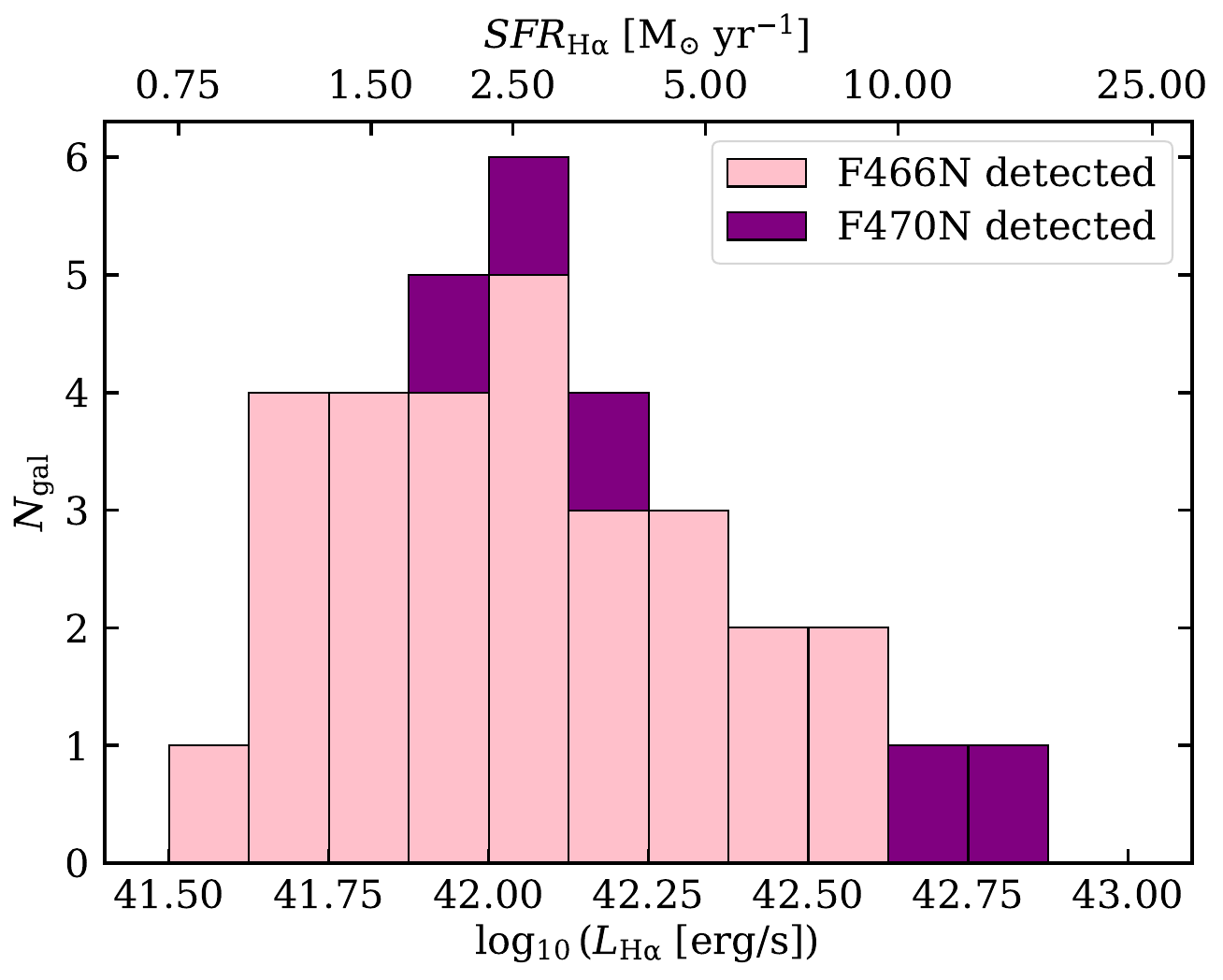}
    \end{subfigure}\hfil

    \medskip
    \begin{subfigure}{0.485\textwidth}
        \includegraphics[width=0.97\linewidth]{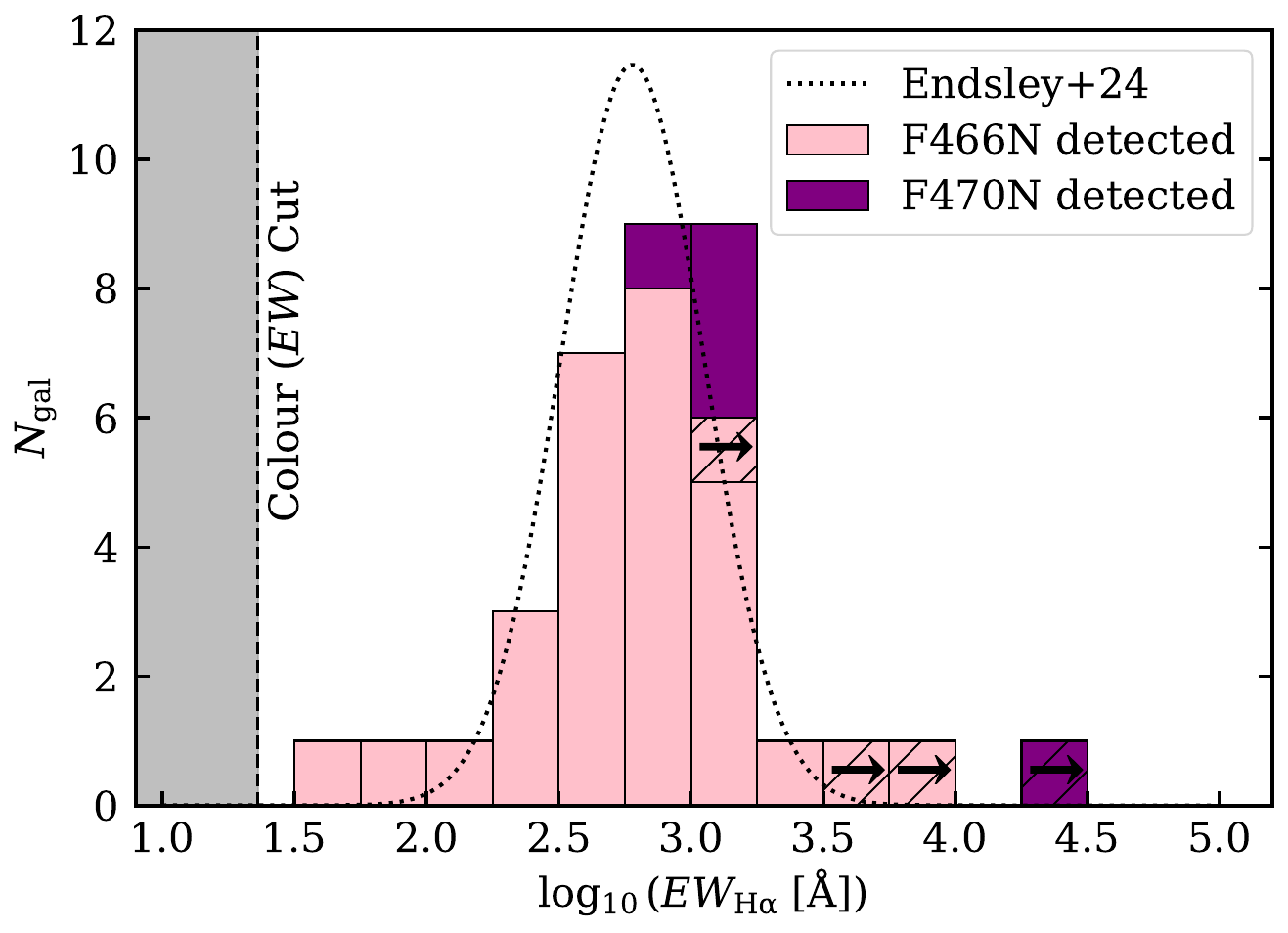}
    \end{subfigure}\hfil

    \caption{Top panel: Histogram of the dust-corrected H$\alpha$ emission line luminosities $L_{\rm{H\alpha}}$ (and corresponding SFRs - see Section~\ref{sec:halpha_prop}) for the final sample of $z$ $>$ 6 H$\alpha$ emitters. The H$\alpha$ sources in each bin were colour coded pink and purple for detections in the F466N and F470N filters respectively. Note that the two potential AGN candidates are excluded from this plot as their H$\alpha$ luminosities lie above the range plotted. Bottom panel: Histograms of the H$\alpha$ rest-frame equivalent widths $EW_{\rm{H\alpha}}$ for the final sample of $z$ $>$ 6 H$\alpha$ emitters. The sources in each bin are coloured coded in the same way as in the top panel. The hatched bins with the right-pointing arrows show objects whose SNR(F444W) $<$ 3 and so 3$\sigma$ upper limits on the F444W flux density measurements were used to calculate the lower limits on $EW_{\rm{H\alpha}}$. The grey shaded region shows the $EW_{\rm{H\alpha}}$ region below the translated colour cuts (see Fig. \ref{fig:colour_mag_plots}) corresponding to the excess source selection criteria ($EW_{\rm{H\alpha}} \lesssim 20 \AA$) which has no effect on our sample selection. For comparison, overplotted in the dotted line is $z \sim 6$ $EW_{\rm{H\alpha}}$ log-normal distribution (normalised to the same area as our histogram) from \citet{2024MNRAS.533.1111E}, which shows good agreement generally.}
    
    \label{fig:l_sfr_halpha}

\end{figure}

We show in Fig. \ref{fig:l_sfr_halpha} that dust-corrected H$\alpha$ luminosities for our sample of H$\alpha$ emitters at $z$ $>$ 6 span a range of 41.6 $\lesssim$ $\log_{10}(L_{\rm{H\alpha}}/\text{erg s}^{-1}$) $\lesssim$ 42.8, corresponding to 0.9 $\lesssim$ $\rm{SFR_{H\alpha}}$ [$\rm{M_{\odot} \ yr^{-1}}$] $\lesssim$ 15. In addition, we show that the H$\alpha$ equivalent width distribution clearly peaks between $\sim$300 and $\sim$2000 $\AA$.   This is in good agreement with the distribution derived in \citet{2024MNRAS.533.1111E} for \emph{JWST}/\emph{HST} drop-out selected sources at $z \sim 6$, although our selection includes sources with a larger dynamic range in $EW$. We do not over-interpret this finding due to the low number statistics, but we do note that the vast majority of our sample are found to have $EW$s well above the observational limit, indicating that the observed distribution is not driven by observational biases.

\subsubsection{UV-continuum luminosity and star-formation rate}
\label{sec:uv_slope}

We then compared the H$\alpha$ emission line derived properties for our $z$ $>$ 6 sample to those derived from the UV-continuum to investigate both the reddening and star-formation timescales of these sources. Our photometric data covered rest-frame UV-continuum wavelengths for our sample of $z$ $\sim$ 6.1 H$\alpha$ candidates from below the Lyman break feature to above 3000$\AA$, and so allows us to measure both the UV-continuum slope and the UV luminosity.

For our sample, the following \emph{HST}/WFC3IR and \emph{JWST}/NIRCam photometric measurements were utilised in fitting the rest-frame UV spectrum: F115W, F125W, F140W, F150W, F160W and F200W (see Fig. \ref{fig:filters}). These measurements probed only the rest-frame UV-continuum where $\lambda_{\rm{rest}}$ $\leq$ 3000$\rm{\AA}$ and also avoided contamination from the Ly$\alpha$ emission line and intergalactic medium (IGM) absorption at $\lambda_{\rm{rest}}$ $\leq$ 1216$\rm{\AA}$. We fitted the rest-frame UV-continuum spectrum for each source using the photometric measurements with SNR $>$ 1. The fit utilised a non-linear least squares approach, weighted by the absolute errors in the photometric measurements using the  \textsc{SciPy} \texttt{curve$\_$fit} function \citep{2020zndo...4100507V}. We fitted the UV-continuum slope ($\beta$) and spectrum normalisation factor ($f_{0}$) as follows: $f_{\lambda}$ = $f_{0}$ $(\lambda / \lambda_{0})^{\beta}$, where $\lambda_{0}$ = 1500$\AA$. The 1$\sigma$ errors in both parameters were extracted from the output fitted co-variance matrix.

For the UV-continuum, we extracted the flux density at rest-frame 1500$\rm{\AA}$ ($f_{1500\AA}$) directly from the fitted spectrum of each source (accounting for the (1+$z$) factor for redshifting to the observed-frame wavelength). We again utilised the inferred luminosity distances to measure $L_{\lambda}$ at 1500$\AA$: $L_{1500 \AA } = 4 \pi \ D^{2}_{L} \ f_{1500 \AA }$ [erg s$^{-1}$ $\rm{\AA}^{-1}$]. We then converted  $L_{1500 \AA }$ (uncorrected for dust) to $L_{\nu}$ [erg s$^{-1}$ $\rm{Hz}^{-1}$] to calculate the observed absolute UV-continuum magnitude, using the relation from \citet{1983ApJ...266..713O}:

\begin{equation}
M_{\rm{UV}} = -2.5 \ \log_{10} (L_{\nu}) \ + \ 51.63
\label{eq:M_uv}
\end{equation}

\noindent As in Section~\ref{sec:halpha_prop}, we utilised the measured $V$-band stellar continuum attenuation $A_{\rm{V}}$ from SED fitting to correct the luminosity measurements. Given the results of Section~\ref{sec:sed_fitting} preferring a \citet{2000ApJ...533..682C} attenuation curve, we scaled the measured $A_{V}$ values using this attenuation curve to obtain stellar continuum attenuation at 1500$\AA$ ($A_{\rm{1500\AA}}$) to dust-correct $L_{1500 \AA }$. We calculate the SFR for our sample from the UV-continuum using Eq. \ref{eq:sfr_uv}:

\begin{equation}
SFR_{\rm{UV}} = \kappa_{\rm{UV}} \ L_{\rm{1500 \AA}}
\label{eq:sfr_uv}
\end{equation}

\noindent Here we adopted $\kappa_{\rm{UV}}$ = 10$^{-43.46}$ ($\rm{M_{\odot} \ yr^{-1}}$)/(erg s$^{-1}$) SFR calibration conversion factor utilising the same set of assumptions for the choice of SPS model, metallicity and IMF as in our $\rm{SFR_{H\alpha}}$ calculations (see Section~\ref{sec:halpha_prop}).

\subsection{Combined Results from SED Fitting and Empirical Calculations}
\label{sec:combined_results}

\subsubsection{Stellar masses}
\label{sec:Stellar_masses}

Both our narrow-band selected H$\alpha$ emission line galaxy sample and the F356W-detected $z$ $\sim$ 6 sample span the following stellar mass range: 7.4 $\lesssim$ $\log_{10}(M_{\star}/\rm{M_{\odot}})$ $\lesssim$ 11.0. For our H$\alpha$ sample, most of our sources have stellar masses $\lesssim$ 10$^{9.5}$ $\rm{M_{\odot}}$ but we identified two candidates (see source ID 2768 and 7810 in Fig. \ref{fig:halpha_sample_cutouts_v1}) with fitted stellar masses of 10$^{10.8}$ and 10$^{9.8}$ $\rm{M_{\odot}}$, respectively, which have very compact and PSF-like morphologies (Stephenson et al., in preparation). Given this early epoch, these stellar mass measurements are large and their morphologies suggest that there may be significant AGN activity within these galaxies contributing to their emission which, at least for source 2768, is probably leading to an over-estimation of the stellar mass; this may also affect other properties derived from the SEDs of these two objects. The remaining objects in the sample were more extended, although we cannot rule out that they contain some AGN activity.

\subsubsection{Star-formation rates and timescales}
\label{sec:SFRs}

For this analysis, we excluded the two most massive sources described in Section~\ref{sec:Stellar_masses} which have $\rm{SFR_{H\alpha}}$ $\sim$ 10$^{2}$ and 10$^{3}$ M$_{\odot}$ yr$^{-1}$ for the 10$^{9.8}$ and 10$^{10.8}$ $\rm{M_{\odot}}$ sources, respectively. Our empirical measurements ($\rm{SFR_{H\alpha}}$ and $\rm{SFR_{UV}}$), coupled with the SED-derived SFRs over different timescales ($\rm{SFR_{10}}$ and $\rm{SFR_{100}}$), allowed us to investigate both the agreement between different star-formation indicators and the timescales of star-formation activity in our H$\alpha$ sample. As discussed in Section~\ref{sec:uv_slope}, the SFRs measured from the rest-frame UV-continuum and the H$\alpha$ emission line respond to instantaneous changes in star-formation activity over differing timescales. 

\begin{figure}
    \centering
    \includegraphics[width=0.48\textwidth]{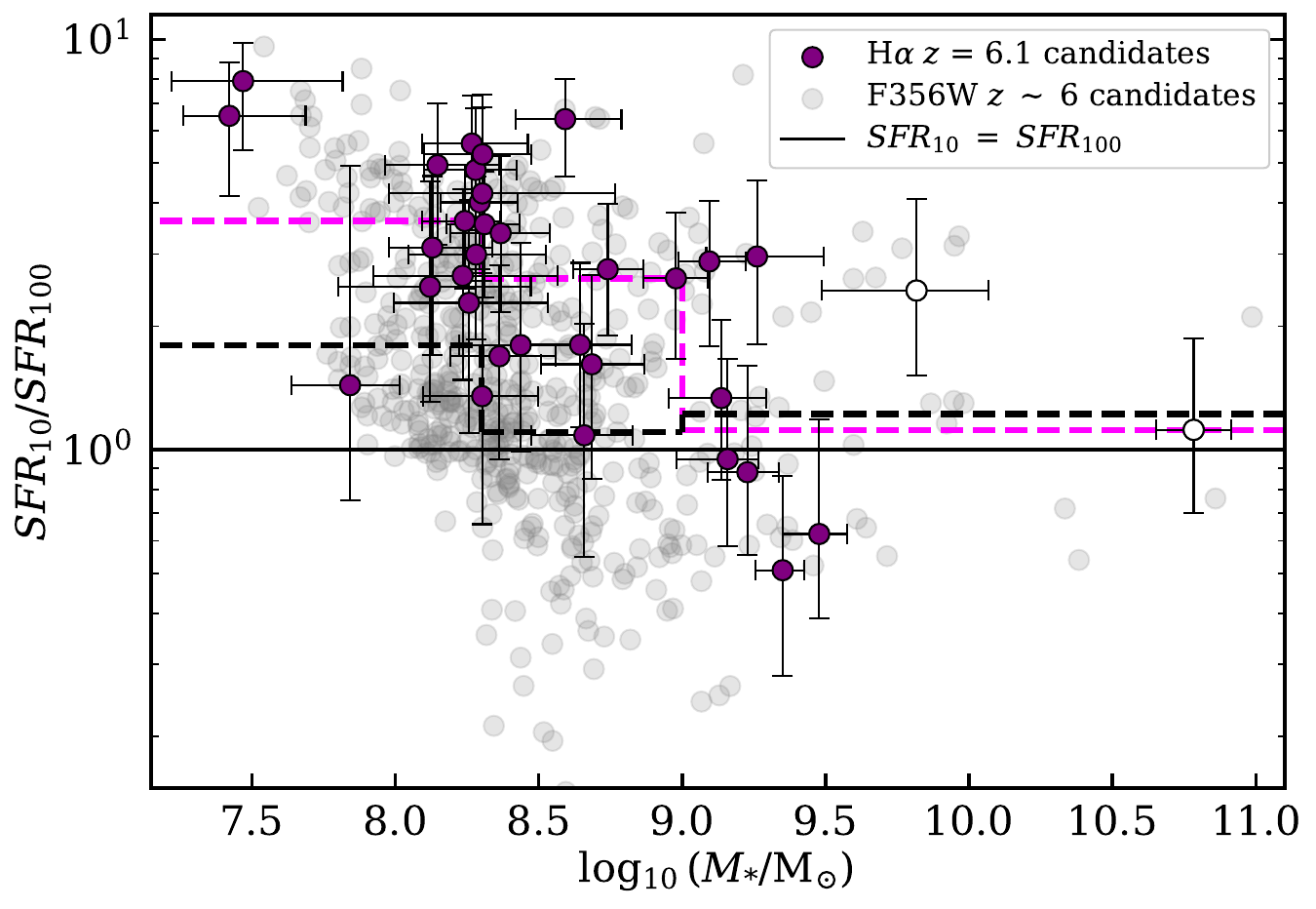}

    \caption{The $\rm{SFR_{10}/SFR_{100}}$ as a function of stellar mass for our H$\alpha$ sample (purple points) and for the F356W $z$ $\sim$ 6 sample (transparent grey points), with all quantities derived from SED fitting. The error bars show the 16th and 84th percentiles for the posterior distributions in $\rm{SFR_{10}/SFR_{100}}$ and stellar mass for our H$\alpha$ sample. The stellar mass uncertainties are similar in magnitude for the F356W sample and the $\rm{SFR_{10}/SFR_{100}}$ error bars are much larger compared to our H$\alpha$ sample (see stacked posterior distribution in Fig. \ref{fig:sed_post_dist}) but we do not show these to keep the figure clear. The two potential AGN candidates are plotted with open symbols.}
    
    \label{fig:sfr_10_100_vs_mass}

\end{figure}

When we investigated the robustness of our SED fitting (Appendix~\ref{app:posteriors}), we found that the typical $\rm{SFR_{10}/SFR_{100}}$ ratios of our H$\alpha$ emitters was above unity, and that this was even more prevalent at lower stellar mass, $\lesssim$ 10$^{9.0}$ $\rm{M_{\odot}}$. This indicates that our sample of H$\alpha$ emitters are exhibiting heightened recent star-formation activity, particularly at lower stellar masses. To illustrate this, in Fig. \ref{fig:sfr_10_100_vs_mass} we show the $\rm{SFR_{10}/SFR_{100}}$ ratio plotted against stellar mass for our samples and the median $\rm{SFR_{10}/SFR_{100}}$ split into the same stellar mass bins as the stacked posterior distributions described in Fig. \ref{fig:sed_post_dist}. This figure clearly shows both the offset above unity and the stellar mass trend, for H$\alpha$ emitters. For comparison, on the same plot we show the F356W-detected $z$ $\sim$ 6 sample. This displays a much larger scatter in $\rm{SFR_{10}/SFR_{100}}$ values with median values around unity (apart from a slight rise in the lowest stellar mass bin), which highlights the difficulty in constraining the recent SFR for this sample in the absence of the narrow-band H$\alpha$ measurement. The F356W-detected sample did not show the same trend of increasing $\rm{SFR_{10}/SFR_{100}}$ ratio towards lower stellar masses, giving additional confidence that the trend for H$\alpha$ emitters was not driven by our fitting procedure.

\begin{figure}

\centering
\begin{subfigure}{0.48\textwidth}
  \includegraphics[width=0.97\linewidth]{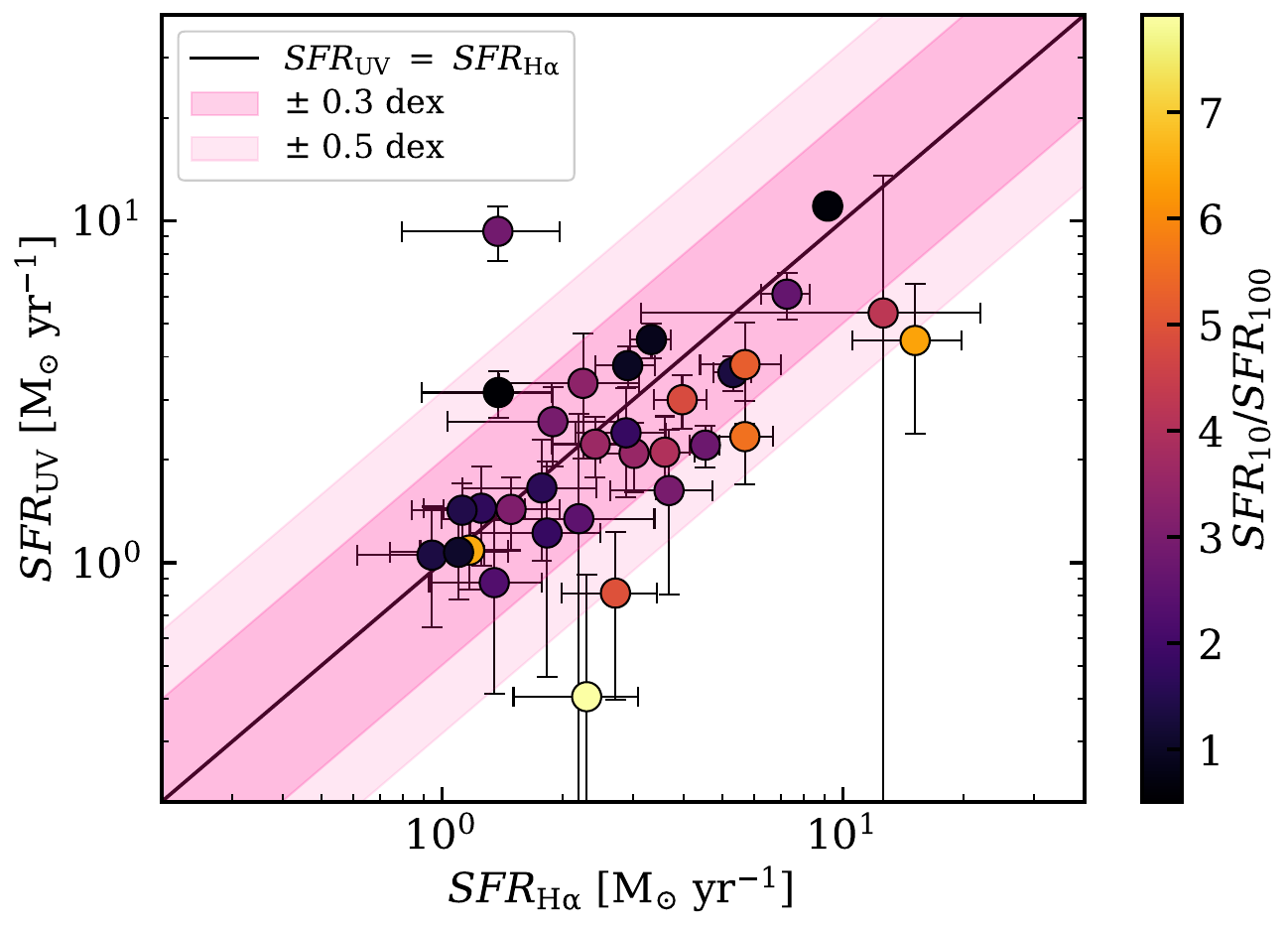}
\end{subfigure}\hfil

\medskip
\begin{subfigure}{0.48\textwidth}
    \includegraphics[width=0.97\linewidth]{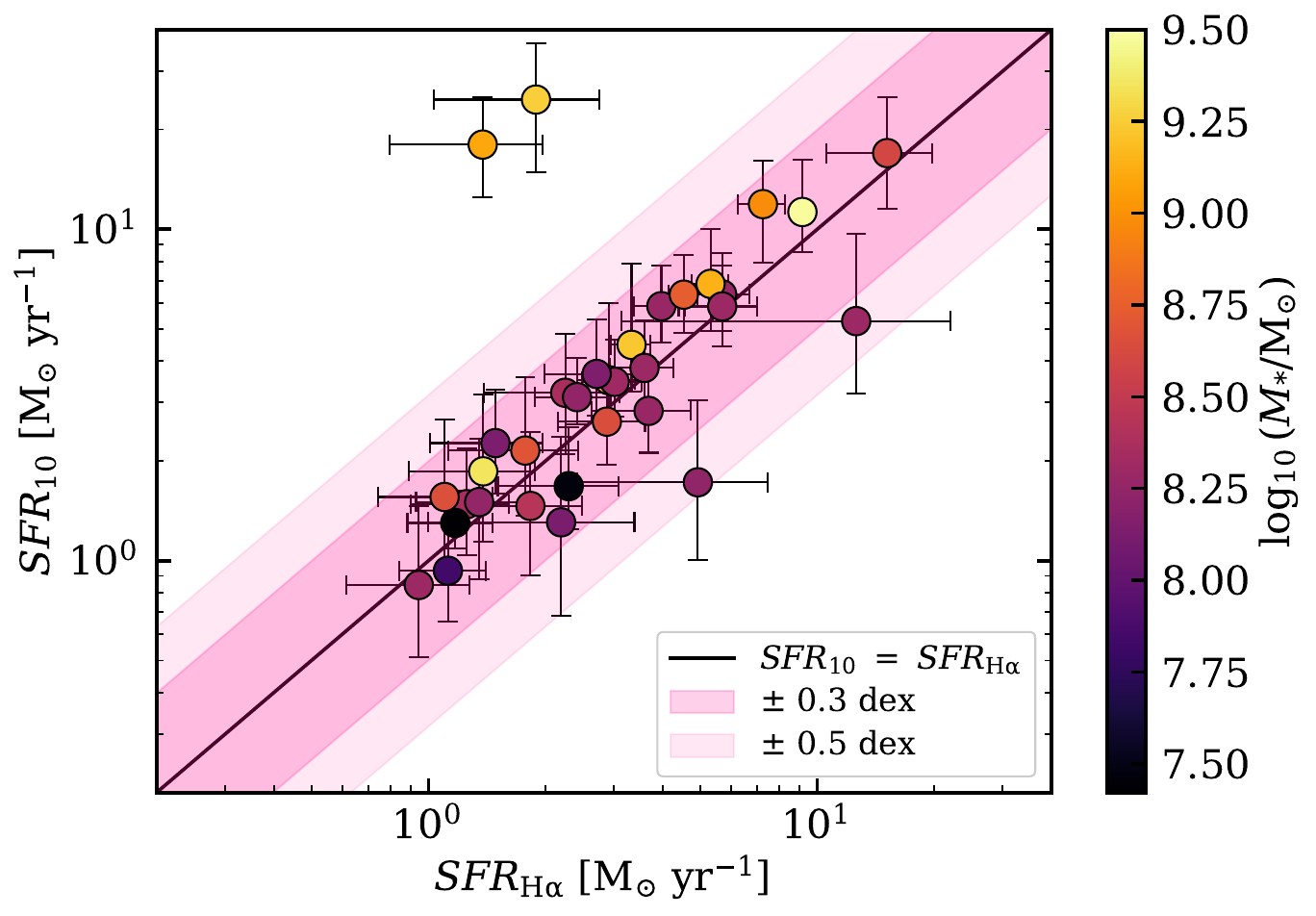}
\end{subfigure}\hfil

\medskip
\begin{subfigure}{0.48\textwidth}
  \includegraphics[width=0.97\linewidth]{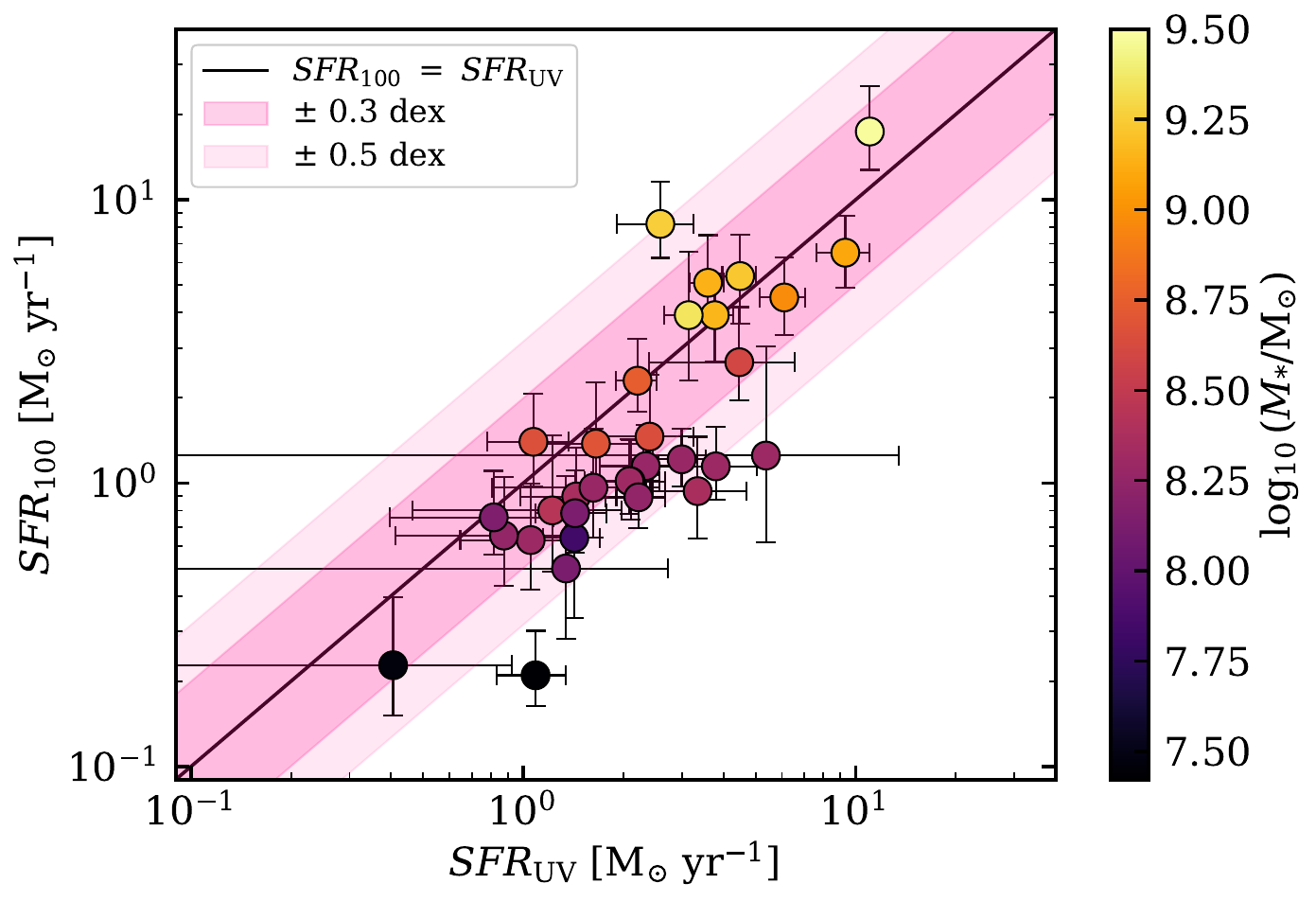}
\end{subfigure}\hfil

    \caption{Top panel: empirically measured $\rm{SFR_{H\alpha}}$ against $\rm{SFR_{UV}}$ (see Section~\ref{sec:empirical}) colour coded by their $\rm{SFR_{10}/SFR_{100}}$ value inferred from their SED fitting (see Section~\ref{sec:sed_fitting}). The two indicators broadly agree, but with a tendency towards higher $\rm{SFR_{H\alpha}}$ values, especially in galaxies with $\rm{SFR_{10}/SFR_{100}} > 1$. Middle panel: empirically measured $\rm{SFR_{H\alpha}}$ against SED fitted $\rm{SFR_{10}}$, colour coded by their stellar mass value inferred from their SED fitting: the two values generally agree well, confirming that H$\alpha$ traces star formation on 10\,Myr timescales. Bottom panel: Empirically measured $\rm{SFR_{UV}}$ against SED fitted $\rm{SFR_{100}}$, also colour coded by their stellar mass values. At high stellar masses (and SFRs) the UV luminosity traces the 100-Myr SFR, but at lower masses where the H$\alpha$ emitters have been shown to be in a burst phase (see Figure~\ref{fig:sfr_10_100_vs_mass}), the UV luminosity is enhanced, tracing also the shorter timescale star formation. The solid black lines in each plot show where the SFRs are equal and the darker and lighter pink regions show regions within 0.3 and 0.5 dex, respectively, from the 1:1 line. Note that the two potential AGN candidates are excluded from this plot as their calculated SFRs lie outside of the plotted ranges.}
    
    \label{fig:sfr_uv_halpha}

\end{figure}

In Fig. \ref{fig:sfr_uv_halpha} we investigated the relationship between the SED fitted SFRs ($\rm{SFR_{10}}$ and $\rm{SFR_{100}}$) and the empirically calculated SFRs ($\rm{SFR_{{\rm{H}}\alpha}}$ and $\rm{SFR_{UV}}$). The top panel compares the empirically-derived H$\alpha$ and UV SFRs. It shows that our H$\alpha$ candidates broadly cluster around $\rm{SFR_{H\alpha}}$ = $\rm{SFR_{UV}}$, but with sources scattering preferentially towards higher $\rm{SFR_{{\rm{H}}\alpha}}$ values (where the median $\rm{SFR_{H\alpha}/SFR_{UV}}$ $\sim$ 1.13). Sources in the plot are colour-coded by their $\rm{SFR_{10}/SFR_{100}}$ ratio, and it was clear that the sources offset towards higher $\rm{SFR_{H\alpha}}$ than $\rm{SFR_{UV}}$ were those which also exhibit the highest $\rm{SFR_{10}/SFR_{100}}$ ratios. This makes broad sense because, canonically, the H$\alpha$ emission line traces changes in star-formation activity over shorter timescales of $\sim$ 10 Myr compared to the UV-continuum which traces activity of $\sim$ 100 Myr \citep[e.g.][and references therein]{1998ARA&A..36..189K,2012ARA&A..50..531K,2013seg..book..419C}. Both the empirical and SED-derived SFRs suggest that our H$\alpha$ emitter sample has mainly experienced a recent rise of star-formation activity over shorter timescales, on average. At the lowest stellar masses (for the faintest sources), we note that selection effects could be biasing the results where sources with equivalent $\rm{SFR_{H\alpha}}$ and $\rm{SFR_{UV}}$ (or $\rm{SFR_{10}}$ and $\rm{SFR_{100}}$) may not be detected. Despite this, we still see a rise in recent star formation activity for the H$\alpha$ sample as shown in Fig. \ref{fig:sfr_10_100_vs_mass} at typical stellar masses for our sample (above the detection limit). In addition, the $EW_{\rm{H\alpha}}$ distribution shown in Fig. \ref{fig:l_sfr_halpha} is well above the lower limit imposed by colour excess criteria (see Fig. \ref{fig:colour_mag_plots}). Therefore, we do not think any selection bias would be significant for our H$\alpha$ sample.

To explicitly explore SFR timescales of the empirical measurements, we show the relationship between $\rm{SFR_{H\alpha}}$ and $\rm{SFR_{10}}$ in the middle panel and between $\rm{SFR_{UV}}$ and $\rm{SFR_{100}}$ in the bottom panel of Fig. \ref{fig:sfr_uv_halpha}. We found that $\rm{SFR_{H\alpha}}$ generally agree well with $\rm{SFR_{10}}$, confirming that the H$\alpha$ emission line traces changes in star-formation activity on timescales of the order of $\sim$ 10 Myr. However, it appears that $\rm{SFR_{UV}}$ is not traced as effectively by $\rm{SFR_{100}}$, with the UV SFR being enhanced by up to a factor 2-3 compared to the SED-derived 100\,Myr SFR in our H$\alpha$ emitter sample. This offset is most pronounced at lower stellar masses, where Fig. \ref{fig:sfr_10_100_vs_mass} had indicated that the H$\alpha$ emitters tended to have enhanced recent star formation. This result therefore suggests that our H$\alpha$ sample could be exhibiting bursty star formation activity where the UV continuum is dominated by star formation on shorter timescales $<$ 100 Myr. However, assumptions on the SPS models could impact the empirically measured SFRs (see discussion in Section~\ref{sec:bursty_star_formation}).

\begin{figure}
    \centering
    \includegraphics[width=0.48\textwidth]{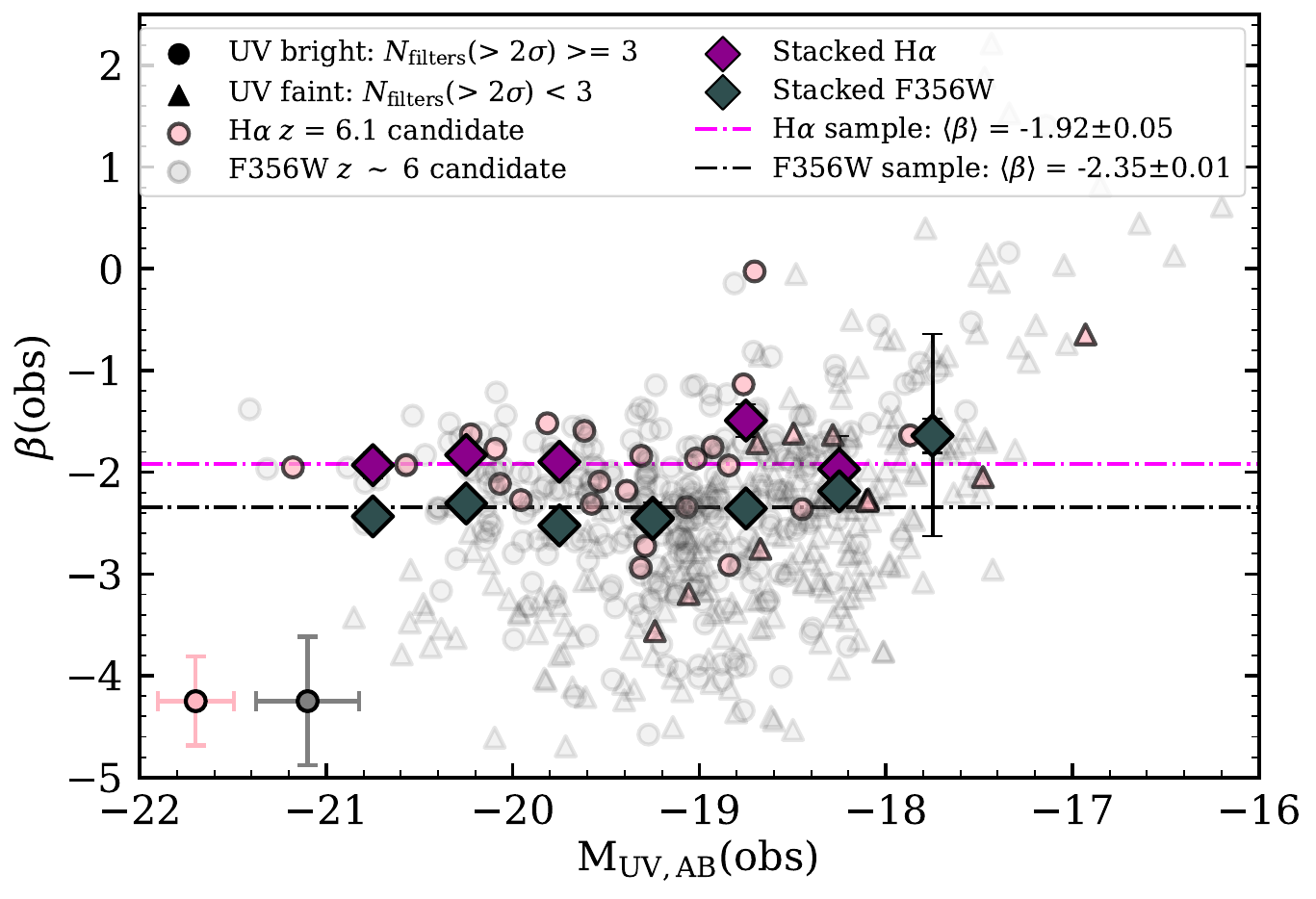}

    \caption{The fitted UV-continuum slopes ($\beta$) as a function of measured $M_{\rm{UV}}$ for the final sample of $z$ $>$ 6 H$\alpha$ emitters (faint pink data points) and our $z \approx 6$ F356W detected sources (faint grey data points), derived as described in Section~\ref{sec:uv_slope}. The triangular data points with the same colours for the relevant samples show the sources that have less than three filters with SNR $>$ 2 probing the rest-frame UV-continuum (denoted `UV-faint') whereas the circular data points show the sources that have 3 or more filters with SNR $>$ 2 (denoted `UV-bright'). Typical error bars sizes for both samples are similar, and shown in the bottom left of the plot. The purple and dark grey diamonds show the inverse-variance weighted average values of $\beta$ for the H$\alpha$ and F356W detected sample, respectively, in bins of width 0.5 magnitude; this shows little dependence on the measured $M_{\rm{UV}}$. The pink and grey dot-dashed lines show the inverse-variance weighted average $\beta$ values for the entire H$\alpha$ and F356W detected sample, respectively.}
    
    \label{fig:beta_slopes}

\end{figure}

\subsubsection{UV-continuum slopes and dust attenuation}
\label{sec:dust_attenutation}

Fig. \ref{fig:beta_slopes} shows the observed UV-continuum slope, $\beta$, as a function of observed $M_{\rm{UV}}$ for the H$\alpha$ emission line galaxy sample and for the F356W detected $z \sim 6$ sources for comparison (see Section~\ref{sec:f356w_z=6_sources}). The plot also shows the inverse-variance weighted average $\beta$ values for the two samples in bins of $M_{\rm{UV}}$, demonstrating that this value is consistently higher for the H$\alpha$ emitters compared to the F356W-selected $z$ $\sim$ 6 selected sources, with inverse-variance weighted averages of $\langle \beta \rangle$ = $-$1.92 and $\langle \beta \rangle$ = $-$2.35, respectively. This indicates redder UV-continuum slopes on average for the H$\alpha$ candidates. We also found no relation between the weighted average $\beta$ and the binned observed $M_{\rm{UV}}$, showing that the H$\alpha$ candidates have systematically redder UV-continuum slopes across the same dynamic range of $M_{\rm{UV}}$. This result holds despite the inverse-variance weighted averages favouring the H$\alpha$ candidates with higher SNR rest-frame UV emission, which would bias the H$\alpha$ sample towards bluer UV-continuum slopes. In addition, we matched the F356W-detected $z$ $\sim$ 6 sample to the same F356W magnitude and then stellar mass range as for the H$\alpha$ sample and found the weighted average value $\langle \beta \rangle$ = $-$2.37 (a decrease of 0.02 compared to the full sample). Therefore, we conclude that our H$\alpha$ sample shows redder $\beta$ values, on average, compared to the rest-optical sample selected at the same epoch.

To investigate the cause of the systematically redder $\beta$ values, we turned to the results from our SED fitting, which investigated the dust attenuation in our sample of H$\alpha$ candidates. As shown in the testing of SED-fitting robustness in Appendix~\ref{app:posteriors}, the $V$-band attenuation in the stellar continuum ($A_{V}$) for our H$\alpha$ emitters was well defined (see the narrow width of the stacked posterior distribution in the top left panel of Fig. \ref{fig:sed_post_dist}) where we derived a median $A_{V}$ = 0.23. The top left panel of Fig. \ref{fig:sed_post_dist} also shows the stacked posterior distributions separated into different stellar mass bins; we see consistently low attenuation values for our sample and no significant trend with stellar mass. Our tests also showed that the $A_{V}$ distributions were unaffected by changes in the dust model assumptions and priors \citep[e.g. different prior choices in $A_{V}$, the slope deviation $\delta$ from ][and the factor on $A_{V}$ for the stars in birth clouds $\eta$]{2000ApJ...533..682C}, implying that the results are robust. 

These results agree well with the stacked posterior distributions of $A_{V}$ for our F356W-detected $z$ $\sim$ 6 sample. These galaxies show similar dust attenuation properties, with a median $A_{V}$ = 0.15 and no evolution with stellar mass. The difference of 0.08 in $A_{V}$ for our H$\alpha$ emitters compared to the F356W-selected sample would give rise to only $\approx 0.1$ difference in $\beta$, far less than the 0.4 difference observed between the two samples. We therefore conclude that our sample of H$\alpha$ emission line galaxies don't exhibit significant dust attenuation, broadly in line with our F356W-selected $z$ $\sim$ 6 population of galaxies. We discuss the probable cause of the different $\beta$ values in Section~\ref{sec:discussion}.

\section{Discussion}
\label{sec:discussion}

\subsection{Reddened and faint UV-continuum}
\label{sec:reddened_uv_continuum}

Results from the empirically calculated UV-continuum slopes (see Section~\ref{sec:uv_slope}) show that our sample of H$\alpha$ emission line galaxy candidates exhibit systematically redder slopes ($\langle \beta \rangle$ = $-$1.92) than the F356W-detected sample of $z$ $\sim$ 6 sources ($\langle \beta \rangle$ = $-$2.35), utilising only the BB filters in a manner analogous to other photo-$z$ selected samples in the literature. The results suggest that our H$\alpha$ narrow-band selections identify a redder population of star-forming galaxies compared to samples based on rest-frame UV/optical-continuum selections. Interestingly, the weighted average $\beta$ for our H$\alpha$ sample is comparable to those found in studies utilising rest-frame UV \emph{HST} selections of $z$ $\sim$ 6 star-forming galaxies \citep[e.g.][]{2012MNRAS.420..901D,2011MNRAS.418.2074M} and in the same $M_{\rm{UV}}$ range: $M_{\rm{UV}}$ $\lesssim$ $-$18. However, other studies reaching $M_{\rm{UV}}$ $\lesssim$ -17 show bluer UV-continuum slopes, $\langle \beta \rangle$ $\sim$ - 2.2 \citep[e.g.][]{2013MNRAS.432.3520D,2014ApJ...793..115B} which highlights the potential biases from selection effects that could occur from different rest-UV/optical selections of galaxies. This is further demonstrated from \emph{JWST} results (which utilised deeper NIR photometry) which showed average UV-continuum slopes in comparable $M_{\rm{UV}}$ and redshift space \citep[e.g.][]{2023ApJ...947L..26N,2024arXiv240410751A} similar to the weighted average $\beta$ value for our F356W-selected $z$ $\sim$ 6 sources.

We note that our narrow-band selected H$\alpha$ sample extends into the `UV-faint' regime, where 12 out of 35 H$\alpha$ candidates have $\leq$ 2 filters with SNR $>$ 2 constraining the rest-frame UV-continuum. This suggests that a significant fraction of these sources would not be selected from rest-frame UV photometry alone (despite the increased depth from the \emph{JWST} imaging) and this could be influencing the results. Studies from \emph{HST} observations have shown bias towards bluer $\beta$ for galaxies at faint magnitudes \citep[e.g.][]{2010ApJ...708L..69B,2012MNRAS.420..901D,2013MNRAS.429.2456R} and this has also been observed with recent \emph{JWST} studies \cite[e.g.][]{2023MNRAS.520...14C}. It is therefore plausible that the average $\beta$ for our H$\alpha$ sample (and our F356W-detected $z \sim 6$ sample) could in reality be redder than measured, given this bias, particularly for the `UV-faint' sample where the slopes are less constrained (though the uncertainties in the photometry are also consistent with the true slopes being bluer). However, the key point is that any bias in the $\beta$ measurements should impact both our H$\alpha$ and F356W-selected $z$ $\sim$ 6 samples (which were calculated using the same method) at faint magnitudes and we still observe redder UV-continuum slopes for our H$\alpha$ candidate sample on average. 

From our \textsc{Bagpipes} SED fitting (see Section~\ref{sec:sed_fitting}) we have constraints on the dust attenuation, $A_{V}$, and note that, on average, galaxies within both our H$\alpha$-selected and F356W-detected samples are rather dust poor. As discussed in Section~\ref{sec:dust_attenutation}, studies have obtained inconsistent constraints on the dust attenuation slopes of galaxies beyond cosmic noon \citep[e.g][]{2016A&A...587A.122A,2018ApJ...853...56R,2018MNRAS.479.4355K,2018MNRAS.479...25M}, and so the \citet{2018ApJ...859...11S} dust model was implemented to allow for steeper and shallower slopes compared to \citet{2000ApJ...533..682C} dust attenuation curve. We tested and implemented both the \citet{2000ApJ...533..682C} and \citet{2018ApJ...859...11S} dust models in our SED fits and found that the former model would predict higher $A_{V}$ values for a subset of sources. This was because we allowed the slope deviation factor ($\delta$) in the \citet{2018ApJ...859...11S} dust model to vary in the SED fit and even though the sample converges to a \citet{2000ApJ...533..682C} slope on average, there was still scatter in the individual measurements. Allowing this slope to vary for our full sample eliminates most $A_{V}$ $>$ 1.0 estimations, as seen in the left panels of Fig. \ref{fig:sed_post_dist} where the posterior distributions flatten significantly beyond this value.

An alternative explanation for the observed reddened $\beta$ values, while accounting for low dust attenuation for the H$\alpha$ sample, could be strong nebular continuum emission (in addition to nebular line emission) due to young stellar populations and high ionising photon production efficiencies \citep[$\xi_{\rm{ion}}$;][]{2010ApJ...708...26R,2017ApJ...840...44B,2022ApJ...941..153T}. Nebular continuum emission is caused by free-free, free-bound and two-photon processes, where free-free emission can boost the flux at $\lambda_{\rm{rest}}$ $\sim$ 3000$\AA$, which is potentially important for reddening the slope of the UV-continuum. Studies have shown that there is a theoretical limit to the measured $\beta$, where the bluest UV-continuum is $\beta$ $>=$ $−$2.6 \citep[e.g.][]{2016MNRAS.456..485S} due to nebular continuum emission, with observations of galaxies at $z$ $>$ 7 showing $\langle \beta \rangle$ $\simeq$ $-$2.6 \citep[e.g.][]{2022ApJ...941..153T,2023ApJ...952L...7A,2024MNRAS.531..997C,2024ApJ...964L..24M}. 

\begin{figure*}
    \centering
    \includegraphics[width=\textwidth]{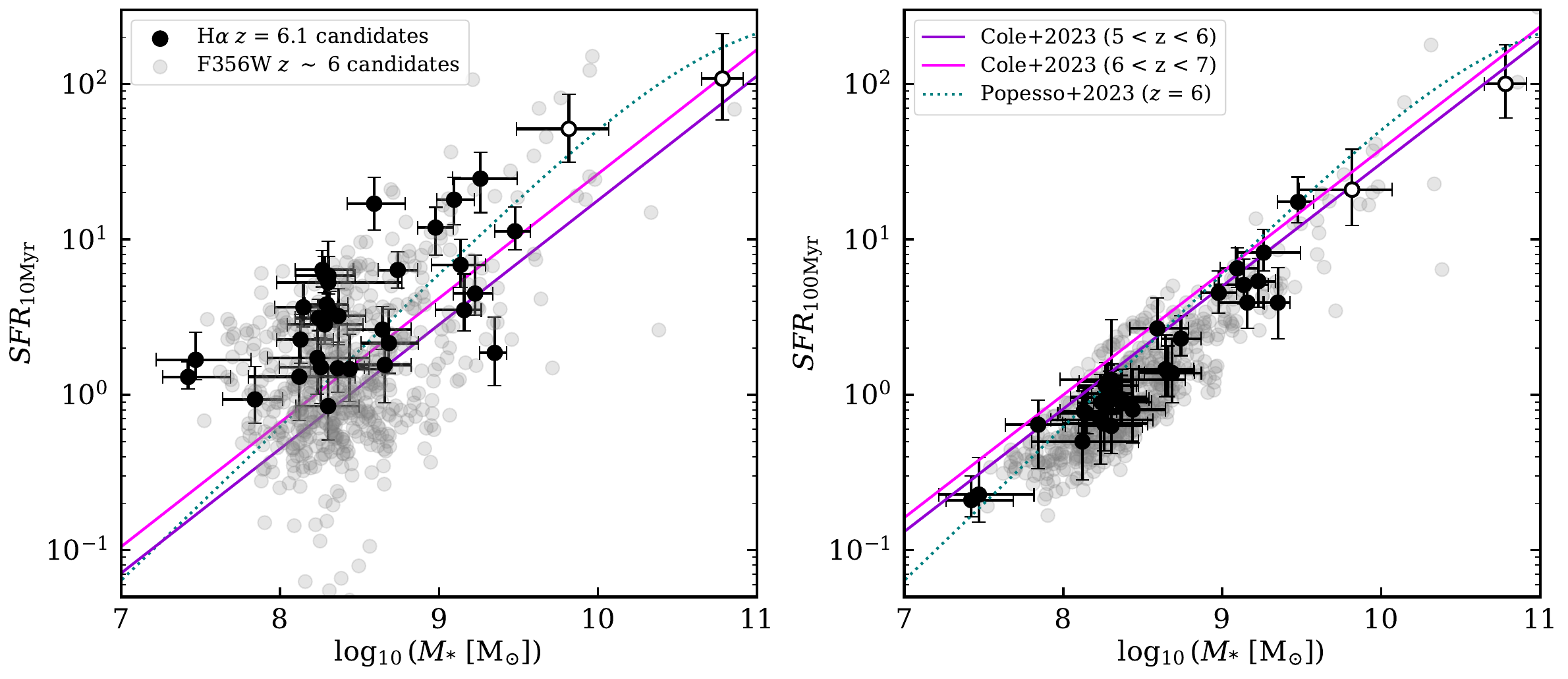}

    \caption{Star-forming main sequence (SFMS) plots showing inferred $\rm{SFR_{10}}$ (left panel) and $\rm{SFR_{100}}$ (right panel) as a function of stellar mass for our H$\alpha$ emission line galaxy sample (black data points) and our F356W-detected $z$ $\sim$ 6 sample (grey data points) from the SED fitted results (described in Section~\ref{sec:sed_fitting}). For the H$\alpha$ candidates, the error bars show the 16th and 84th percentiles of the parameter posterior distributions. In addition, functional forms of the SFMS at the same epoch are also plotted. The pink and purple lines are taken from \citet{2025ApJ...979..193C} and correspond to their $5 \ < \ z \ < \ 6$ and $6 \ < \ z \ < \ 7$ samples respectively measured from SED fitting and averaged over 10 Myr and 100 Myr for the relevant plots. The dotted teal line shows the functional form to the SFMS from \citet{2023MNRAS.519.1526P} using a variety of SFR indicators from calibrated empirical measurements and from SED fitting. Our samples provide a tight SFMS based on the 100\,Myr SFR (right panel), with both H$\alpha$ and photo-z-selected samples in good agreement with the literature relations. For SFRs estimated over 10\,Myr (left panel), our F356W-selected population broadly agrees with previous determinations on average, albeit with more scatter due to the noisier measurements, but our H$\alpha$ sample is systematically offset to higher SFRs, consistent with these galaxies showing evidence of recent starburst activity. Again, the two potential AGN candidates are plotted with open symbols.}
    
    \label{fig:sfms}

\end{figure*}

Pre-\emph{JWST}, spectral features indicating strong nebular continuum emission (including the `Balmer Jump at 3646 $\AA$) were only detected in low-redshift highly star-forming galaxies \citep{1969BOTT....5....3P,2007A&A...464..885G}. 
However, simulations have predicted that Balmer jumps and hence strong nebular continuum should be common at high-redshift \citep[e.g.][]{2023OJAp....6E..44K,2024MNRAS.527.7965W}. In addition, SED fitting of high-redshift galaxy candidates identified by \emph{JWST} have required the addition of these spectral features into the modelling to obtain good fits \citep[e.g.][]{2024MNRAS.533.1111E,2024MNRAS.529.4087T} and these features have now been spectroscopically confirmed at high-redshift \citep[e.g.][]{2024MNRAS.534..523C}. It could be expected that galaxies selected on the basis of their nebular emission lines (like our sample) would be likely to show strong nebular continuum. \citet{2025ApJ...982....7N} show that nebular continuum could be responsible for reddening in $\beta$ by 0.2--0.4 on average for high-redshift galaxies using zoom-in hydrodynamic simulations; this is the scale of the change needed to explain the different $\beta$ values of our H$\alpha$ emitters compared to the F356W-selected sample, and so nebular continuum is therefore a plausible explanation of the redder slopes that we observe in our H$\alpha$ sample. Further investigation, including stacking analysis and/or follow-up spectroscopy is needed to conclusively explain this trend.

\subsection{Burstiness of Star formation}
\label{sec:bursty_star_formation}

Results from our SED fitting indicate that our sample of H$\alpha$ candidates had elevated recent star-formation activity (traced by inferring $\rm{SFR_{10}/SFR_{100}}$). This is not surprising, as the H$\alpha$ emission line better traces instantaneous changes in star-formation activity over smaller timescales compared to the UV-continuum (as discussed in Section \ref{sec:SFRs}). Both the empirical and SED-fitted SFR ratios ($\rm{SFR_{H\alpha}/SFR_{UV}}$ and $\rm{SFR_{10}/SFR_{100}}$ respectively) show good qualitative agreement (see top panel of Fig. \ref{fig:sfr_uv_halpha}) that star-formation activity in our H$\alpha$ sample seems to be recently elevated compared to further back in their star-formation histories. This indicates that these galaxies are exhibiting rather stochastic and `bursty' star-formation activity, particularly for decreasing galaxy stellar masses (as seen in the right panel of Fig. \ref{fig:sed_post_dist}). In addition, this aligns with the fact that our sample predominately contains lower stellar mass galaxies (mostly in the dwarf galaxy regime in the local Universe: $\log_{10}(M_{\star} [\rm{M_{\odot}}])$ $\lesssim$ 9.0) which exhibit higher recent SFRs. 

Local dwarf galaxies have been observed to have `bursty' SFHs \citep[e.g. through measurements of their $\rm{H\alpha}$-to-UV ratio;][]{2012ApJ...744...44W,2019ApJ...884..133F,2019ApJ...881...71E,2022MNRAS.511.4464A} agreeing with simulations \citep[e.g.][]{2014MNRAS.445..581H,2014ApJ...792...99S} where intense star formation creates supernovae feedback, heats and expels the gas resulting in temporary quenching. Recycled and new gas from the intergalactic/circumgalactic medium (IGM/CGM) is then accreted to form new stars, creating bursts and quenching cycles on 1 to $>$10 Myr timescales. However, there is also evidence of dwarf galaxies exhibiting longer lasting ($\sim$few hundred Myr) bursts and steady-state star formation \citep[e.g.][]{2010ApJ...724...49M,2018ApJ...856...62C,2019ApJ...887..112C} and irregular bursts likely due to mergers rather than stellar feedback \citep[e.g.][]{2001A&A...374..800O,2008MNRAS.388L..10B,2012ApJ...748L..24M,2024A&A...692A..51S}. We believe that the SFHs for our sample are likely `bursty' over shorter timescales producing $\sim$few Myr cycles of star-formation and quenching activity. This is because galaxies at the Epoch of Reionization tend to have higher specific star-formation rates (sSFRs) due to the increased gas fractions and accretion rates at higher redshift, leading to increased stellar feedback \citep[e.g.][]{2015MNRAS.451..839D,2018MNRAS.478.1694M,2018MNRAS.473.3717F,2023OJAp....6E..44K,2025MNRAS.537..629D}. In addition, JWST observations have shown that the increase in galaxy interactions \citep[e.g.][]{2024MNRAS.52711372A} aligning with the increase in the average galaxy merger rate from the local Universe to the Epoch of Reionization \citep[e.g.][]{2019ApJ...876..110D,2025MNRAS.540..774D} could also be further enhancing burstiness in galaxy SFHs over shorter timescales at this epoch. These findings add to other recent studies which have found downturns in the recent SFHs in UV-faint ($M_{\rm{UV}}$ $>$ $-$18) and low stellar mass selected galaxies into the Epoch of Reionization \citep[e.g.][]{2024arXiv241001905E,2024MNRAS.533.1111E} along with other studies that have built empirical evidence for bursty SFHs among high-redshift galaxies undergoing bursts and declines in recent star-formation \citep[e.g.][]{2023ApJ...949L..23S,2024ApJ...964..150D,2024MNRAS.527.2139D,2025A&A...697A..88L}. The mass-dependence of the stochasticity is not surprising as high mass sources are expected to have higher SFRs (as shown by the star-forming main sequence). Therefore, higher-mass sources are more likely to make our sample selection flux limit without requiring them to be going through a burst of star-formation activity. Lower mass galaxies (which typically have lower SFRs) only reach the threshold for H$\alpha$ detection if they are selected when they are at the peak of a burst in star-formation activity.

We investigated the star-forming main sequence for our H$\alpha$ emitters in Fig. \ref{fig:sfms}, for both our SED-fitted 10\,Myr (left) and 100\,Myr (right) SFR timescales, and compared this to previous literature determinations. For star-formation on 100\,Myr timescales, we found a tight star-forming main sequence for our H$\alpha$ emitters, that is broadly in line with a variety of recent literature determinations at the same redshifts. We over-plotted the F356W-selected $z\sim6$ sample and also found good agreement. On 10\,Myr timescales, we found that the H$\alpha$-selected galaxies showed enhanced star-formation rates compared to the main sequence, again indicating that they are observed during or just after a burst of star formation. They are also offset from the F356W-selected galaxies (like in Fig. \ref{fig:sfr_10_100_vs_mass}) with the median $\rm{SFR_{10}}$ for the H$\alpha$ sample being a factor of 2.75, 2.31 and 1.46 higher than the F356W-detected sample in the lowest, middle and highest stellar mass bins respectively (shown in Fig. \ref{fig:sed_post_dist}). The F356W-detected sample agree broadly with the star-forming main sequence determinations from the literature, on average, but show considerably larger scatter in their measurements around this. This agrees with work from \citet[][]{2025ApJ...979..193C} utilising a similar procedure which also reports an increased scatter in SFR averaged over 10\,Myr timescales at high-redshift. The scatter in the main-sequence is also re-produced in simulations utilising complex feedback models at this Epoch \citep[e.g.][]{2025MNRAS.537..629D}. However, this could also be reflective of the difficulty in constraining SFRs on these shorter timescales using only broad-band measurements.  

As discussed in Section~\ref{sec:SFRs}, the empirical and SED-fitted SFR ratios show the same overall trends but differ in detail, with the median $\rm{SFR_{10}/SFR_{100}}$ $\sim$ 2.7 and the median $\rm{SFR_{H\alpha}/SFR_{UV}}$ $\sim$ 1.3. The UV-continuum and H$\alpha$ emission line are often quoted in the literature to be tracing star-formation timescales of $\sim$10 Myr \citep{2011ApJ...737...67M} and $\sim$100 Myr \citep{2011ApJ...741..124H} respectively, and if this is the case then a broad agreement between $\rm{SFR_{H\alpha}}$ and $\rm{SFR_{10}}$ and between $\rm{SFR_{UV}}$ and $\rm{SFR_{100}}$ would be expected. These calibrations assume steady state star-formation activity over an assumed timescale but the timescales that the H$\alpha$ and UV emission trace depend on recent star-formation history (SFH) and so these indicators can be better described as a convolved measurement of the intrinsic `true' SFH with a time-delayed response function. Work utilising the FIRE simulations \citep{2021MNRAS.501.4812F} suggests that the H$\alpha$ emission line traces star-formation activity on $\sim$5 Myr timescales for both bursty or constant SFHs. In line with this, our results suggest that $\rm{SFR_{H\alpha}}$ traces star-formation activity over the canonical $\sim$10 Myr timescales with small scatter (see middle panel of Fig. \ref{fig:sfr_uv_halpha}). However, these simulations also show that bursty phases of star formation cause the UV-continuum to trace timescales that range from $\sim$ 10 Myr to $>$ 100 Myr (particularly for extreme bursts of star formation). Stellar populations emit significant UV-continuum emission for ages $>$ 100 Myr, however, the total integrated UV emission is strongly weighted towards massive and hence younger and shorter lived stellar populations tracing star formation over shorter timescales. This would explain why we are seeing $\rm{SFR_{UV}}$ fail to trace star-formation activity over the canonical $\sim$100 Myr timescales (see bottom panel of Fig. \ref{fig:sfr_uv_halpha}), particularly at lower stellar masses where we see higher values of $\rm{SFR_{UV}}$ suggesting this indicator is tracing timescales $\lesssim$100 Myr. Again, this makes sense as the canonical SFR calibrations assume steady state SFHs and these results suggest that the SFHs of these galaxies (particularly at lower stellar masses) are not steady-state and instead are likely `bursty'.

However, caution must be exercised with this interpretation as calibrations of $\rm{SFR_{H\alpha}}$ and $\rm{SFR_{UV}}$ depend on the choice of SPS models, IMF (and mass cutoff) and metallicity (particularly true for the H$\alpha$ calibration). For example, \textsc{BPASS} v2.2 \citep{2018MNRAS.479...75S} $Z$ = 0.002 binary models used in this study \citep[and in][]{2019ApJ...871..128T}, lowers $\rm{SFR_{H\alpha}}$ by $\sim$0.35 dex and $\rm{SFR_{UV}}$ by $\sim$0.1 dex relative to the canonical conversion factors discussed in \citet{2012ARA&A..50..531K} which assumes solar metallicity nebulae and single star SPS models more appropriate for ISM conditions at $z$ $\sim$ 0 but not at higher-redshift. The reduction in SFR per H$\alpha$ luminosity will be more prominent for younger stellar populations where massive stars are still present compared to stellar populations after 100 Myr timescales (hence the smaller difference in $\rm{SFR_{UV}}$ compared to $\rm{SFR_{H\alpha}}$). We controlled this by assuming the same IMF (and cutoff) and SPS models between both empirical and SED measurements. However, determining the metallicity of these sources is difficult from only photometric data, and so if the true metallicity, $Z$ $<$ 0.002 for our sample then $\rm{SFR_{H\alpha}}$/$\rm{SFR_{UV}}$ will be even higher than what we have measured (and the reverse is also true).

We investigated the impact of dust corrections to $\rm{SFR_{H\alpha}}$ and $\rm{SFR_{UV}}$, which could have an impact on the $\rm{SFR_{H\alpha}}$/$\rm{SFR_{UV}}$ ratios. We assumed a \citet{2000ApJ...533..682C} dust attenuation slope firstly to scale our V-band attenuation in the stellar continuum $A_{V}$ to obtain the attenuation at 1500$\AA$ ($A_{1500\AA}$) and at the H$\alpha$ emission line wavelength ($\lambda_{\rm{rest}}$ = 6563 $\AA$). Using the median $A_{V}$ $\sim$ 0.23 (see Section~\ref{sec:dust_attenutation}) for our H$\alpha$ sample and scaling by either the \citet{2000ApJ...533..682C} or \citet{1989ApJ...345..245C} slopes gives differences in $A_{1500 \AA}$ of $\sim$ 4 per cent, which is negligible compared to the photometric uncertainties and isn't a dominant factor in the dust-corrected UV luminosities. However, assuming a \citet{2003ApJ...594..279G} SMC attenuation slope could increase the dust-corrections to the UV luminosities by $\sim$ 85-90 per cent compared to the other dust slopes decreasing $\rm{SFR_{H\alpha}}$/$\rm{SFR_{UV}}$ by $\sim$ 0.3 dex. However, as discussed in Section~\ref{sec:sed_fitting}, our sources exhibit, on average, \citet{2000ApJ...533..682C} slopes and so it was unlikely that our dust corrections are underestimated significantly on average.

The choice of the above attenuation slopes has negligible impact on the attenuation at $\lambda_{\rm{rest}}$ = 6563 $\AA$, but the assumed factor between the nebular reddening ($E(B \ − \ V)_{\rm{neb}}$) and reddening due to the stellar continuum ($E(B \ − \ V)_{\rm{cont}}$), used to obtain $A_{\rm{H}\alpha}$, will have an impact on the dust-corrections applied to $\rm{SFR_{H\alpha}}$. \citet{2020ApJ...902..123R} showed that the average $E(B \ − \ V)_{\rm{cont}}$/$E(B \ − \ V)_{\rm{neb}}$ ratios for star-forming galaxies at `cosmic noon' were 2.070, 2.273, 2.712 and 4.331 for the \citet{2015ApJ...806..259R}, \citet{2000ApJ...533..682C}, Calzetti+SMC and SMC \citep{2003ApJ...594..279G} curves, where these ratios increased with the steepness of the assumed stellar attenuation curve. The scatter in the relation between $E(B \ − \ V)_{\rm{cont}}$ and $E(B \ − \ V)_{\rm{neb}}$ \citep[see Fig. 5 of][]{2020ApJ...902..123R} is thought to be driven by the longer molecular cloud crossing timescales seen in high-redshift galaxies compared to local molecular clouds, the constant or rising SFRs where newly formed and dustier OB associations always dominate the ionising flux, and/or that the dust responsible for reddening the nebular emission may be associated with non-molecular (i.e., ionised and neutral) phases of the ISM \citep{2015ApJ...806..259R}. At the lowest stellar masses and extrapolating the evolution of the size-mass relation to high-redshift \citep[e.g.][]{2024ApJ...962..176W}, it is possible that galaxy sizes are small such that they consist of a single HII region. In this scenario, $E(B \ − \ V)_{\rm{cont}}$ = $E(B \ − \ V)_{\rm{neb}}$ is possible and the \citet{2000ApJ...533..682C} factor in this case would not be a good assumption. We tested the assumption of $E(B \ − \ V)_{\rm{cont}}$ = $E(B \ − \ V)_{\rm{neb}}$ and found that the resultant median empirical SFR ratio $\rm{SFR_{H\alpha}}$/$\rm{SFR_{UV}}$ $\sim$ 1; this disagrees even further with the $\rm{SFR_{10}}$/$\rm{SFR_{100}}$ $\sim$ 2.7 seen from our SED fitting. Given the above work at higher-redshift and our SED fitting results, it is likely there are additional dust corrections required for the nebular lines (like H$\alpha$), but a large spectroscopic follow-up of emission line selected samples would be required to constrain dust properties further (e.g. via the Balmer decrement).

Recent results have shown that non-parametric models are better able to re-construct sharp changes to the SFH compared to parametric models \citep[e.g.][]{2019ApJ...876....3L,2019ApJ...873...44C,2022ApJ...935..146S,2024ApJ...961...73N} and this is also true for the assumption on the prior distribution for the allowed change in SFR between adjacent time bins. We fit our sample with both a `continuity' prior and a `bursty continuity' prior \citep[e.g.][]{2022ApJ...927..170T} which are weighted towards smooth and `bursty' changes in the SFH, respectively. We initially adopted the latter prior (given we expect our sample to have sudden rises in recent SFR) and found that the posterior distributions in the sample $A_{V}$ and $\rm{SFR_{10}/SFR_{100}}$ converged to the prior. Therefore, we adopted a `continuity' prior which did not converge to the prior and was data driven (see results in Fig. \ref{fig:sed_post_dist}). Despite our choice of prior for changes in SFR between time bins being weighted towards smoother changes, our sample still exhibits sudden increases in recent SFR on average. However, given this prior choice, these ratios could be higher, on average, especially when compared to what is measured for galaxies with `bursty' SFHs shown in simulations with large stochasticity over short timescales \citep[e.g.][]{2023ApJ...949L..23S,2024ApJ...964..150D,2024MNRAS.527.2139D,2025A&A...697A..88L}. Despite the challenges of modelling `bursty' SFHs, our H$\alpha$ sample still suggest that these are a population of star-forming galaxies under-going recent bursts of star formation and we are catching these galaxies in a rise in SFR within the last 5-10 Myr (even if we could be under-estimating the degree of this activity).

\section{CONCLUSIONS}
\label{sec:conclusion}

We analysed data from the \emph{JWST} Emission Line Survey (JELS; proposal 2321; PI: Philip Best) which utilised the \emph{JWST}/NIRCam instrument to perform narrow-band imaging at $\lambda$ $\sim$ 4.7 $\rm{\mu}$m, using the F466N and F470N filters. We then compiled the available multi-wavelength ancillary data from \emph{HST} (e.g. CANDELS) and \emph{JWST}/NIRCam (PRIMER) and created multi-wavelength catalogues detected on the F466N and F470N narrow-band images. We performed `excess source' selection criteria to identify a total of 609 emission line galaxy candidates. 

We selected a sample of candidate H$\alpha$ emitters, as sources meeting the `excess source' criteria and in the photometric redshift range 5.5 $<$ $z_{\rm{phot}}$ $<$ 6.5. After visual inspection of these sources, we obtained a secure sample of 35 H$\alpha$ emission line galaxies at $z \approx 6.1$ ($6.03 \lesssim z_{\rm{phot}} \lesssim 6.17)$, into the Epoch of Reionization. We found that these galaxies occupied a dust-corrected H$\alpha$ luminosity range of $\sim$ 10$^{41.6}$ $-$ 10$^{42.8}$ erg s$^{-1}$ and a rest-frame $EW$ distribution that peaks between $\sim$300 and $\sim$2000 $\AA$ \citep[in good agreement with the $EW$ distribution derived for \emph{JWST}/\emph{HST} drop-out selected sources at $z \sim 6$ in][] {2024MNRAS.533.1111E}. We obtained physical properties of these H$\alpha$ emission line galaxy candidates using empirical relations measured directly from the photometry, in combination with SED fitting analysis using the \textsc{Bagpipes} spectral fitting code. 

Specifically, we used an appropriate conversion factor (applicable for high-redshift star-forming galaxies into the Epoch of Reionization) to transform our $L_{\rm{H\alpha}}$ measurements into H$\alpha$-derived SFRs ($\rm{SFR_{H\alpha}}$); these were in the range of $\sim$ 0.9 $-$ 15 $\rm{M_{\odot} \ yr^{-1}}$. The rest-frame UV-continuum spectra were fitted using a power-law slope to measure the UV-continuum slope ($\beta$) and both the UV luminosity and absolute magnitude ($M_{\rm{UV}}$) of these sources. We then utilised an appropriate conversion factor to convert the UV luminosity into the $\rm{SFR_{UV}}$. From our \textsc{Bagpipes} fitting, we obtained 10 Myr and 100 Myr averaged SFRs ($\rm{SFR_{10}}$ and $\rm{SFR_{100}}$ respectively), stellar masses and $V$-band dust attenuation values ($A_{V}$). We explored these values and also compared our narrow-band selected H$\alpha$ emitters to a sample of F356W $z$ $\sim$ 6 candidates whose $z_{\rm{phot}}$ were primarily driven by the Lyman-break spectral features. Here, we summarise the key results from these measurements:

\begin{enumerate}
  \item We found that the measured $\beta$ were systematically redder for our narrow-band selected sample of H$\alpha$ emission line galaxies ($\langle \beta \rangle$ = $-$1.92) compared to our F356W-selected $z$ $\sim$ 6 galaxies ($\langle \beta \rangle$ = $-$2.35). This result is robust as both samples were prone to the same biases when obtaining measurements of $\beta$ (including bluer sources generally having higher SNR rest-frame UV photometry and therefore driving the inverse-variance weighted averages for both samples).
  
  \item Our SED fitting results, constrained by photometric data, suggested that our H$\alpha$ emitters were quite dust-poor, with a median $A_V = 0.23$. This is not substantially different from the median $A_V = 0.15$ of the F356W-selected $z\sim 6$ galaxies, indicating that the different $\beta$ values were not driven by dust attenuation. Instead, we argue that the reddened slopes could be due to nebular continuum emission, assumed to be more common at high-redshift and particularly in samples selected based on their nebular emission lines.
  
  \item We found qualitative agreement for elevated empirical and SED-fitted SFR ratios ($\rm{SFR_{\rm{H\alpha}}/SFR_{\rm{UV}}}$ and $\rm{SFR_{10}/SFR_{100}}$, respectively) for our H$\alpha$ candidates, particularly at low stellar masses ($\log_{10}(M_{\star}/\rm{M_{\odot}})$ < 9.0). 

  \item We found that the SFR measured directly from H$\alpha$ emission line traced star-formation timescales $\sim$10 Myr, within scatter. However, the UV-continuum appears be weighted towards star formation on timescales shorter than $\sim$100 Myr timescales, particularly at lower stellar masses. Caution must be exercised, however, as these results could also be driven by the calibration choice made when empirically measuring SFRs (which depend on the choice of SPS model, IMF shape and upper-mass cut off and metallicity), and on the assumed factor of the nebular reddening and the reddening due to the stellar continuum. 

  \item The combination of the elevated SFR ratios ($\rm{SFR_{\rm{H\alpha}}/SFR_{\rm{UV}}}$ and $\rm{SFR_{10}/SFR_{100}}$) and the low-mass disagreement between $SFR_{\rm{UV}}$ and $SFR_{100}$ measurements along with previous evidence from observations and simulations of galaxies at the Epoch of Reionization means that we infer `bursty' SFHs for our sample of H$\alpha$ emission line galaxies.
  
\end{enumerate} 

\noindent In future work, we will derive the first narrow-band estimated H$\alpha$ $z$ $\sim$ 6 luminosity function (LF) and thus quantify $\rho_{\rm{SFR}}$; we will then compare to previous measurements based on rest-frame UV-continuum selected samples. In addition, NIRCam's resolving power will allow the ionised gas morphologies to be measured to a high degree of accuracy and compared to morphologies in the stellar continuum. Thanks to the narrow-band filter widths, our new sample of H$\alpha$ emitters (and other emission line galaxy candidates from our sample) are ripe for clustering analyses. 

To obtain a greater understanding of the physical properties and answer the questions generated from this study for our narrow-band selected H$\alpha$ emitters, follow-up targeted spectroscopy from \emph{JWST}/NIRCam will be essential; this would enable further constraints on the dust attenuation, metallicity and hence the degree of star formation in these sources, as well as on the nebular continuum emission. We can then go further and obtain chemical abundances, study their nebular attenuation and investigate the kinematics of these sources. Given all these exciting results, we have demonstrated that narrow-band imaging continues to be a powerful tool to select emission-line star-forming galaxies over cosmic history, and now specifically into the Epoch of Reionization. Even better, we can select previously undetected and UV-faint sources, revealing a new population of star-forming galaxies at high-redshift and so giving a more complete picture of cosmic star formation and hence galaxy evolution.

\section*{ACKNOWLEDGMENTS}

The authors would like to thank Adam Carnall and Joel Leja for their helpful advice with the SED fitting of our sample, Callum Donnan for his advice on the selection techniques of high-redshift galaxies, Alice Shapley for discussion around H$\alpha$ SFR calibrations at high-redshift, Fred Jennings for providing insight into the interpretation of SFR ratios, and the anonymous referee for their helpful comments -- all of which have greatly improved this paper. Several other authors acknowledge the support of the UK Science and Technology Facilities Council (STFC) via grants ST/W507441/1 (CAP), ST/V000594/1 (DJM, PNB, RK and RJM), ST/Y000951/1 (PNB and RK) and ST/X001075/1 (AMS and IRS), and through an Ernest Rutherford Fellowship (KJD; grant number ST/W003120/1). RKC was funded by support for programme \#02321, provided by NASA through a grant from the Space Telescope Science Institute, which is operated by the Association of Universities for Research in Astronomy, Inc., under NASA contract NAS5-03127. RKC and CLH are both grateful for support from the Leverhulme Trust via a Leverhulme Early Career Fellowship, and CLH also acknowledges support from the Oxford Hintze Centre for Astrophysical Surveys which is funded through generous support from the Hintze Family Charitable Foundation. JSD acknowledges the support of the Royal Society via a Royal Society Research Professorship. EI gratefully acknowledge financial support from ANID - MILENIO - NCN2024$\_$112 and ANID FONDECYT Regular 1221846. LOF acknowledges funding by ANID BECAS/DOCTORADO NACIONAL 21220499. For the purpose of open access, the author has applied a Creative Commons Attribution (CC BY) licence to any Author Accepted Manuscript version arising from this submission.

\section*{Data Availability}

The data underlying this article are available in the Mikulski Archives for Space Telescopes (MAST: \url{https://mast.stsci.ed}) Portal under proposal ID number 2321 (JELS imaging). Higher level data products, including reduced mosaics in the JELS narrow and other \emph{JWST} and \emph{HST} and broad-band filters, as well as associated catalogues is publicy available through the University of Edinburgh \href{https://datashare.ed.ac.uk}{DataShare}. Any other data produced for the article will be shared on reasonable request to the corresponding author.



\bibliographystyle{mnras}
\bibliography{ref} 




\onecolumn
\appendix

\section{Postage Stamp Images of the H$\alpha$ emission line galaxy sample}
\label{app:postage}

The postage-stamp images of all 35 visually confirmed $z$ $>$ 6 H$\alpha$ emission line candidates are provided in this appendix and are shown in Fig.~\ref{fig:halpha_sample_cutouts_v1}.

\begin{figure}
\centering
    \includegraphics[width=\textwidth]{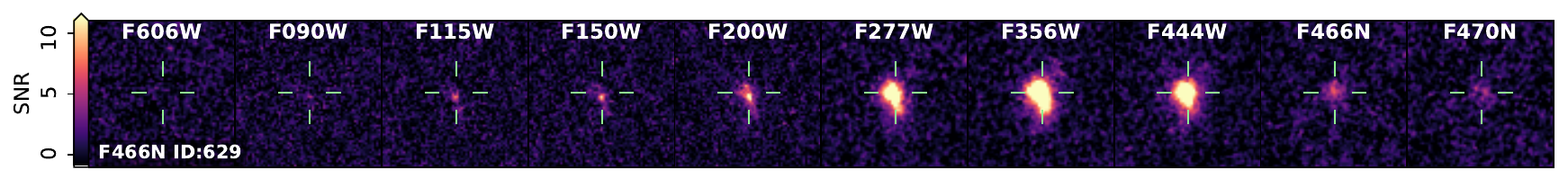}\vspace{-0.4cm}
  \includegraphics[width=\textwidth]{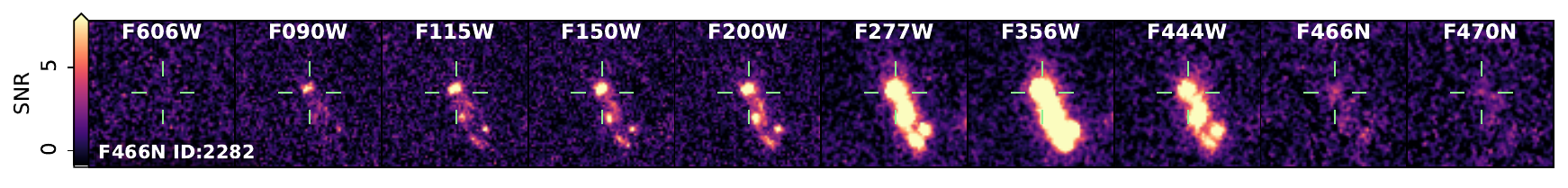}\vspace{-0.4cm}
  \includegraphics[width=\textwidth]{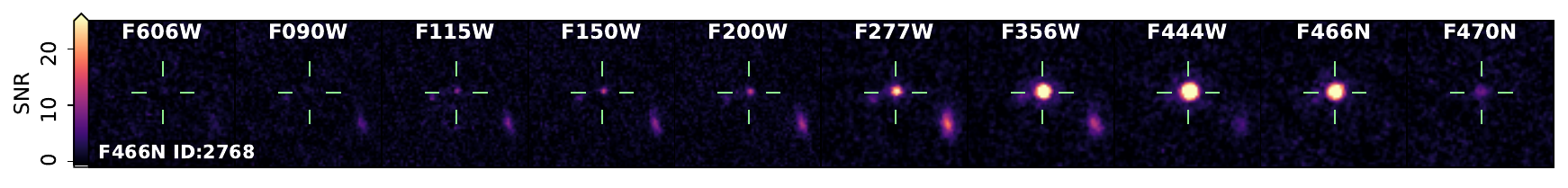}\vspace{-0.4cm}
  \includegraphics[width=\textwidth]{images/cutouts/row/f466n_excess_2976_cutouts_row.pdf}\vspace{-0.4cm}
  \includegraphics[width=\textwidth]{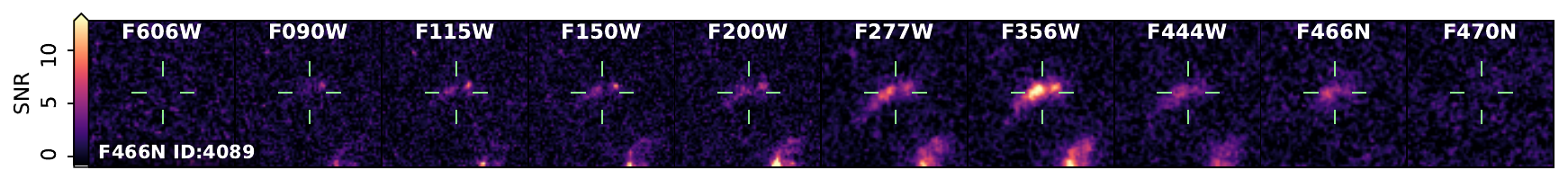}\vspace{-0.4cm}
  \includegraphics[width=\textwidth]{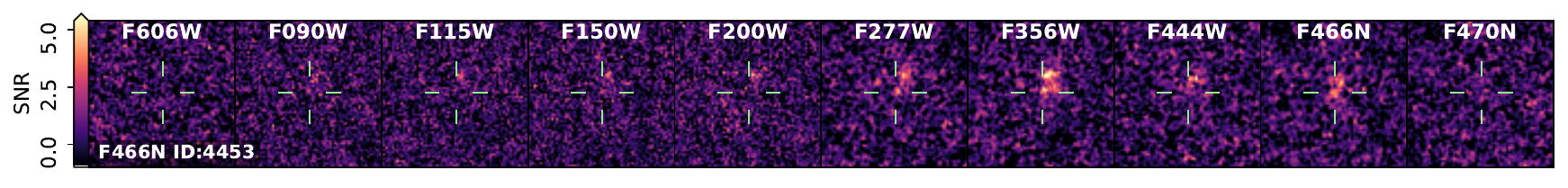}\vspace{-0.4cm}
  \includegraphics[width=\textwidth]{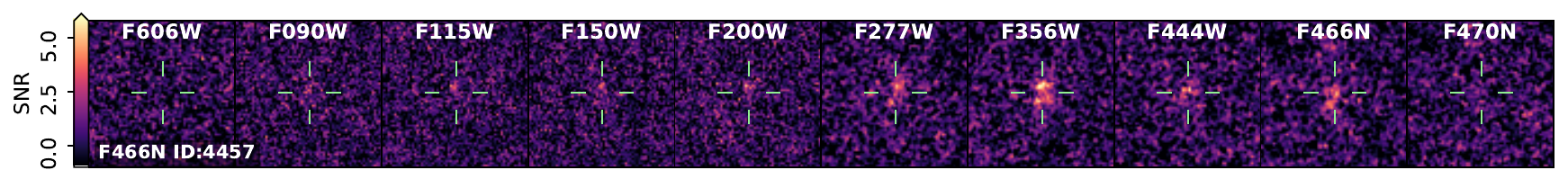}\vspace{-0.4cm}
  \includegraphics[width=\textwidth]{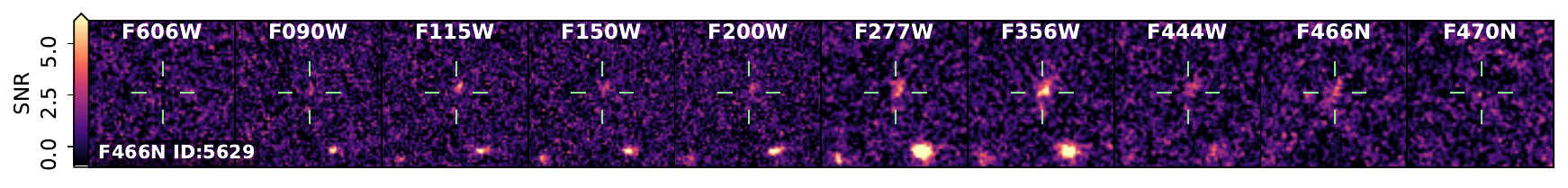}\vspace{-0.4cm}
  \includegraphics[width=\textwidth]{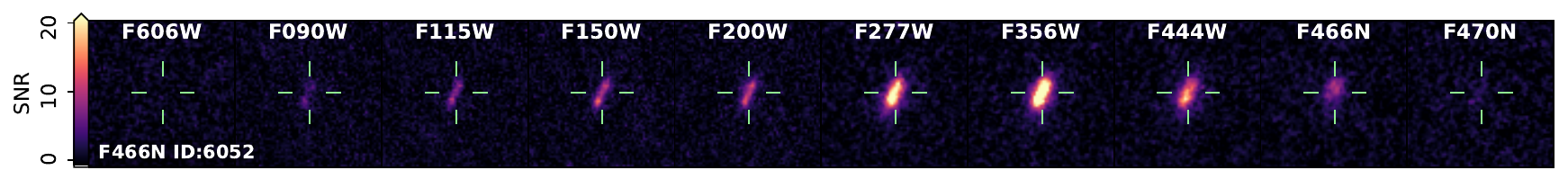}\vspace{-0.4cm}
  \includegraphics[width=\textwidth]{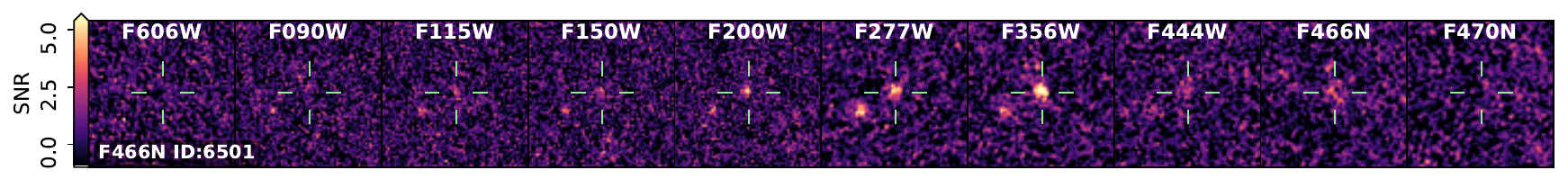}\vspace{-0.4cm}

\caption{Cutouts of the multi-wavelength imaging centred on the F466N and F470N detected sources passing the `excess source' (Section~\ref{sec:narrowband_cat}), H$\alpha$ photo-$z$ (Section~\ref{sec:halpha_sample}) and visual inspection (Section~\ref{sec:halpha_sample}) criteria. The colour bar again shows the range of SNR in the image cutouts for the following filters: F606W, F090W, F115W, F150W, F200W, F277W, F356W, F444W, F466N and F470N.}
\label{fig:halpha_sample_cutouts_v1}
\end{figure}

\begin{figure}\ContinuedFloat
\centering
  \includegraphics[width=\textwidth]{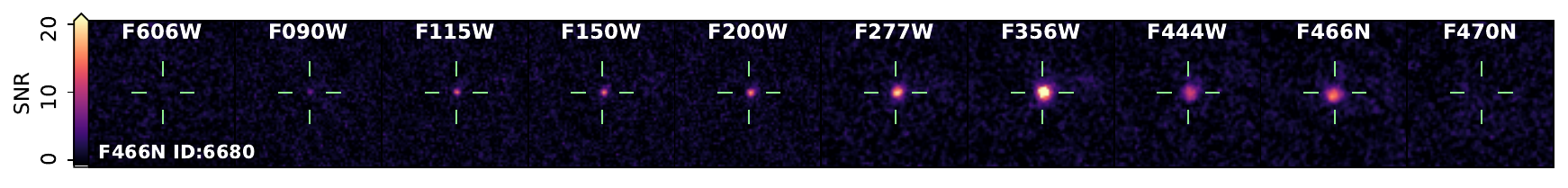}\vspace{-0.4cm}
  \includegraphics[width=\textwidth]{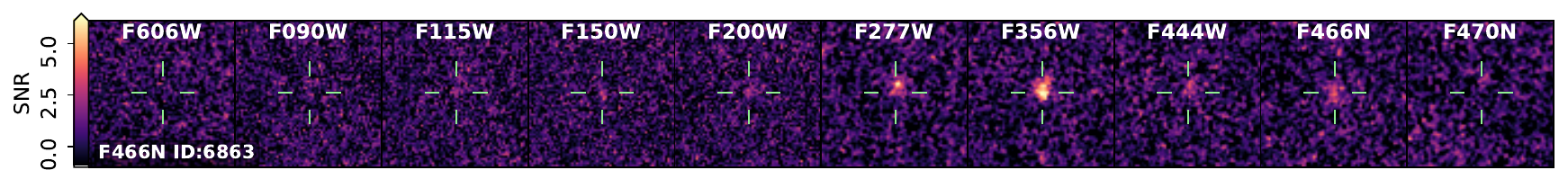}\vspace{-0.4cm}
  \includegraphics[width=\textwidth]{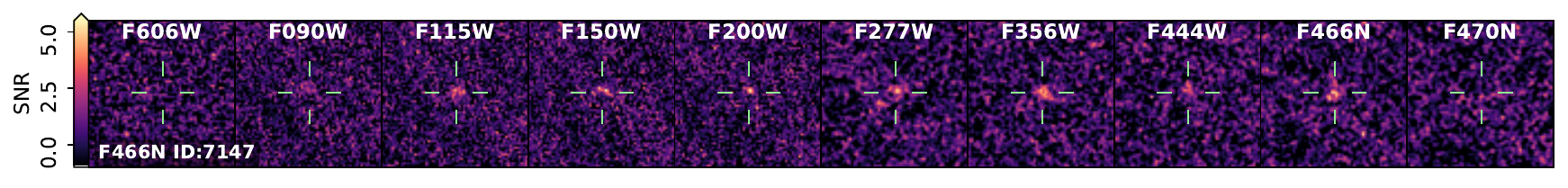}\vspace{-0.4cm}
  \includegraphics[width=\textwidth]{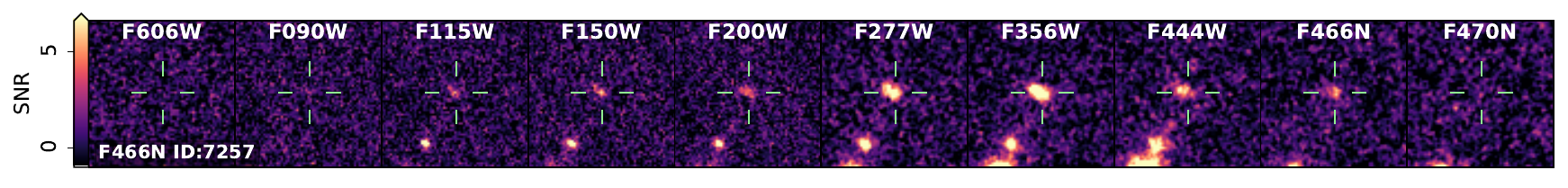}\vspace{-0.4cm}
  \includegraphics[width=\textwidth]{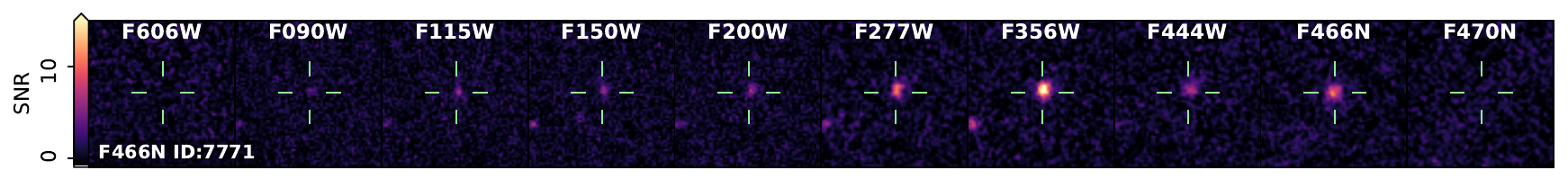}\vspace{-0.4cm}
  \includegraphics[width=\textwidth]{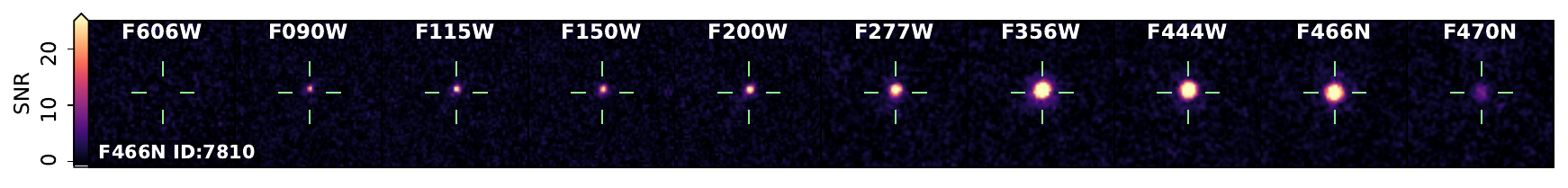}\vspace{-0.4cm}
  \includegraphics[width=\textwidth]{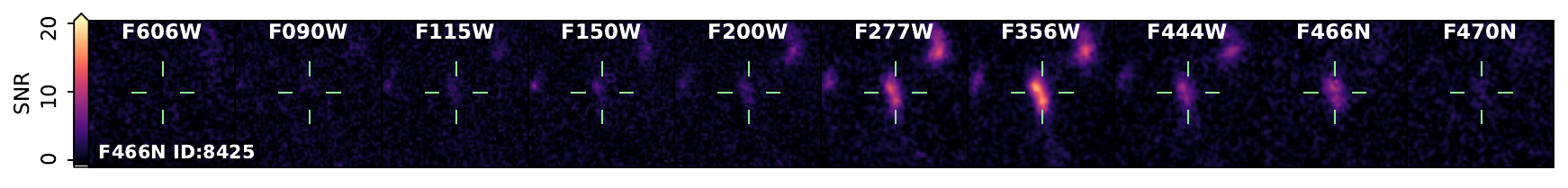}\vspace{-0.4cm}
  \includegraphics[width=\textwidth]{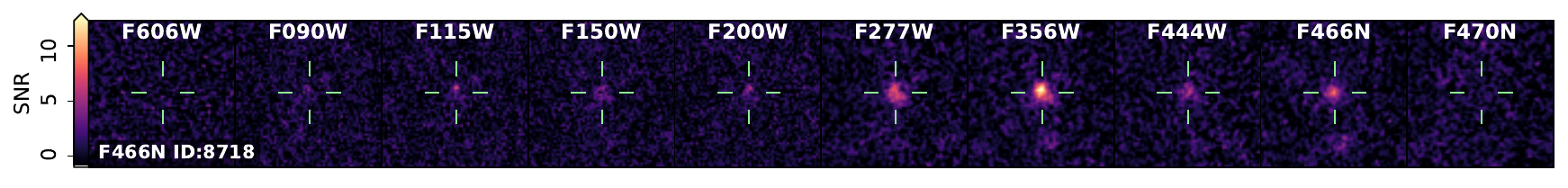}\vspace{-0.4cm}
      \includegraphics[width=\textwidth]{images/cutouts/row/f466n_excess_9123_cutouts_row.pdf}\vspace{-0.4cm}
  \includegraphics[width=\textwidth]{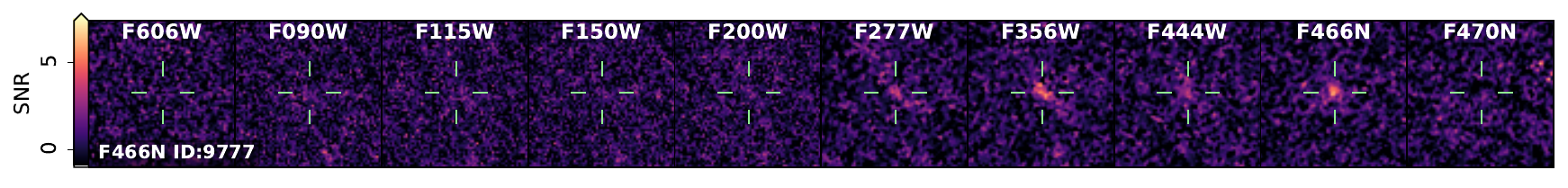}\vspace{-0.4cm}
  \includegraphics[width=\textwidth]{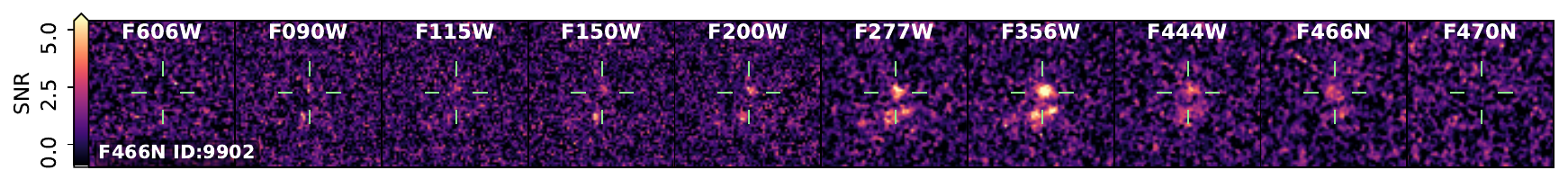}\vspace{-0.4cm}
  \includegraphics[width=\textwidth]{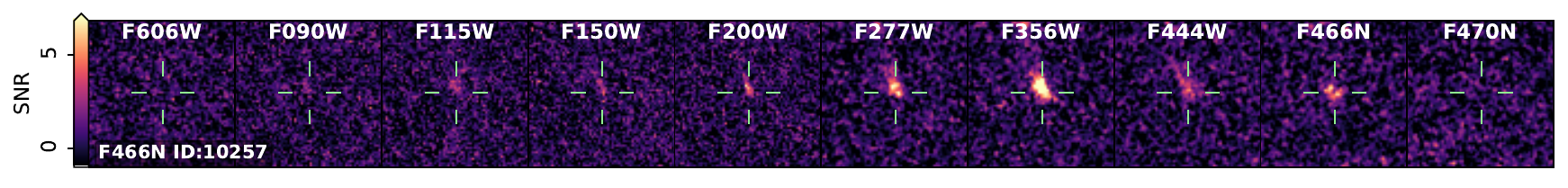}\vspace{-0.4cm}
  \includegraphics[width=\textwidth]{images/cutouts/row/f466n_excess_10983_cutouts_row.pdf}\vspace{-0.4cm}
\caption{Continued.}
\end{figure}

\begin{figure}\ContinuedFloat
\centering

  \includegraphics[width=\textwidth]{images/cutouts/row/f466n_excess_11272_cutouts_row.pdf}\vspace{-0.4cm}
  \includegraphics[width=\textwidth]{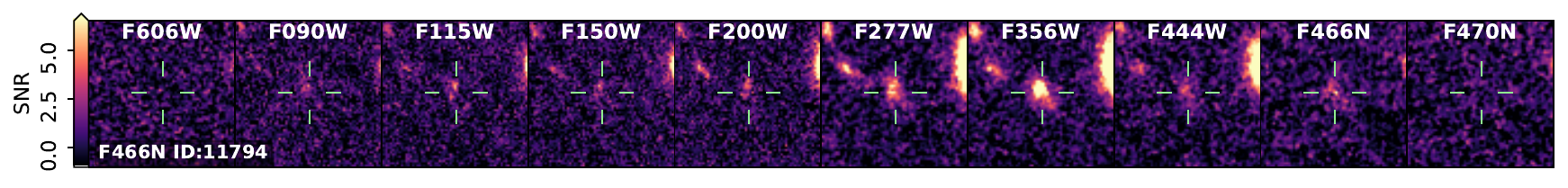}\vspace{-0.4cm}
  \includegraphics[width=\textwidth]{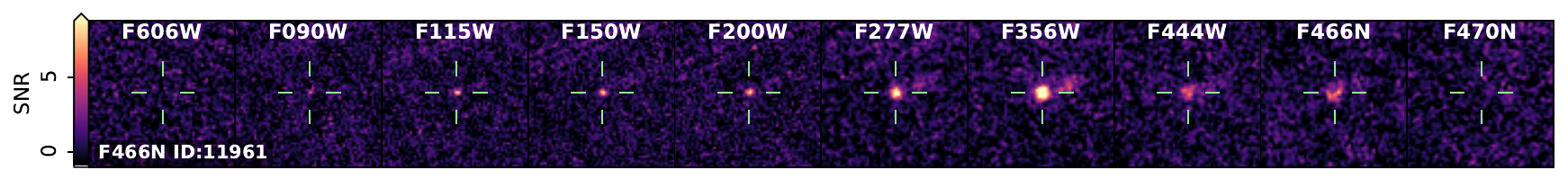}\vspace{-0.4cm}
  \includegraphics[width=\textwidth]{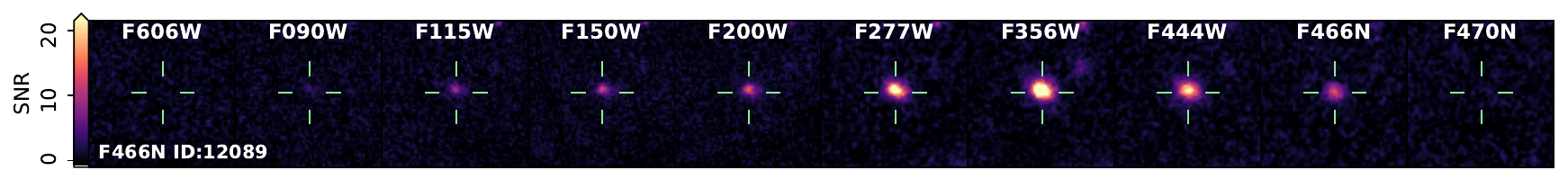}\vspace{-0.4cm}
  \includegraphics[width=\textwidth]{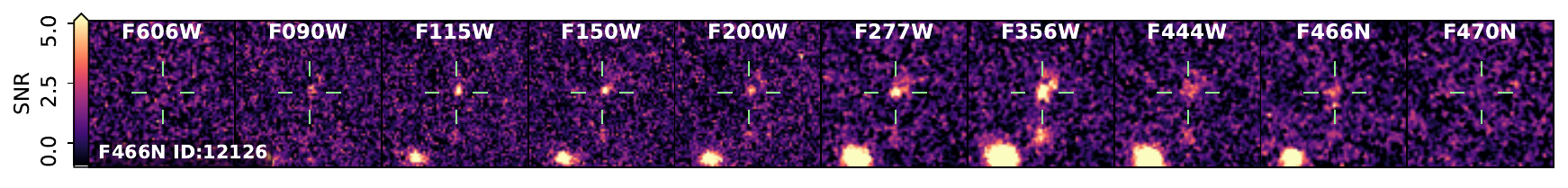}\vspace{-0.4cm}
  \includegraphics[width=\textwidth]{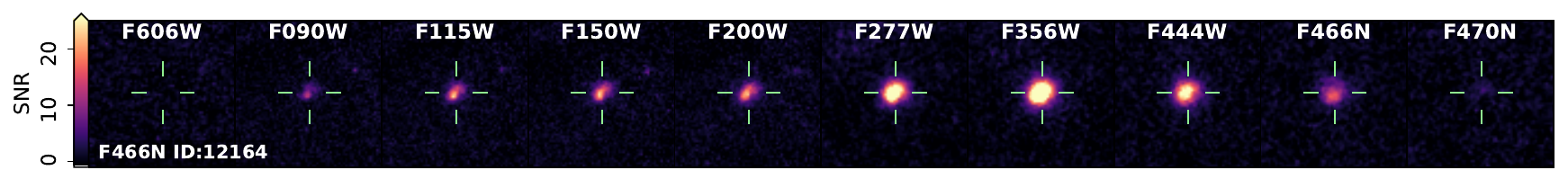}\vspace{-0.4cm}
  \includegraphics[width=\textwidth]{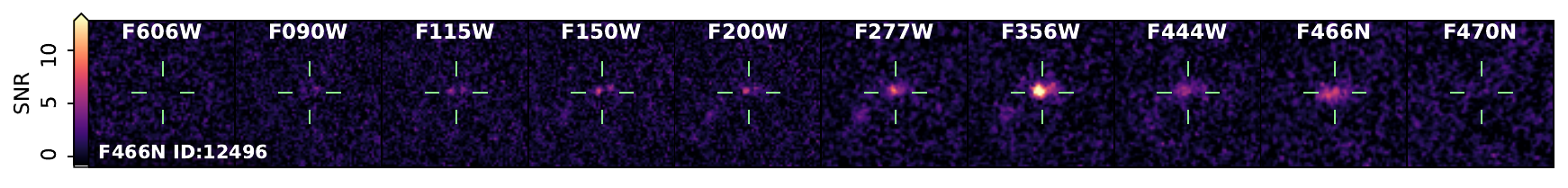}\vspace{-0.4cm}
    \includegraphics[width=\textwidth]{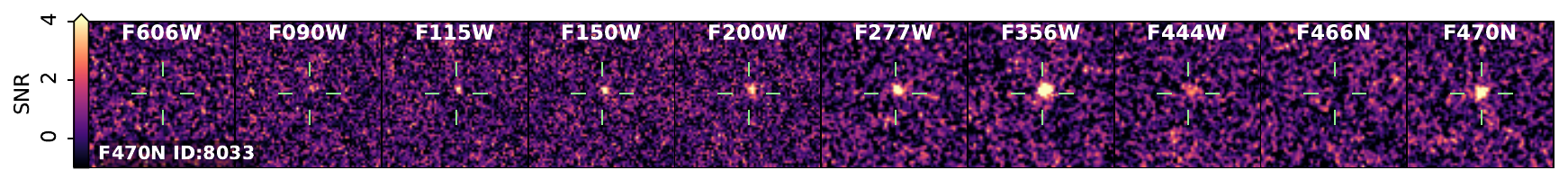}\vspace{-0.4cm}
  \includegraphics[width=\textwidth]{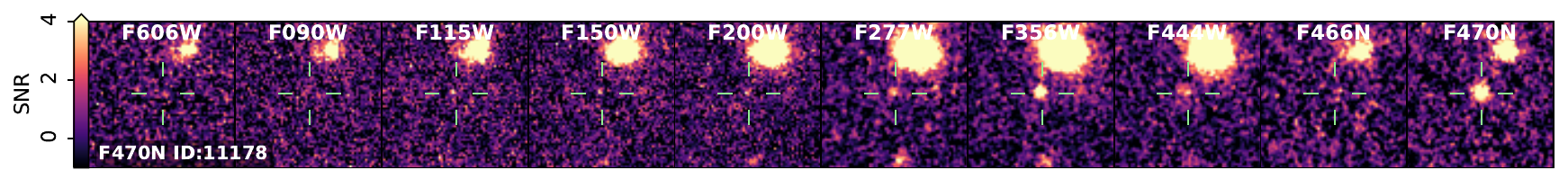}\vspace{-0.4cm}
    \includegraphics[width=\textwidth]{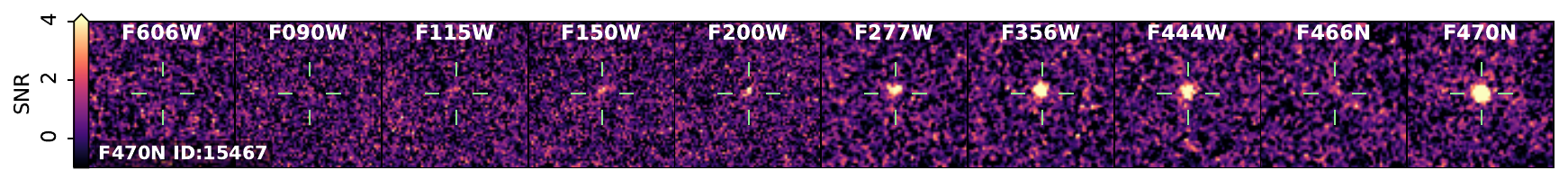}\vspace{-0.4cm}
  \includegraphics[width=\textwidth]{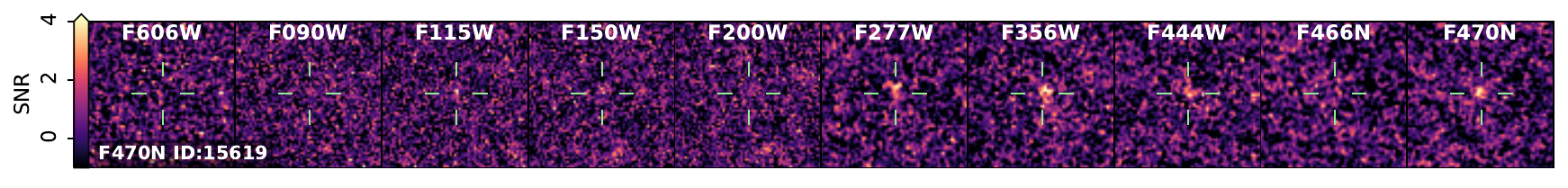}\vspace{-0.4cm}
    \includegraphics[width=\textwidth]{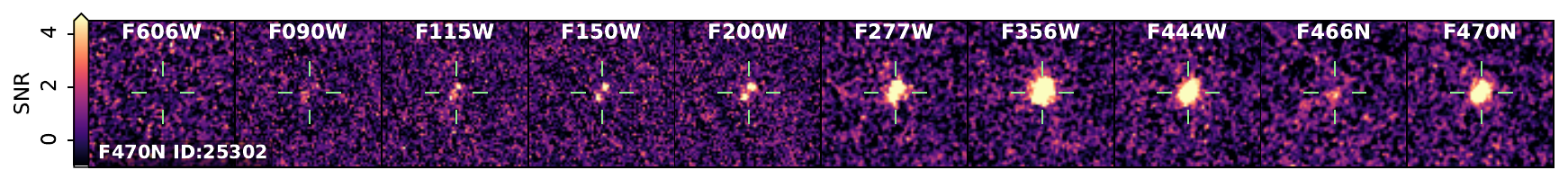}\vspace{-0.4cm}
\caption{Continued}
\end{figure}

\section{Testing the impact of priors on the derived SED parameters}
\label{app:posteriors}

\begin{figure*}
\centering
\begin{subfigure}{0.5\textwidth}
    \includegraphics[width=\linewidth]{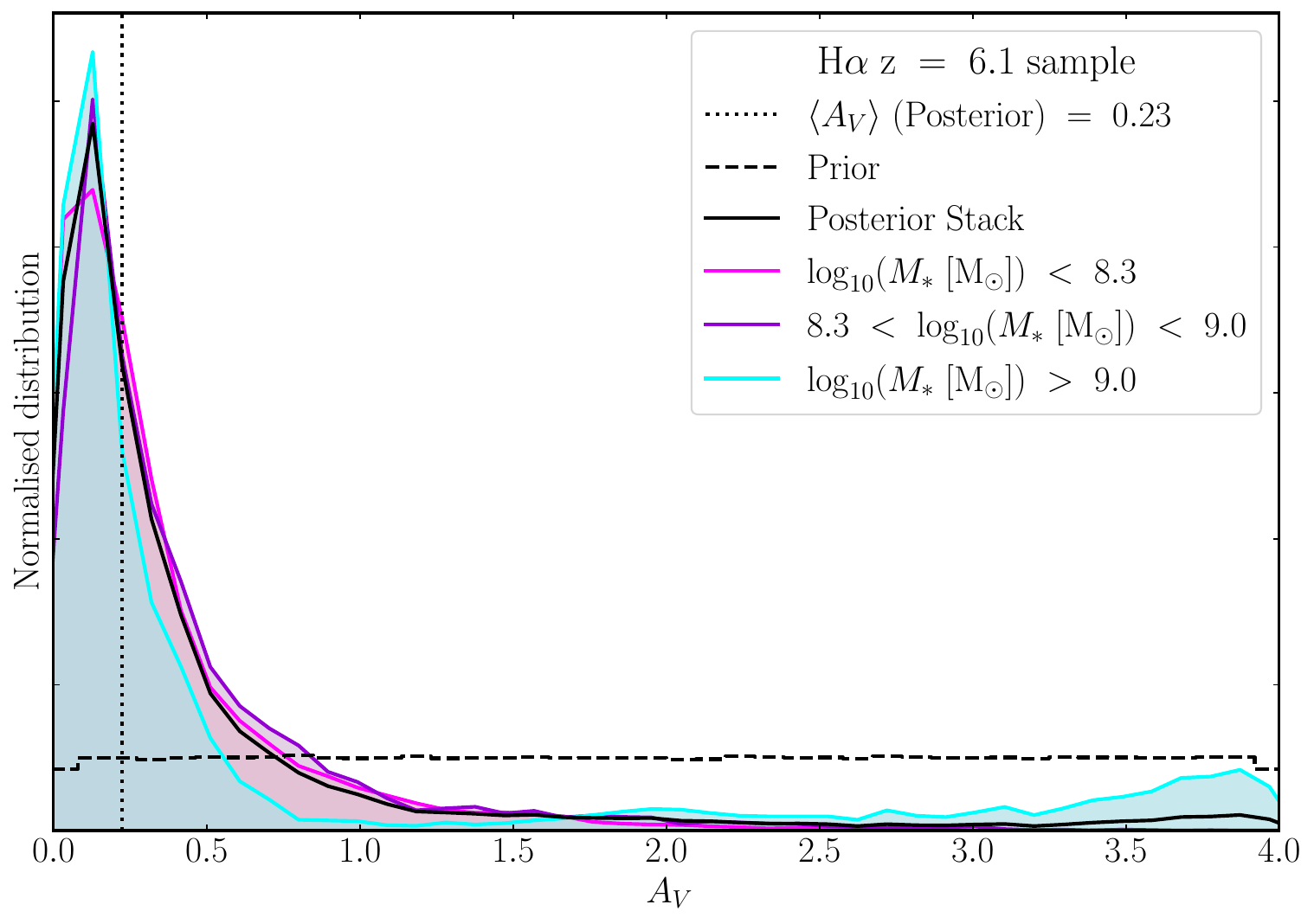}
\end{subfigure}\hfil
\begin{subfigure}{0.5\textwidth}
    \includegraphics[width=\linewidth]{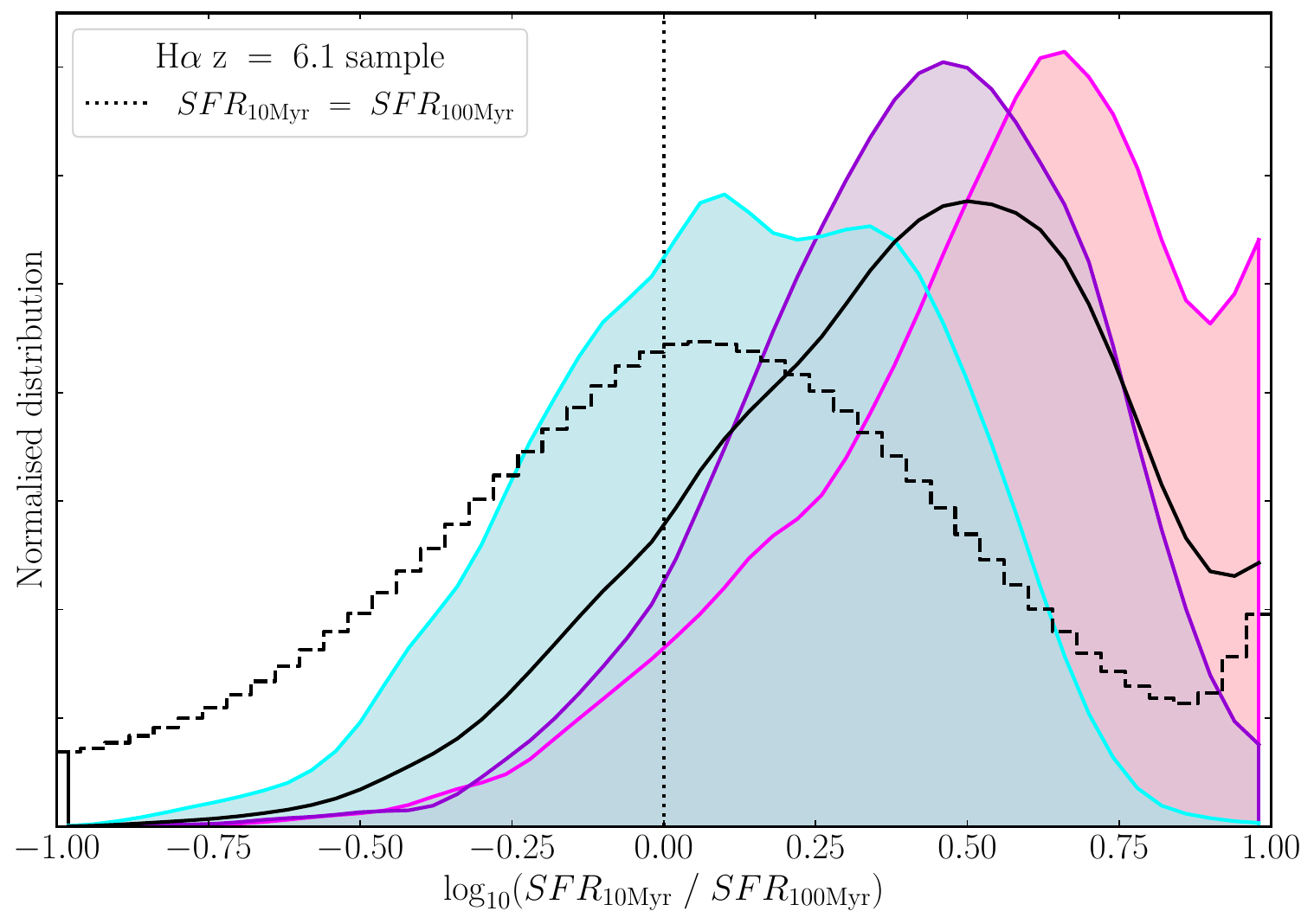}
\end{subfigure}\hfil

\medskip
\begin{subfigure}{0.5\textwidth}
  \includegraphics[width=\linewidth]{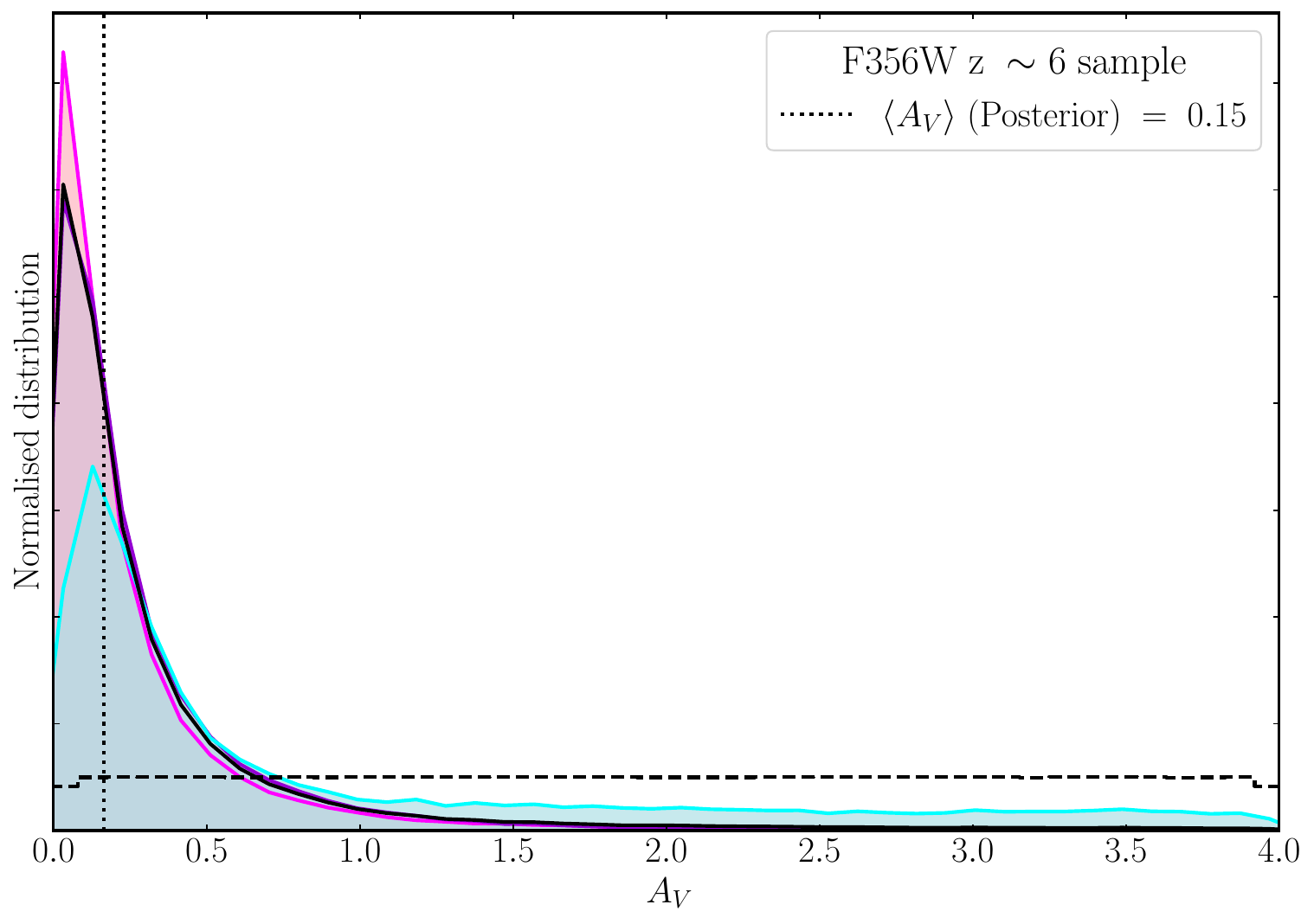}
\end{subfigure}\hfil
\begin{subfigure}{0.5\textwidth}
  \includegraphics[width=\linewidth]{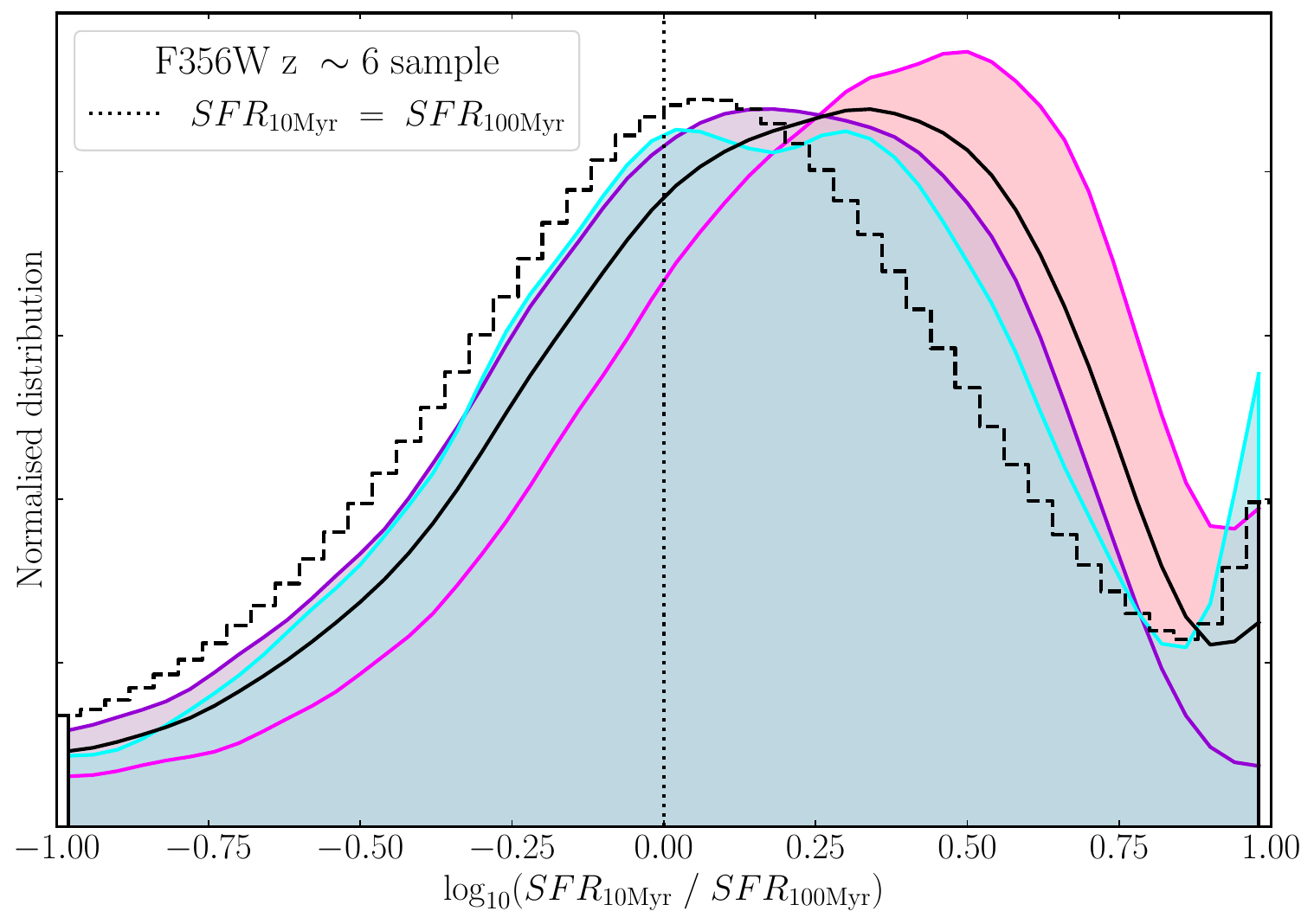}
\end{subfigure}\hfil

\caption{Left panels: Stacked posterior distributions from our \textsc{Bagpipes} SED fitting results for the $V$-band attenutation $A_{V}$ for the full sample of candidate H$\alpha$ emission line galaxies (top panel) and for the F356W-selected $z \sim 6$ sample ((lower panel). In both panels the full sample distribution is shown (solid black line) along with this sample separated by stellar mass (colour lines - see legend) and the distribution generated directly from the priors (black dashed line). The vertical black dotted line represents the median of the stacked posterior distribution for the V-band attenuation $A_{V}$. Right panels: The same as the left panels but for the ratio of the SFRs measured over 10Myr and 100Myr respectively: $\rm{SFR_{10}/SFR_{100}}$. This time, the vertical black dotted line represents $\rm{SFR_{10}/SFR_{100}}$ = 1 (showing continuous SFR over 100 Myr).}

\label{fig:sed_post_dist}

\end{figure*}

To ensure that our SED fitting results were not being driven by the choice of input priors, we considered the stacked and normalised posterior distributions for the dust attenuation, $A_V$, and for the SFR ratio, $\rm{SFR_{10}/SFR_{100}}$. These are shown in Fig. \ref{fig:sed_post_dist}.  

Specifically, we obtained the stacked and normalised posterior distributions for $A_V$ and for the SFR ratio ($\rm{SFR_{10}/SFR_{100}}$) from our H$\alpha$ emission line galaxy sample, and then also separated this into the following stellar mass bins (defined using the median of the stellar mass posterior distribution): i) $\log_{10}(M_{\star} \ [\rm{M_{\odot}}])$ $<$ 8.3; ii) 8.3 $<$ $\log_{10}(M_{\star} \ [\rm{M_{\odot}}])$ $<$ 9.0; and iii) $\log_{10}(M_{\star} \ [\rm{M_{\odot}}])$ $>$ 9.0. These mass ranges were chosen to capture roughly the same number of H$\alpha$ candidates in each mass bin (13, 13 and 9 galaxies respectively) to limit biases due to low number counts. The results can be found in the upper panels of Fig. \ref{fig:sed_post_dist} where we plot the distributions in $A_V$ and in $\log_{10}(SFR_{10}/SFR_{100})$. 

For $A_V$, we observe a posterior distribution which peaks sharply around $A_V \approx 0.2$, with a median $A_V = 0.23$. This is substantially different from the input prior distribution which was flat over the following range: $0 < A_V < 4$. For the SFR ratio, we observe the peak of the stacked posterior distribution to be above unity. This tells us that our sample of sources exhibit, on average, elevated star formation in the last 10 Myr with respect to the previous 100 Myr, with median $\rm{SFR_{10}/SFR_{100}}$ $\sim$ 2.7. When we separated the sample out by stellar mass, we found that the lowest stellar-mass bin exhibits a posterior distribution for $\rm{SFR_{10}/SFR_{100}}$ systematically higher than for the higher stellar mass bins and the average of the full sample. In all cases, but especially at lower masses where the input data is typically of lower SNR and therefore more likely to be influenced by choice of prior, the derived SFR ratio distributions differed significantly from the input prior distribution. 

We repeated the same analysis but for our F356W-selected $z$ $\sim$ 6 sample; these results are shown in the lower panels of Fig. \ref{fig:sed_post_dist}. From this, we found that the stacked posterior distributions for $A_V$ were largely similar to the H$\alpha$ sample, but with a lower median value of $A_V = 0.15$. The SFR ratio distributions were broader in shape (signifying the increased difficulty in constraining recent SFR without the narrow-band H$\alpha$ measurement) with a lower median of $\rm{SFR_{10}/SFR_{100}}$ $\sim$ 1.3 (closer to a constant SFR). The lowest stellar mass bin is still peaked towards higher $\rm{SFR_{10}/SFR_{100}}$ but much less so than what is observed for our H$\alpha$ sample. 

Overall, we concluded from this analysis that our SED fitting was able to derive reliable output parameters that were not driven by the input priors. The trends observed in these parameters, and the differences between the H$\alpha$ and F356W-selected samples, were further explored in the main text. 

\bsp	
\label{lastpage}
\end{document}